# Multi-Carrier Thermal Transport in Electronic and Energy Conversion Devices

Te-Huan Liu[1], Tianyu Wang[1], Jun Zhou[2,a], Xin Qian[1,b], and Ronggui Yang[1,3,c]

[1]School of Energy and Power Engineering,

Huazhong University of Science and Technology, Wuhan 430074, China

[2]School of Physics and Technology, Nanjing Normal University, Nanjing 210023, China

[3]College of Engineering, Peking University, Beijing 100871, China

**ABSTRACT**

Nonequilibrium multi-carrier thermal transport is essential for both scientific research and technological applications in electronic, spintronic, and energy conversion devices. This article reviews the fundamentals of phonon, electron, spin, and ion transport driven by temperature gradients in solid-state and soft condensed matters, and the microscopic interactions between energy/charge carriers that can be leveraged for manipulating electrical and thermal transport in energy conversion devices, such as electron-phonon coupling, spin-phonon interaction, and ion-solvent interactions, *etc*. In coupled electron-phonon transport, we discuss the basics of electron-phonon interactions and their effects on phonon dynamics, thermalization, and nonequilibrium thermal transport. For the phonon-spin interaction, nonequilibrium transport formulation is introduced first, followed by the physics of spin thermoelectric effect and strategies to manipulate them. Contributions to thermal conductivity from magnons as heat carriers are also reviewed. For coupled transport of heat and ions/molecules, we highlight the importance of local molecular configurations that determine the magnitude of the electrochemical gradient, which is the key to improving the efficiency of low-grade heat energy conversion.

Authors to whom correspondence should be addressed:

a) zhoujunzhou@njnu.edu.cn; b) xinqian21@hust.edu.cn; c) ronggui@pku.edu.cn

TABLE OF CONTENTS

# I. INTRODUCTION

Multi-carrier thermal transport is the key process in a wide range of electronic and energy devices, determining the heat generation and dissipation in electronics,[1-4] the performance of information processing and quantum computing,[5-8] safety and efficiency of thermal energy storage and conversions.[9-11] Understanding and tailoring the multi-carrier interaction have also been a long-standing challenge in physical sciences. According to nonequilibrium thermodynamics, fluxes of various extensive quantities, such as charge currents, mass fluxes, spin currents, energy flux *etc.* can be induced when exerting a generalized driving force upon the system.[12,13] If the system where the generalized force can be treated as a perturbation, the relation between generalized forces and fluxes can be written in the linear response form:

$$J_i = \sum_{ij} L_{ij} f_j . \qquad (1.1)$$

where $i$ and $j$ denote species of fluxes and thermodynamic forces, and $L_{ij}$ is the corresponding Onsager linear transport coefficient correlating the flux $J_i$ with the force $f_j$. The generalized forces $f$ can include not only mechanical forces such as electromagnetic fields, pressure gradients, but also thermodynamic forces such as gradients in temperature, concentrations (chemical potential), *etc.* The diagonal transport coefficients ($L_{ii}$) features the well known Fourier's law, Ohm's law, and Fick's law of heat conduction, electrical conduction, and diffusion, correspondingly. The famous Onsager's reciprocal relation dictates that off-diagonal components are symmetrical $L_{ij} = L_{ji}$ in a system with time-reversal symmetry.[14] For example, the famous Kelvin's relation correlating the Peltier coefficient and the Seebeck coefficient $\Pi = ST$ is a manifestation of Onsager reciprocity in thermoelectrics.[15] Similar correlation between cross-transport coefficients as a result of the Onsager reciprocity also exists in coupled transport of phonon and spin,[16] and ionic thermodiffusion,[17,18] and thermo-osmosis.[19]

While Onsager's framework provides a phenomenological and macroscopic framework for studying coupled transport processes, recent years have witnessed significant advancements in studying the microscopic dynamics of multicarrier thermal transport. In solid-state material systems, electrons and phonons are the most well-studied energy and charge carriers. Owing to the development in *ab initio* modeling, electron-phonon interaction (EPI) and its effect on transport properties can be predicted with high fidelity.[1,20,21] Ultrafast pump-probe measurements also provide femtosecond to picosecond resolution for directly observing scattering of electrons and phonons, and the relaxation dynamics to local thermal equilibrium.[2,22,23] In addition to electron-phonon coupling, coupled transport of phonons and spins (quantized spin wave) received increasing interest in recent years, due to the technological relevance in the field of quantum magnonics for information processing and quantum computing, where the control of magnon-phonon interaction to preserve the magnon coherence over extended length and time scales is essential.[5-7] Similar to thermoelectric effects originating from coupled electron-phonon transport, magnon-phonon interaction also gives rise to phenomena such as the spin Seebeck effect[24-27] and the spin caloric effects.[28,29] Recently, phonon-activated spin Seebeck effect has been discovered in nonmagnetic material with chiral structures under thermal nonequilibrium, which provides opportunities to leverage often undesired thermal excitation and fluctuations for information processing.[30] In addition to multi-carrier thermal transport in solid-state systems, coupled transport of heat and ions/molecules in soft condensed matter has been much less studied, due to the intrinsically complicated atomic structures and interatomic forces between the molecules and ions. Despite the long history of discovering the thermodiffusion and thermo-osmosis effects of ions and molecules,[31,32] it was not until recent few years the research interests were revitalized, due to the opportunities in harvesting low-grade heat using ionic thermoelectric and thermo-osmotic energy conversion devices.[9,10]

In this review, we focus on discussing the fundamentals and methods for tailoring multi-carrier cross-transport phenomena associated with temperature gradients. While there have been several reviews on transport of heat carriers, focusing on first-principles modeling or experimental methodologies,[4,33,34] or specific material systems, such as inorganic materials,[35] metals, polymers,[36] magnetic materials,[28,29] low-dimensional materials,[37,38] and even quantum materials,[39] there yet exists a comprehensive discussion on multi-carrier transport with a parallel treatment of phonons, electrons, magnons, ions, and molecules in both solid-state materials and soft condensed matters. This article will summarize recent advancements in studying the effect of multi-carrier thermal transport and perspectives for future developments. This work is organized as follows. In section II, the coupled electron and phonon transport driven by temperature gradient is reviewed. Section III reviews the spin current driven by temperature gradient and spin-dependent thermal conductivity where spin current and energy flow carried by both conduction electrons and magnons are considered. In section IV, the ionic thermoelectric effect which describes chemical transport driven by temperature gradient is reviewed. Section V concludes this review.

## II. COUPLED ELECTRON AND PHONON TRANSPORT

The interaction between electrons and phonons, known as EPI, is one of the cornerstones of condensed matter physics. The advancement has been closely linked with the historical progress of quantum mechanics and solid-state physics. The origin of the EPI concept dates back to 1921, when Bloch introduced the interaction of the "lattice elastic wave", now known as phonons, with electrons into the Schrödinger equation in search for a dropped item corresponding to the coupling between electrons and lattice vibration under the Born-Oppenheimer approximation.[40] The initial Bloch theory was followed by a large body of subsequent works that elucidate the effects of EPI on a rich variety of physical phenomena, including superconductivity,[41-43] temperature dependence of electrical and thermal conductivities,[44-46] *etc*. Numerous classic textbooks and review articles provide detailed descriptions of classic EPI models, and the research scope encompasses metals,[45,47-49] semiconductors,[44,49,50] ionic crystals,[51,52] among others. In particular, this section delves into the thermal and thermoelectric properties associated with EPI, highlighting their applications in fields such as electronic thermal management and thermoelectric. In Section II.A, we discuss the basics of EPIs, followed by an exploration of their influence on thermal transport. In Section II.B, we review the role of long-range interactions in modulating both EPIs and phonon transport. Section II.C delves into phonon renormalization, examining how EPI alters phonon dynamics and material properties. In Section II.D, we analyze the complex phenomena of nonequilibrium electron-phonon transport, focusing on their implications for material performance. Finally, Section II.E discusses the persistent challenges and emerging research opportunities in the field of long-standing EPI.

### A. Effect of electron-phonon interaction on thermal transport

#### 1. Basics of electron-phonon interaction

EPI can be conceptualized as the process in which electrons traverse the lattice's periodic potential field, encountering perturbations that transfer energy and momentum to them, often accompanied by phonon emission or absorption. Phonon, the quantized mode of lattice waves, is thus influenced by their interactions with electrons. This mutual influence results in the redistribution and modulation of phonon energy. The mathematical description of EPI is represented in a Hamiltonian model, comprising three components: the electron Hamiltonian ($\hat{H}_e$), the phonon Hamiltonian ($\hat{H}_p$), and the electron-phonon coupling Hamiltonian ($\hat{H}_{ep}$).

$$\hat{H} = \hat{H}_e + \hat{H}_p + \hat{H}_{ep}$$

$$= \sum_{n\mathbf{k}} \varepsilon_{n\mathbf{k}} \hat{c}_{n\mathbf{k}}^\dagger \hat{c}_{n\mathbf{k}} + \sum_{\mathbf{q}v} \hbar\omega_{\mathbf{q}v}\left(\hat{a}_{\mathbf{q}v}^\dagger \hat{a}_{\mathbf{q}v} + \frac{1}{2}\right) + N_p^{-\frac{1}{2}} \sum_{\mathbf{k},\mathbf{q},mnv} g_{mnv}(\mathbf{k},\mathbf{q}) \hat{c}_{m\mathbf{k}+\mathbf{q}}^\dagger \hat{c}_{n\mathbf{k}} (\hat{a}_{\mathbf{q}v} + \hat{a}_{-\mathbf{q}v}^\dagger), \qquad (2.1)$$

where the $\varepsilon_{n\mathbf{k}}$ is the energy of an electron at wavevector $\mathbf{k}$ in band index $n$, $\hbar\omega_{\mathbf{q}v}$ is the energy of lattice vibration at wavevector $\mathbf{q}$ in branch index $v$, and $N_p$ is the number of unit cells under Born–von Kármán boundary condition. $\hat{c}_{n\mathbf{k}}^\dagger$ and $\hat{c}_{n\mathbf{k}}$ ($\hat{a}_{\mathbf{q}v}^\dagger$ and $\hat{a}_{\mathbf{q}v}$) are the operators of the creation and annihilation of electrons (phonons). $g_{mnv}(\mathbf{k},\mathbf{q})$ is the EPI matrix element that evaluates the strength of interaction between electrons and phonons. The fundamental nature of the EPI matrix element $g_{mnv}(\mathbf{k},\mathbf{q})$ lies in the fact that the potential felt by electrons is disturbed when the atoms move around due to lattice vibrations.[40,44,53] The different types of scattering mechanisms introduce various formalisms of the $g_{mnv}(\mathbf{k},\mathbf{q})$. In non-polar materials, EPI is predominantly governed by deformation potential scattering. This mechanism, primarily a short-range interaction, was initially described by Bardeen and Shockley,[54] where the strength of EPI is related to local variations in band edges induced by strain. In this context, long-wavelength acoustic phonons cause dilation within the crystal structure, altering atomic spacings and creating a perturbation potential that shifts electron band energies, thus facilitating electron coupling, described as acoustic deformation potential scattering, for which the EPI matrix element is denoted as $g_{mnv}^{\mathrm{ADP}}(\mathbf{k},\mathbf{q})$. Extending this theory, Herring and Vogt[55] applied the deformation potential concept to optical phonons, namely optical deformation potential scattering, for which the EPI matrix element is denoted as $g_{mnv}^{\mathrm{ODP}}(\mathbf{k},\mathbf{q})$. With long-wavelength optical phonons, changes in interatomic distances among base atoms directly modify the surrounding lattice potential, acting as a scattering mechanism for electrons. However, in polar materials, the EPI can be driven by the generation of internal long-range electric fields due to phonon activities.[50,55,56] Long-range EPI refers to the coupling between electrons and phonons that occurs over larger distances within a material, beyond the immediate vicinity of the electrons. Interactions among ions generate electric dipoles that engage in long-range interactions with both acoustic and optical phonons. The interactions with acoustic phonons are classified as piezoelectric scatterings with their EPI matrix denoted by $g_{mnv}^{\mathrm{PZ}}(\mathbf{k},\mathbf{q})$, while those with optical phonons are known as polar optical phonon scatterings with their EPI matrix denoted by $g_{mnv}^{\mathrm{POP}}(\mathbf{k},\mathbf{q})$.[56] The different types of scattering mechanisms introduce significant mathematical complexity into the formalism of the EPI matrix element. There have been significant efforts in accurately determining the EPI matrix element over the past many decades.

The development of the density-functional theory (DFT) undoubtedly opens a new chapter in the analysis of EPI. Within the DFT formalism, the advent of the Kohn-Sham equation[57] and pseudopotential method[58,59] enable accurate calculations of the electronic structure, lattice dynamic, and the strength of electron-phonon coupling. This advancement allows for precise parametrization of the complex components $g_{mnv}(\mathbf{k},\mathbf{q})$. The EPI matrix element $g_{mnv}(\mathbf{k},\mathbf{q})$ can be described by the Kohn-Sham effective potential as

$$g_{mnv}(\mathbf{k},\mathbf{q}) = \langle u_{m\mathbf{k}+\mathbf{q}} | \Delta_{\mathbf{q}v} v^{\mathrm{KS}} | u_{n\mathbf{k}} \rangle_{\mathrm{uc}}, \qquad (2.2)$$

in which $u_{n\mathbf{k}}$ and $u_{m\mathbf{k}+\mathbf{q}}$ are the Bloch functions underlying Kohn-Sham electron wave functions, $\Delta_{\mathbf{q}v} v^{\mathrm{KS}}$ is a lattice-periodic function, which indicates the variations of the Kohn-Sham self-consistent potential of electrons induced by phonons. The Kohn-Sham effective potential can be calculated by density-functional perturbation theory (DFPT)[60,61] and the frozen-phonon method.[62,63] Thus, the EPI matrix element $g_{mnv}(\mathbf{k},\mathbf{q})$ can be accurate parametrizations within the DFT theoretical framework. While these matrix elements have been employed to evaluate the impact of EPI on electronic transport properties,[64,65] achieving convergence in EPI calculations often requires a very dense phonon mesh, which can be computationally intensive. Recent advancements in interpolation schemes utilizing maximally localized Wannier functions[66,67] have enabled EPI calculations on much finer meshes, thereby facilitating more precise and efficient analyses.

After the EPI matrix element $g_{mn\nu}(\mathbf{k},\mathbf{q})$ is determined, the Green's function method enables the calculation of phonon self-energy due to EPIs, which can be written as

$$\Pi_\nu(\mathbf{q}, \omega_{\mathbf{q}\nu}) = 2 \sum_{m,n} \int \frac{d\mathbf{k}}{\Omega_{\text{BZ}}} |g_{mn\nu}(\mathbf{k},\mathbf{q})|^2 \frac{f_{m\mathbf{k}+\mathbf{q}} - f_{n\mathbf{k}}}{\varepsilon_{m\mathbf{k}+\mathbf{q}} - \varepsilon_{n\mathbf{k}} - \hbar\omega_{\mathbf{q}\nu} - i\delta}. \quad (2.3)$$

$f_{n\mathbf{k}}$ is the distribution function for electrons. $\Omega_{\text{BZ}}$ is the volume of the first Brillouin zone, and $\delta$ is an infinitesimal regularization factor. On the one hand, the imaginary part of phonon self-energy, $\text{Im}\Pi_\nu(\mathbf{q}, \omega_{\mathbf{q}\nu})$, is directly related to the scattering rates of phonons by electrons and can be given by:

$$\frac{1}{\tau^{ep}_{\mathbf{q}\nu}} = \frac{2\text{Im}\Pi_\nu(\mathbf{q},\omega_{\mathbf{q}\nu})}{\hbar} = \frac{4\pi}{\hbar} \sum_{mn,\mathbf{k}} |g_{mn\nu}(\mathbf{k},\mathbf{q})|^2 (f_{n\mathbf{k}} - f_{m\mathbf{k}+\mathbf{q}}) \times \delta(\varepsilon_{m\mathbf{k}+\mathbf{q}} - \varepsilon_{n\mathbf{k}} - \hbar\omega_{\mathbf{q}\nu}). \quad (2.4)$$

The Dirac $\delta$-function enforces the conservation conditions for both energy and momentum. Sections II. A and II. B of this paper primarily examines the influence of scattering rates on phonon properties, focusing on both short-range and long-range interactions, respectively.

On the other hand, the real part of the phonon self-energy $\text{Re}\,\Pi_\nu(\mathbf{q},\omega_{\mathbf{q}\nu})$ is associated with renormalized phonon frequency $\Omega_{\mathbf{q}\nu}$, which can be written as[45,68]

$$\Omega_{\mathbf{q}\nu}^2 = \omega_{\mathbf{q}\nu}^2 + 2\omega_{\mathbf{q}\nu}\,\text{Re}\,\Pi_\nu(\mathbf{q},\omega_{\mathbf{q}\nu}). \quad (2.5)$$

The influence of phonon transport on phonon frequency renormalization caused by EPI is discussed in Section II.C of this paper.

## 2. Effect of electron-phonon interaction on phonon dynamics

In this section, we undertake an examination of the complex dynamics inherent in EPI across a spectrum of materials, including the most well-studied semiconductor silicon (Si), various metals, and 2D systems. The focal point of our analysis is the pivotal role of EPI in influencing thermal and thermoelectric properties, explored within the context of deformation potential theory.

The influence of EPI on the electronic thermal conductivity in metals is apparent and studied most.[69,70] The lattice thermal conductivity arising from EPI has often been disregarded in semiconductors. Since intrinsic semiconductors have fewer free electrons, usually with the carrier concentration below $10^{18}$ cm$^{-3}$, imposing the effect of EPI on phonons is likely an encumbrance. The developments of thermoelectric and high-power electronics have changed this situation. These devices require a heavy doping to achieve high performance, usually with a carrier concentration well above $10^{19}$ cm$^{-3}$.[71-75] The high carrier concentration induces more scattering processes, whereby the electrons scatter off other phonons. Therefore, a systematic review of the impact of the EPI effect on phonon transport is essential.

Silicon characterized as a pivotal semiconductor in electronics and exhibiting relatively high thermal conductivity, has emerged as the primary subject in the study of EPI. Liao et al.[20] first proposed systematic first-principles calculations to examine the effects of EPI on lattice thermal conductivity in Si. Within the deformation potential approximation, the EPI matrix elements are typically replaced by constant deformation potentials, namely $D_A$ for acoustic phonons and $D_O$ for optical phonons. The EPI matrix elements are replaced by constant deformation potentials, namely $D_A$ $(\hbar q^2/2m_0\omega_{\mathbf{q}\nu})^{1/2}$ for acoustic phonons and $D_O(\hbar q^2/2m_0\omega_{\mathbf{q}\nu})^{1/2}$ for optical phonons,[49,76] thereby allowing the calculation of phonon lifetimes without further approximations. As shown in Fig. 1(a), EPI significantly decreases phonon lifetime in $p$-type Si with a carrier concentration as high as $10^{21}$ cm$^{-3}$, subsequently suppressing lattice thermal conductivity by up to 45%. The deformation potential approximation within the first-principles framework provides a theoretical basis for understanding the influence of EPI on lattice thermal conductivity, thus providing a foundation for subsequent studies on thermal conductivity. To validate the theoretical predictions, Liao et al.[23] quantitatively analyzed the scattering rate of a phonon mode associated with the EPI in Si using three-pulse photoacoustic spectroscopy. Fig. 1(b) shows that their experiment measured the lifetime of 250 GHz coherent longitudinal acoustic phonons, which scattered due to photo-excited carriers at room temperature, and confirmed that EPI dominates the decay of phonon lifetime when the carrier concentration exceeds $2 \times 10^{19}$ cm$^{-3}$. Such an investigations differ from the previous understanding, which typically attributed the reduction of lattice thermal conductivity in Si to phonon-phonon interaction processes, providing a more nuanced understanding of the role of EPI in influencing thermal transport.[77,78] The following year, Zhu et al.[79] synthesized the grain-refined heavily doped polycrystalline $Si_{0.94}P_{0.06}$ via proper doping of P atoms, and observed a significant decrease of lattice thermal conductivity. After excluding the effect of other scattering sources, the influence of phonon transport by EPI was quantitatively analyzed. Lattice thermal conductivity was reduced by about 36% in fine-grained heavily doped bulk $Si_{0.94}P_{0.06}$.

The influence of EPI on phonon transport is also observed in pure metals such as Al, Ag, and Au although the electronic thermal conductivity is dominant in these metals.[80] It was found that the lattice thermal conductivity of Al, Ag, and Au decreased by 50%, 14%, and 6%, respectively, when considering the EPI at 100 K. The different strengths of EPI can be partially quantified by the electron-phonon coupling constant, which measures the rate of heat transfer from hot electrons to cold phonons. The relatively high electron-phonon coupling constant of Al significantly reduces the phonon lifetime and subsequently leads to a considerable reduction in lattice thermal conductivity. It is noteworthy that the strength of EPI in these metals is suppressed with increasing temperatures. This suppression occurs because phonon-phonon scattering intensifies due to the higher phonon population at elevated temperatures. Consequently, the influence of EPI on phonon thermal conductivity ($k_L$) in these metals decreases with rising temperature, with $k_L \sim 1/T$ above the Debye temperature of the material. However, a strong EPI effect remains observable at higher temperatures in d-band metals such as Pt and Ni.[81] This seems to contradict with long-standing belief that the effect of EPI can be ignored at high temperatures. The occurrence of a strong effect in Pt and Ni because the Fermi level in these metals is located within the d-band, which overlaps with the s-band and p-band, resulting in a high electronic density of states (DOS) at the Fermi level, as shown in Fig. 1(c). The strength of the EPI is strongly dependent on DOS.[82] High DOS near the Fermi level facilitates the satisfaction of both energy and momentum conservation conditions in Eq. (2.4), thereby promoting significant electron involvement in EPI processes.

The temperature dependence of the thermal conductivity undergoes significant changes when the EPI is substantial. Even in pure metals, such as Pt and Ni, the complex electronic structures near the Fermi level, reveal a degree of freedom to affect the strength of EPI, leading to the change of lattice thermal conductivity deviates from the intrinsic $1/T$ trend dominated by phonon-phonon interaction. When the EPI is further enhanced and significantly stronger than the mutual interactions between phonons, fundamentally different behavior in lattice thermal conductivity can be observed, which can even become nearly independent of temperature under these conditions. Due to the extensive Fermi surfaces combined with a high density of electronic states, certain materials have a high possibility to form Fermi surface nesting that enhances EPIs. Fermi surface nesting describes a scenario where large parallel segments of the Fermi surface can be aligned or "nested" with each other through a specific phonon wave vector. When a large portion of the Fermi surface can be perfectly mapped onto another portion by translating it through a vector, there is a high density of electronic states that can interact via this vector, significantly enhancing the strength of EPI. The strong electron-phonon scatterings tend to introduce additional thermal resistance particularly in the low-temperature regime, resulting in a thermal conductivity that is weakly dependent on temperature. First-principles calculations for group V transition-metal carbides (VC, NbC, and TaC)[83] and transition-metal nitrides (TiN and HfN)[84] showed that these metals possess Fermi surface nesting, characterized by large parallel sections within the Brillouin zone that form a six-armed nested shape, as shown in Fig. 1(d). This unique structure near the Fermi level leads to strong electron-phonon scattering, resulting in a striking decrease in lattice thermal conductivity by two orders of magnitude. The temperature-independent EPI dominates phonon thermal transport, which is responsible for the temperature-independent lattice thermal conductivity around room temperature and beyond. The Fermi surface nesting was also observed in transition metal borides $ZrB_2$ and $TiB_2$.[85] Unlike six-armed Fermi surface nesting discussed above, the wrinkled dumbbell-like hole sheet and a ring-like electron sheet underpinning the Fermi surface nesting in $ZrB_2$ and $TiB_2$ induce a relatively weaker EPI effect at room temperature, due to the absence of nested Fermi surface regions interacting with phonon wave vectors. Following the theoretical studies, subsequent experimental investigations of charge-density-wave $TaS_2$ confirmed the inherent relationship between the anomalously suppressed thermal conduction and Fermi surface nesting.[86] The strong scattering of phonons by this specific Fermi surface, leading to a reduction in thermal conductivity, underscores the pivotal role of electronic states at the Fermi level in phonon scattering.

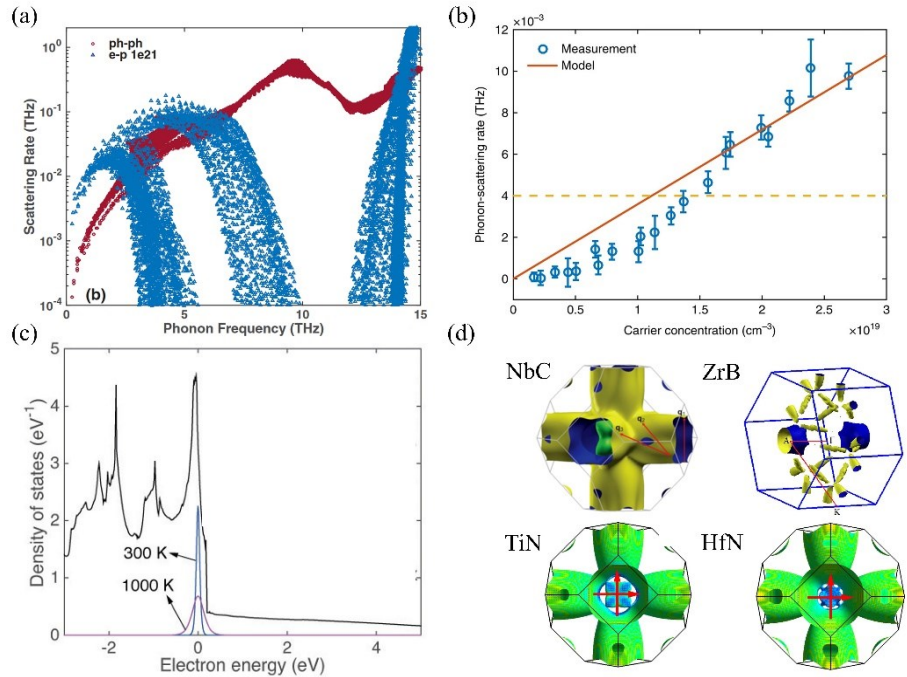

FIG. 1. The phonon scattering rates resulting from intrinsic phonon-phonon interactions, alongside EPI with holes at a carrier concentration of $10^{21}$ cm$^{-3}$,[20] Copyright 2015 with the permission of APS. (b) phonon scattering rate due to EPI as a function of carrier concentration,[23] compared with the theoretical prediction in Ref. 20.[20] (c) Electronic density of states and the Fermi window for Ni at temperatures of 300 K and 1000 K,[81] Copyright 2020, with permission from Elsevier. (d) Fermi surface of NbC, ZrB, TiN, and HfN calculated from first principles. The red arrows show the nesting vectors, Reprinted from Ref. 83-85,[83-85] with permission from APS (2018) and Elsevier (2020).

In addition to conventional metals and semiconductors, unique electronic band structures observed in many low-dimensional materials, such as the Dirac energy band structure or van Hove singularities, promise opportunities for active regulation of the EPI effect. Particularly in two-dimensional (2D) semiconductors, the carrier concentration can be precisely modulated over a broad range through electric gating,[87-89] enabling the isolated regulation of EPI without the influence of other scattering mechanisms, thereby motivating further studies into EPI within 2D materials. In 2D silicene, which features a Dirac cone structure,[90] long-wavelength phonons at the Brillouin zone's center are strongly scattered by electrons. Additionally, intervalley electron scattering between the two Dirac cones of silicene interacts with phonons at the Brillouin zone's boundary. This interaction significantly reduces the phonon lifetime in silicene and consequently decreases the lattice thermal conductivity, as shown in Fig. 2(a). Furthermore, the flat conduction band structure near the Fermi level in 2D semiconductor hexagonal boron phosphorus (h-BP)[91] gives rise to sharply localized electron states, known as van Hove singularities, within DOS. Tuning the Fermi level to align with these singularities facilitates additional scattering channels for EPI. This tuning not only enhances the interaction possibilities but also significantly impacts the thermal and electronic properties through modified phonon dynamics.

Unlike three-dimensional (3D) crystals with spatially extended Fermi surfaces near the Fermi level, where the three-particle EPI process (an electron and phonon decay and a new electron being created) is emphasized due to the high Fermi energy relative to $k_BT$, the scenario in 2D Dirac materials is starkly distinct.[92] In undoped 2D Dirac crystals, the Fermi surface area collapses to the Dirac point, leading to an infinite screening length. Given the limited concentration of free electrons, the Fermi energy cannot be assumed to be significantly larger than $k_BT$. To accurately express the thermal transport in 2D Dirac crystals, the phonon scattering rate should be formulated for phonons with short wavelengths and significant dispersive characteristics. In this case, the three-particle EPI process is often disregarded in favor of the four-particle EPI process, where an electron and a phonon decay to create new electron and phonon pairs, as shown in Fig. 2(b). This is because the phonons generated in the four-particle EPI process can partially or completely offset those eliminated in the three-particle EPI process, thereby altering the thermal conductivity. Kazemian and Fanchini highlighted a notable shift in the primary EPI processes within 2D Dirac materials.[93] At elevated Fermi levels and lower temperatures, the three-particle processes are predominant. However, as the temperature increases above 300 K and the Fermi level decreases, the four-particle EPIs become more significant. This transition is essential for accurately determining the phonon scattering rates and thermal conductivity in 2D Dirac crystals.

Low-dimensional materials inherently amplify the impact of classical size effects and quantum confinement effects on the strength of EPI. Many materials, including quantum dots, nanowires, and 2D layers, manifest unique physical properties due to their confined electron motion and altered phonon dispersion. The quantum confinement in these systems leads to discrete energy levels, which modify the electron DOS and potentially enhance EPI strength.[94-96] Similarly, size effects result from the scaling down of material dimensions, which can amplify the interactions between electrons and phonons. It is therefore of interest to investigate the effect of nanostructuring on the strength of EPI. It can be seen in Fig. 2(c) that Fu et al.[97] built various models featuring different nanoscale length scales to simulate nanowires, solid thin films, and nanoporous thin films. Subsequently, they explored the combined influence of EPI and grain boundary scattering in the lattice thermal conductivity of Si. The results demonstrated that the reduction of lattice thermal conductivity caused by EPI reaches up to over 10% with higher nanoscale length scales with a high carrier concentration of $10^{21}$ cm$^{-3}$, whereas the reduction can be negligible at smaller feature sizes. The low-dimensional materials also exhibit charge and thermal transport features quite different from their bulk counterpart, arising from the quantum confinement effect. Moreover, the reduction in dimensionality and size of 2D material leads to a competition between quantum well width, namely quantum confinement length, and thermal de Broglie wavelength. The quantum confinement effect would significantly enhance the thermoelectric power factor in dimensionally reduced quantum well. The experimental work on 2D InSe confirmed that quantum confinement induced a shaper edge of the conduction-band DOS, thereby enhancing the Seebeck coefficient and power factor of InSe.[94] Subsequent studies showed that quantum confinement could also affect phonon transport via the tailoring strength of the EPI effect. In both 2D and 3D $XB_2$ (X = Mg and Al), quantum confinement modulates the band structure and phonon dispersion, subsequently affecting the strength of the EPI effect and thus suppressing lattice thermal conductivity.[95] It has been observed that lattice thermal transport exhibits greater sensitivity to EPI in low-dimensional monolayer SnSe compared to its bulk counterpart, which can be attributed to quantum confinement effect.[96] As shown in Fig. 2(d), upon considering electron-phonon scattering, the lattice thermal conductivity exhibits a more rapid decrease, deviating from its typical inverse temperature dependence.

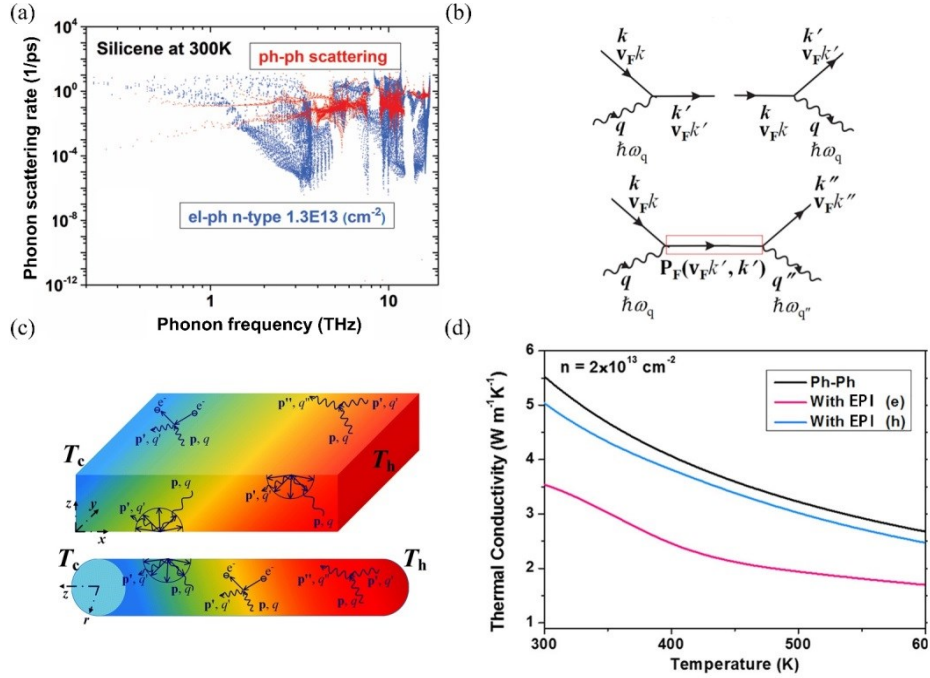

FIG. 2. (a) Comparison of electron-phonon and phonon-phonon scattering rates as a function of phonon frequency in *n*-type silicene,[90] Copyright 2019, with permission from APS. (b) The three-particle and the four-particle scatterings for EPI,[93] Copyright 2024, with permission from APS. (c) Schematic representation of the phonon transport in thin films and nanowires detailing phonon-phonon, phonon-boundary, and electron-phonon scatterings,[97] Copyright 2017, with permission from Royal Society of Chemistry. (d) Variation of thermal conductivity in monolayer SnSe as a function of temperature, considering electron-phonon, hole-phonon, and phonon-phonon interactions, at a carrier concentration of $2 \times 10^{13}$ cm$^{-2}$.[96] Reprinted from Ref. 96,[96] Copyright 2019, with permission from ACS Publications.

**B. Long-range electron-phonon interaction in polar materials**

While the short-range EPI is dominated by the deformation potential scattering, quantitative modeling of long-range EPI in polar materials is a long-standing challenge in condensed matter physics. As shown in Table I, the challenges in handling the EPI matrix elements of long wavelength optical phonons, which diverge as $1/\mathbf{q}$ when $\mathbf{q} \to 0$, result in significant information loss during Wannier interpolation. Specifically, this issue affects the accurate representation of long-range interactions originating from polar optical phonon and piezoelectric interactions in EPI. To address these challenges, recent advancements have included the "polar Wannier interpolation scheme" proposed by Sjakste et al.[98] and Verdi and Giustino,[99] which separates the EPI matrix elements into short-range $g_s$ (**k,q**) and long-range $g_L$ (**k,q**) components. The long-range $g_L$ (**k,q**) characterized by 1/**q** singularity, is evaluated using analytic formulas based on the Vogl model,[50] while $g_s$ (**k,q**) is computed using Wannier interpolation. This approach efficiently matches the results from DFPT across all **q** values with good accuracy. Based on the framework of the polar Wannier interpolation scheme, numerous studies have begun investigations into how piezoelectric scattering and polar optical phonon scattering—the main types of long-range interactions—affect the thermal transport and thermoelectric properties of polar materials.[100-103]

Initial studies focused on investigating the electron mobility in polar materials. Within the DFT framework, Zhou et al.[104] calculated the mobility of GaAs by employing the Boltzmann transport equations (BTE) with state-dependent relaxation times. However, in polar semiconductors, the applicability of the relaxation time approximation is questionable due to strong inelastic scattering induced by longitudinal optical phonons.[49,76] Liu et al.[105] addressed this limitation by solving the linearized electron-phonon BTE instead of relying on the relaxation time approximation to study the mobility of GaAs. Through mode-by-mode analysis, it was found that in GaAs, piezoelectric scattering is comparable to deformation potential scattering of electrons with acoustic phonons, even at room temperature, significantly contributing to mobility, which in turn explains the influence of long-range interactions on the thermoelectric properties of polar materials. Subsequently, Yang et al.[106] identified a significant role of long-range interactions in phonon thermal transport. They found that the significant difference in electronegativity between Ga and N atoms (~1.4) in wurtzite GaN results in a substantial non-zero Born effective charge. Such large polarity leads to exceptionally strong Fröhlich coupling between electrons and longitudinal optical phonons in wurtzite GaN.[107,108] As a result, considering EPI, the lattice thermal conductivity of natural wurtzite GaN decreases by 24% to 34% at room temperature, demonstrating good agreement with experimental results. Zhao et al.[109] employed circularly polarized Raman spectroscopy to differentiate the contributions of deformation potential and Fröhlich interaction in the $E_{2g}^1$ mode of strained $MoS_2$. Their results indicated that the relative contributions of deformation potential and Fröhlich interaction vary with applied strain and are predominantly modulated by excitonic effects. Excitons, which localize charge carriers and enhance the overlap between electron and phonon wave functions, play a crucial role in influencing both deformation potential and Fröhlich interactions. The presence of strain amplifies excitonic effects, leading to significant alterations in exciton-phonon coupling.

To further elucidate the role of dipoles in long-range interactions within polar materials, Liu et al.[110] employed both *ab initio* calculations and analytical modeling to investigate the impact of electron-phonon-induced dipole coupling on phonon transport. Their study demonstrated that the scattering rate of phonons under electron-phonon-induced dipole interaction varies asymptotically with the reciprocal of the phonon wavevector, surpassing short-range deformation potential scattering as the wavevector decreases to several reciprocal centimeters. As shown in Fig. 3(a), a unique upturn in the scattering rate of low-frequency acoustic phonons caused by electron-phonon-induced dipole interaction was found, underscoring the critical importance of considering long-range dipole effects in the phonon scattering phase space of polar materials. They introduced a characteristic phonon frequency, $\omega_c = v_q e_{PZ} e_0 (\epsilon_0 D_A)^{-1}$, where $v_q$ is the sound velocity, $e_{PZ}$ represents the piezoelectric constant, $e_0$ is the elementary charge, and $\epsilon_0$ is the static dielectric constant. This characteristic phonon frequency is used to determine whether EPI is dominated by electric dipoles or deformation potential. Furthermore, their research revealed that breaking centrosymmetry affects long-range dipoles and consequently changes phonon transport. In wurtzite ZnO, the absence of a mirror plane perpendicular to the *c*-axis results in lacking inversion symmetry of out-of-plane phonons, thereby disrupting the inversion symmetry of out-of-plane vibrational modes. This disruption allows a large number of vibrational modes to induce a dipole field, leading to stronger suppression of long-wavelength phonons compared to zinc-blende crystals. Consequently, a significant reduction was found in the thermal conductivity of wurtzite ZnO.

In addition to the long-range scattering affecting the thermal transport properties of the system, thermal expansion of the lattice in turn influences the long-range Fröhlich coupling. Caruso *et al.*[111] found that soft polar phonons significantly enhance the EPI at elevated temperatures in SnSe, resulting in complex band-structure renormalization and increased electron linewidth. Figure 3(b) shows spectral functions along the **k** direction at temperatures of 80 K and 600 K, respectively. At 80 K, the quasiparticle peaks exhibit slight broadening but remain closely aligned with the DFT bands, indicating minimal renormalization at lower temperatures. At 600 K, significant broadening and shifts from the DFT bands (light blue line) are observed, indicative of increased EPIs and phonon scattering at elevated temperatures. These effects, when combined with lattice thermal expansion, lead to a pronounced temperature dependence of both the effective mass of electrons and electrical conductivity. This interplay underscores the necessity of considering both EPI and thermal expansion to accurately predict and optimize the thermoelectric properties of materials like SnSe.

TABLE I. Derived semiempirical equations describe the EPI matrix element and scattering rates for various EPIs in 3D crystals.[105,112,113] Here $e$ is the electron charge, $m_e$ is the electron elective mass, $\Omega$ is the volume of the unit cell, and $\rho$ is the mass density of the material. $\omega_\mathbf{q}$ and $n_\mathbf{q}^0$ are the phonon frequency and distribution function at equilibrium state, respectively. The screening effect is shown by $\epsilon_\infty$ and $\epsilon_0$, which are the high frequency (no lattice response) and static (including lattice response) dielectric constants, respectively. In the piezoelectric scattering, $\lambda_\text{D}$ is the Debye screening length.

| Mechanism | EPI matrix element | Scattering rate $\tau^{-1}(\varepsilon_\mathbf{k})$ |
|---|---|---|
| Acoustic deformation potential | $\sqrt{\dfrac{\Xi_\text{ADP}^2 \hbar |\mathbf{q}|}{2\rho\Omega v_\mathbf{q}}}$ | $\dfrac{\Xi_\text{ADP}^2 (2m_e)^{3/2} k_\text{B} T}{2\pi\hbar^3 \rho v_\mathbf{q}^2}\sqrt{\varepsilon_\mathbf{k}}$ |
| Optical deformation potential | $\sqrt{\dfrac{\Xi_\text{ODP}^2 \hbar}{2\rho\Omega \omega_\mathbf{q}}}$ | $\dfrac{\Xi_\text{ODP}^2 (2m_e)^{3/2}}{4\pi\hbar^3 \rho \omega_\mathbf{q}}\left[\left(n_\mathbf{q}^0 + \dfrac{1}{2}\mp\dfrac{1}{2}\right)\sqrt{\varepsilon_\mathbf{k}\pm\hbar\omega_\mathbf{q}}\right]$ |
| Piezoelectric | $\sqrt{\dfrac{e_\text{PZ}^2 \hbar}{2\rho\Omega v_\mathbf{q}}}\dfrac{e^2}{\epsilon_\infty |\mathbf{q}|}$ | $\dfrac{e_\text{PZ}^2 e^2 k_\text{B} T}{\pi\hbar^2 \epsilon_\infty^2 \sqrt{2\varepsilon_\mathbf{k}/m_e}} \ln\left(1 + \dfrac{8 m_e \varepsilon_\mathbf{k}/\hbar^2}{\lambda_\text{D}^2}\right)$ |
| Polar optical phonon | $\dfrac{ie}{|\mathbf{q}|}\sqrt{\dfrac{2\pi\hbar\omega_\mathbf{q}}{\Omega}\left(\dfrac{1}{\epsilon_\infty}-\dfrac{1}{\epsilon_0}\right)}$ | $\dfrac{e^2 \omega_\mathbf{q}\left(\dfrac{1}{\epsilon_\infty}-\dfrac{1}{\epsilon_0}\right)}{2\pi\hbar\epsilon_\infty \sqrt{2\varepsilon_\mathbf{k}/m_e}}\left[\left(n_\mathbf{q}^0+\dfrac{1}{2}\mp\dfrac{1}{2}\right)\sinh^{-1}\sqrt{\dfrac{\varepsilon_\mathbf{k}}{\hbar\omega_\mathbf{q}}-\dfrac{1}{2}\pm\dfrac{1}{2}}\right]$ |

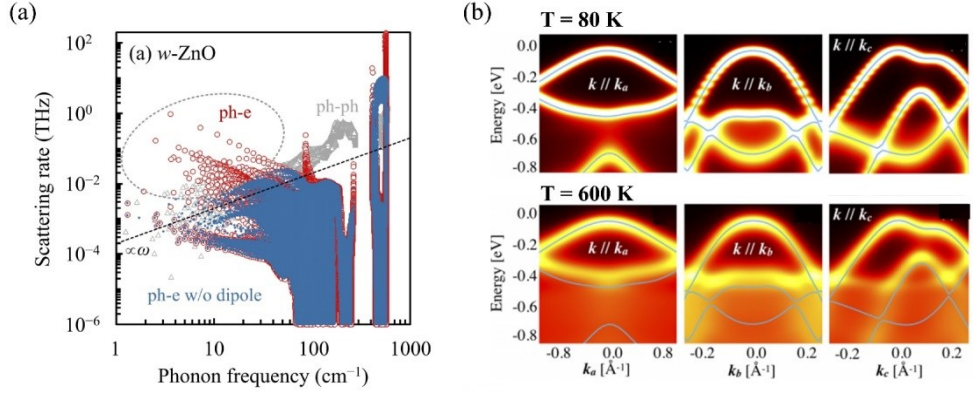

FIG. 3. (a) Calculated electron-phonon (red circles) and phonon-phonon (gray triangles) scattering rates of (a) wurtzite ZnO at a doping concentration of $10^{21}$ cm$^{-3}$ and temperature of 300 K,[110] Copyright 2022, with permission from Elsevier. (b) Temperature dependence of the angle-resolved spectral function for the valence bands along the $\mathbf{k}_a$, $\mathbf{k}_b$, and $\mathbf{k}_c$ crystal momenta within the Brillouin zone, with the corresponding DFT-calculated bands (light blue) superimposed for comparison,[111] Copyright 2019, with permission from APS.

## C. Phonon renormalization due to electron-phonon coupling

EPI leads to significant changes in the phonon self-energy and modify the phonon dispersion predicted by the adiabatic Born-Oppenheimer approximation, as shown in Eq. (2.5). As a result, phonon renormalization occurs with changes in the real part of the phonon self-energy, as evidenced by alterations in the phonon dispersion relations. These renormalization alter phonon group velocities and soften phonon modes, thereby affecting the overall thermal conductivity of the material. Low-dimensional carbon-based materials, such as graphene and carbon (CNT), are ideal candidates for studying EPI-related phonon renormalization. Low-dimensional graphene and CNT can be electrostatically doped through gate voltage control,[87-89] thereby effectively modulating the carrier concentration and screening out other scattering sources on phonon transport. In graphene, researchers have focused on the highest optical branches at Γ and K points ($E_{2g}$ and $A_1$ modes), which correspond to the intensity and position of G and D the peaks in Raman spectroscopy, respectively. These two modes of graphene exhibit a pronounced dependence on EPI.[114-117] Ando[114] utilized a continuum model to calculate the self-energy of the phonon Green's function in graphene, thereby determining the shift and broadening of phonon modes. They manipulated the carrier concentration in 2D graphene via gate voltage and studied the interaction with long-wavelength optical phonons. Figure 4(a) shows that as the Fermi energy approaches half of the optical phonon energy, the frequency shift exhibits a logarithmic singularity, indicating a dramatic change in the frequency shift, with the function value increasing without bound. Beyond this logarithmic singularity, the phonon frequency increases approximately in proportion to the Fermi energy. As the phenomenological relaxation time decreases, the logarithmic singularity of the frequency shift and the sharp drop in the broadening disappear, revealing complex couplings between electronic states and lattice vibrations. Within the DFT framework, Attaccalite et al.[115] calculated the deformation potential energy at the Γ and K points for highest optical branches and demonstrated that the EPI of the highest optical branch at the high-symmetry point K exhibits a strong dependence on the doping level.

By conceptually folding the electron and phonon dispersions of graphite, one can derive the corresponding properties for CNT. This methodological approach enables researchers to systematically classify and analyze the electronic and phonon characteristics of CNT based on their graphite origins. Das et al.[118] investigated the effect of electrochemical doping on the phonon spectrum of single-walled CNT.

Through in situ Raman spectroscopy and transport measurements, the study revealed significant electron-hole asymmetry in the frequency shifts of G⁻ (transverse optical mode at Γ) and G⁺ (longitudinal optical mode at Γ) modes. The G⁻ mode exhibits a larger blue shift compared to the G⁺ mode under hole doping, while the shifts are minimal under electron doping. Additionally, the electronic properties and structure of CNTs influence the extent of phonon renormalization's response to EPI.[119] In metallic CNTs, the abundance of free electrons facilitates strong interactions with phonons. The strong coupling induces significant alterations in both the frequency and linewidth of phonons, often manifesting as a blueshift in phonon frequency and a reduction in linewidth. This effect is attributable to enhanced screening and interactions between densely packed charge carriers and lattice vibrations. Conversely, semiconducting CNTs display distinct interaction mechanisms, owing to their bandgap and the characteristics of their electronic states. In semiconductors, the reduced freedom of electrons, compared to metals, leads to weaker EPI. Consequently, although the phonon frequency may shift due to changes in charge density (often a redshift), the phonon linewidth generally remains unchanged. This behavior indicates that phonons in semiconductors are less perturbed by electron interactions, reflecting the more localized nature of electronic states and reduced screening effects compared to metals. In carbon nanotubes, the introduction of defects also significantly affects the abundance of electrons.[120] In $sp^2$ carbon systems—distinct from traditional systems where defect-induced symmetry-breaking predominantly influences phonon transport—the introduction of defects in carbon nanotubes induces not only symmetry breaking but also the emergence of additional defect charges. Figure 4(b) illustrates the consequential effects on electron and phonon energies near a negatively charged defect, which markedly affects the double-resonance G' scattering process. The upper panel of Fig. 4(b) presents electron dispersion within the hexagonal Brillouin zone, highlighting transitions driven by photon absorption and phonon scattering. The lower panel details the in-plane transverse optical phonon branch and its interactions, with dashed lines indicating the renormalized structures adjacent to the defect. Arrows in both panels depict electron transitions resulting from photon absorption (vertical arrows) and phonon scattering (nearly horizontal arrows), leading to EPI-induced phonon renormalization, a direct consequence of the additional defect charges.

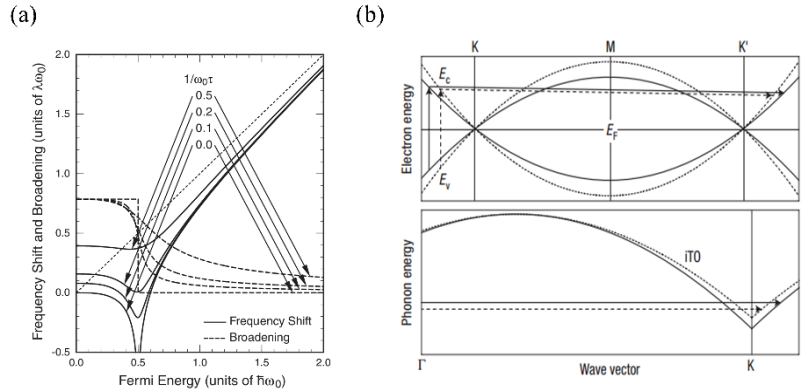

FIG. 4. (a) The dependency of optical phonon frequency shift (solid) and broadening (dashed lines) on the Fermi energy. τ denotes a phenomenological relaxation time that characterizes broadening level induced by disorder,[114] Copyright 2016, with permission from JPSJ. (b) Schematic model showing the renormalization of electron and phonon energies adjacent to a negatively charged defect and delineates its consequent impact on the double-resonance G' scattering process (iTO: in-plane transverse optical-phonon branch)[120], Copyright 2008, with permission from Springer Nature.

Moreover, the phonon wavevector **q** at Γ and K is equal to the difference $\mathbf{k}_F - \mathbf{k}'_F$ between two electronic states on the Fermi surface, resulting in a discontinuity in the derivative of phonon frequency in the highest optical branches for graphene and CNT, known as the Kohn anomaly.[121] Under these conditions, a portion of the Fermi surface can perfectly overlap with another through a specific vector **q**, forming a "nest". At these nested wave vectors, the EPI is exceptionally strong, leading the ratio of the phonon frequency to the logarithmic gradient of the phonon wave vector approaches infinity, culminating in sudden changes in phonon frequency. Graphite represents the first material where a straightforward analytic description of the anomaly is feasible. Piscanec et al.[122] verified the existence of two Kohn anomalies at the Γ-$E_{2g}$ and K-$A_1$ modes in the phonon dispersions of graphite through DFT calculations and analytical derivations, as shown in Fig. 5(a). Furthermore, they demonstrated that the slope of these kinks in the Kohn anomalies is proportional to the square of the EPI, highlighting the critical role of EPI in phonon renormalization. Since the electronic and vibrational properties of CNT can be derived by folding the electron and phonon dispersions of graphite, it is predicted that metallic CNT exhibits a more pronounced Kohn anomaly due to reduced dimensionality. Therefore, they proposed an electronic zone folding method allowing the evaluation of confinement effects on phonon-dispersions and EPI of single-walled CNT of any diameter and chirality.[123] By using the dynamic method, they verified the Kohn anomaly effect in metallic CNT. Farhat et al.[124] also observed the softening and linewidth increase of the M-G peak, corresponding to the LO phonon mode in the phonon spectrum. Figure 5(b) shows M-$G^+$ and M-$G^-$ as the Fermi level varies. These shifts indicate phonon softening due to the Kohn anomaly, which occurs as the Fermi level aligns with half the phonon energy, thereby enhancing EPI. These findings are consistent with theoretical predictions and confirm the existence of Kohn anomalies associated with EPI. Unlike metallic CNT, the Fermi level in semiconducting single-walled CNT is situated within the energy band gap, resulting in essentially no free electrons at room temperature. This implies that even if the phonon wavevector matches a potential electron wavevector, the EPI is insufficient to induce the Kohn anomaly, owing to the absence of free electrons on the Fermi surface. Furthermore, the energy of electrons in semiconducting single-walled CNT typically does not suffice to bridge this energy gap during phonon scattering, preventing electron scattering mechanisms akin to those observed in metallic single-walled CNT. Kohn anomalies have also been observed in other one-dimensional (1D) materials due to the interplay of quasi-1D Fermi surface nesting and hidden nesting of electronic states. In the case of quasi-1D α-Uranium, the ridgelike features in the real part of the electronic susceptibility enable strong modulation of interatomic forces, resulting in multiple Kohn anomalies.[125]

It is worth mentioning that a new class of Kohn anomalies was found in topological Weyl and Dirac semimetals, which were experimentally observed through inelastic x-ray and neutron scattering. As shown in Fig. 5(c), unlike traditional Kohn anomalies, Fermi surfaces in Weyl TaP semimetal exhibit multiple topological singularities at Weyl nodes ($W_1$ and $W_2$), which lead to unique nested conditions characterized by chiral selection, power-law divergence, and significant dynamical effects.[126] Traditionally, the Kohn anomalies exhibit logarithmic divergence in Fermi liquids, whereas the newly identified Kohn anomalies demonstrate a stronger power-law divergence in Weyl semimetals. This pronounced power-law divergence is attributed to the 3D dispersion of the Weyl semimetal and the distinctive electronic structure of the topologically protected Weyl points. In Weyl semimetals such as NbAs[127] and LaAlSi[128], the observed Raman frequency shifts are not induced by the Kohn anomaly. Instead, these shifts are attributed to the anisotropic scattering behavior associated with Fano resonance.[129] As shown in Fig. 5(d), the $B_1$ phonon modes in NbAs exhibit Fano line shapes, with the lower energy $B_1^1$ mode showing more spectral weight on the low-energy side and the higher-energy $B_1^2$ mode having more spectral weight on the high-energy side. This phenomenon arises from the quantum interference between a discrete state (phonon mode) and a continuum of states (electronic states). The study found that the Fano asymmetry parameter (1/**q**) in NbAs remains nearly temperature-independent, suggesting robust EPI over the measured range.[127] The anisotropic

Fano resonance in the $B_1^2$ phonon mode for specific laser excitations (488 nm and 532 nm) was observed in LaAlSi, characterized by asymmetric line shapes.[128] Furthermore, the Raman intensity and frequency of the $B_1^2$ mode exhibits fourfold rotational symmetry as a function of the polarization angle, demonstrating a anisotropic Fano resonance. Further research demonstrates that in Dirac semimetals, the high density of states at the Dirac modes induces Kohn anomalies, leading to the softening of phonons at the Brillouin zone center in $Cd_3As_2$.[130] This phonon softening significantly enhances the scattering phase space for heat-carrying phonons, resulting in $Cd_3As_2$ exhibiting ultralow lattice thermal conductivity. The study on β-$As_2Te_3$ demonstrates that uniaxial strain can induce a quantum topological phase transition near 2 GPa, transforming the material into a topological Dirac semimetal where Kohn anomalies become observable.[131] Specifically, at frequencies below 2 THz, the low-frequency densities dominated by the As and Te elements correspond to the $E_u$ and $E_g$ soft modes, which are related to Kohn anomalies.

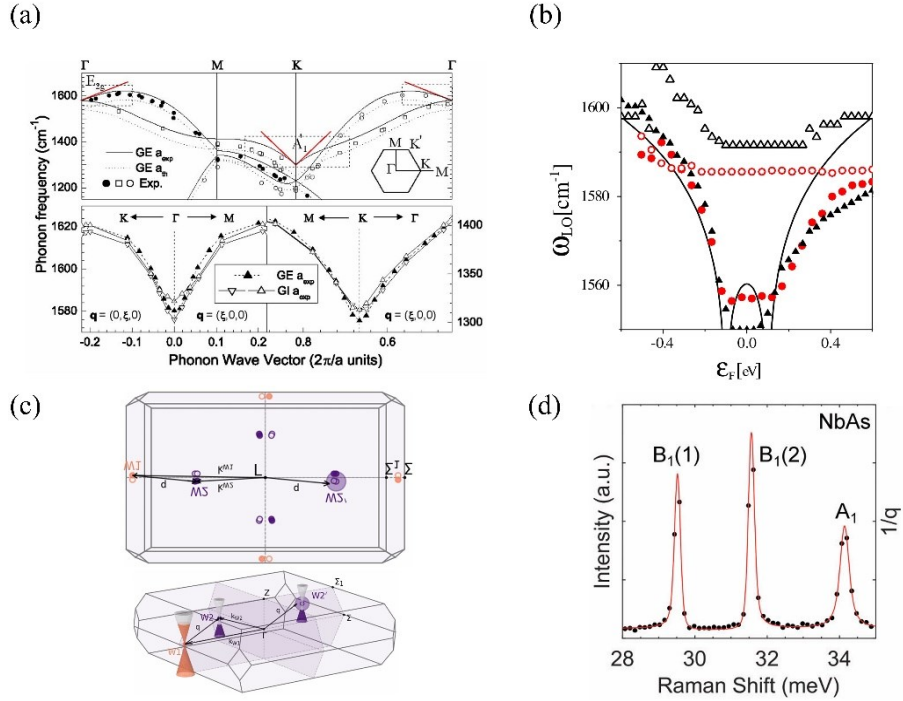

FIG. 5. (a) The highest optical branches around Γ and K of graphene. The red straight lines represent the slopes near the **Γ** and **k**, corresponding to the EPI matrix elements and the energy bands near the Fermi level [122] Copyright 2005, with permission from APS. (b) Frequency of M-G$^+$(open points) and M-G$^-$ (filled points) with solid lines representing the calculated frequencies at absolute zero (T = 0 K) of the LO A symmetry phonon,[124] as reported in Ref. 132.[132] Copyright 2007, with permission from APS. (c) Both 2D and 3D views of the conditions under which the Kohn anomaly occurs, showing a phonon vector $\mathbf{q} = \mathbf{k}_{W2} - \mathbf{k}_{W1} \approx \mathbf{k}_{W2'}$, connecting two Weyl nodes $\mathbf{k}_{W1}$ and $\mathbf{k}_{W2}$,[126] Copyright 2020, with permission from APS. (d) Raman spectra of NbAs at 5 K, where the $B_1$ phonon modes exhibit a subtle Fano line shape not previously reported,[127] Copyright 2019, with permission from APS.

## D. Nonequilibrium electron-phonon transport

When external perturbations, such as ultrafast laser pulses,[133-136] drive the system out of equilibrium, nonequilibrium electron-phonon transport takes place, as illustrated in Fig. 6. A minor fraction of electrons absorbs a substantial amount of laser energy and remains in a nonthermal state, exciting the bulk of the electronic system to achieve a new local thermal equilibrium. Energy relaxation within the electronic subsystem occurs through interactions with various phonon subsystems, which in turn exchange energy among themselves. Electrons and phonons may exhibit different temperatures, with the rate of energy exchange between them varying over time. These conditions necessitate more sophisticated modeling approaches to capture the rapid dynamics.

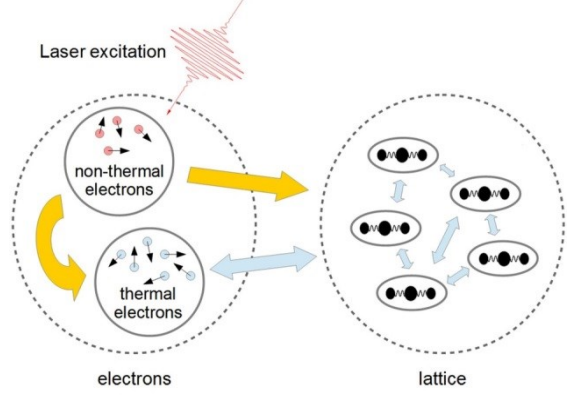

FIG. 6. Scheme depicting the out-of-equilibrium dynamics of electrons and phonons after excitation,[137] Copyright 2017, with permission from APS.

The time-dependent Boltzmann transport equation (TBTE) framework describe the dynamics of electron and phonon where the distribution functions for electrons and phonons are disturbed. The interactions between electron and phonon systems through the TBTE are described as follows:

$$\frac{\partial f_{\alpha \mathbf{k}}}{\partial t} + \mathbf{v}_{\alpha \mathbf{k}} \cdot \nabla_{\mathbf{r}} f_{\alpha \mathbf{k}} - \frac{e\mathbf{E}}{\hbar} \cdot \nabla_{\mathbf{k}} f_{\alpha \mathbf{k}} = \left(\frac{\partial f_{\alpha \mathbf{k}}}{\partial t}\right)_{\text{e-ph}}, \quad (2.6)$$

$$\frac{\partial n_{\lambda \mathbf{q}}}{\partial t} + \boldsymbol{v}_{\lambda \mathbf{q}} \cdot \nabla_{\mathbf{r}} n_{\lambda \mathbf{q}} = \left(\frac{\partial n_{\lambda \mathbf{q}}}{\partial t}\right)_{\text{ph-ph}} + \left(\frac{\partial n_{\lambda \mathbf{q}}}{\partial t}\right)_{\text{e-ph}}. \quad (2.7)$$

Here, $f_{\alpha \mathbf{k}}$ and $n_{\lambda \mathbf{q}}$ represent the nonequilibrium electron and phonon distribution functions with wave vectors $\mathbf{k}$ and $\mathbf{q}$, and electronic band and phonon mode indices $\alpha$ and $\lambda$, respectively. $\mathbf{v}_{\alpha \mathbf{k}}$ and $\boldsymbol{v}_{\lambda \mathbf{q}}$ are drift velocities of the electrons and phonons, respectively. The term $\frac{e\mathbf{E}}{\hbar}$ accounts for the influence of an effective electric field $\mathbf{E}$ on the electrons. The right-hand side terms, $\left(\frac{\partial f_{\alpha \mathbf{k}}}{\partial t}\right)_{\text{e-ph}}$, $\left(\frac{\partial n_{\lambda \mathbf{q}}}{\partial t}\right)_{\text{ph-ph}}$, and $\left(\frac{\partial n_{\lambda \mathbf{q}}}{\partial t}\right)_{\text{e-ph}}$, represent the collision terms. Specifically, $\left(\frac{\partial f_{\alpha \mathbf{k}}}{\partial t}\right)_{\text{e-ph}}$ accounts for electron-phonon collisions affecting the electron distribution function. Under low-fluence excitation and perturbation, EPIs are often weak due to the Pauli exclusion principle, allowing their contribution to be initially neglected.[138] The terms $\left(\frac{\partial n_{\lambda \mathbf{q}}}{\partial t}\right)_{\text{e-ph}}$ and $\left(\frac{\partial n_{\lambda \mathbf{q}}}{\partial t}\right)_{\text{ph-ph}}$ describe changes in the phonon distribution due to phonon-electron and phonon-phonon scattering, respectively. The TBTE describes the nonequilibrium dynamics of electron and phonon populations with a full momentum resolution, effectively capturing the distributions of electron and phonon in the reciprocal space. By solving the coupled TBTEs, one can obtain detailed insights into the ultrafast dynamics of coupled electron-phonon systems under nonequilibrium conditions. The TBTE has used its capability to model heat transport across a wide range of length and time scales, particularly in scenarios involving short-pulse lasers, high energy densities,[139] and ultrafast magnetization dynamics in photo-excited ferromagnets[140], while accurately capturing the anisotropic population of the electronic and vibrational states.[141,142]

It is evident that accurately solving the TBTE is directly linked to correctly determining the distributions of $f_{\alpha\mathbf{k}}$ and $n_{\lambda\mathbf{q}}$, which significantly affects EPI and imparts a high degree of complexity to the behavior of the electronic system. As shown in Fig. 7(a), the conduction and valence bands near the Fermi level of monolayer $MoS_2$ in the excited state do not fully adhere to the Fermi-Dirac distribution.[143] The excited electrons primarily occupy the states near the K and Q high-symmetry points, while the holes are concentrated near the K and Γ points. This highlights the complexity of the electron distribution function in the excited state and underscores the importance of accurately solving the electron distribution function in nonequilibrium systems to fully understand the thermalization and scattering processes. When an ultrashort laser pulse heats a material, the electron distribution function under excitation is closely linked to the DOS. For example, Fig. 7(b) presents the transient electron distribution functional $\phi[f(E,t)] = -\ln[1/f(E,t) - 1]$ for Ni after laser irradiation at a fluence of 0.12 mJ/cm$^2$, with snapshots at 80 fs and 160 fs.[144] The initial distribution represents the thermal equilibrium state before irradiation. At 80 fs, the excited distribution (I) shows steps corresponding to the absorbed photon energy ($\hbar\omega$), heavily influenced by the DOS peak at the Fermi level (II), where a significant number of $d$-electrons reside. Due to the many available states around the chemical potential ($\mu$), the scattering probability increases, leaving a large number of holes at $E = \mu - \hbar\omega$ (II). Additionally, the large number of electrons at the Fermi energy absorbs photons, creating a peak (III) above the chemical potential. This shows how laser irradiation disrupts the electron system, creating a transient nonthermal distribution, and emphasizes the critical role of the DOS in shaping the thermalization dynamics which eventually return to equilibrium at an elevated temperature. Jhalani et al.[145] have demonstrated that the asymmetry observed in the dynamics of hot carriers primarily stems from the distinct distribution of electronic states near the Fermi level. This asymmetry is attributed to several key factors: the degeneracy of the valence band, the relatively higher effective mass of holes compared to electrons, and the specific interactions with various phonon modes within the valence and conduction bands. These factors collectively lead to notable disparities in the scattering and cooling rates between holes and electrons. These findings collectively demonstrate the accuracy of the TBTE method in describing non-equilibrium states and its applicability in advanced heat transport modeling. Zhou et al.[146] proposed an electrohydrodynamics model to account for electron density fluctuations and transient currents, to be the alternative of the coupled TBTE for electrons and phonons. Deng et al.[147] extended the TBTE to weakly coupled systems, proposing the phonon weak coupling model to describe phonon interactions with other energy carriers. Phonon weak coupling model categorizes these interactions into "implicit" couplings within a single structure and "explicit" couplings between different structures, capturing complex nonequilibrium thermal transport.

When studying nonequilibrium electron-phonon transport by solving the TBTE, it is interesting and important to discuss and focus on another key phenomenon: the phonon-electron drag effect.[148] Under nonequilibrium, a temperature gradient induces phonon (electron) drag on electrons (phonons), causing them to deviate from equilibrium. This phonon (electron) drag effect has been shown to significantly influence the transport properties of metals and semiconductors.[149-151] The understanding of phonon drag in this contexts not only helps in accurately modeling thermal and electronic transport but also opens pathways for optimizing thermoelectric materials for better performance under varying thermal conditions. Wang et al.[152] investigate the influence of substrate phonons on the phonon drag effect in thin $Bi_2Te_3$ films. Figure. 7(c) shows the phonon-drag process in a film. Leakage of phonons primarily manifests the lattice dynamics inherent to the substrate, not the film. This phenomenon imprints the substrate's characteristics onto the charge carriers inside the film, thereby altering the EPI. Non-equilibrium phonons are generated in the substrate due to a thermal gradient leak into the 2D layer of the heterostructure, where they interact with the 2D electron system. These phonons transfer their momentum to the electrons, creating an electric current similar to that produced by an applied electric field. By selecting substrates with different Debye temperatures, researchers demonstrated that the position and magnitude of the phonon-drag peak can be controlled. Zhou et al.[153] conduct a first-principles study into optimizing the phonon drag effect in heavily doped Si to enhance thermoelectric performance at lower temperatures. They emphasized that the long-wavelength phonon modes contribute significantly to the phonon drag. As shown in Fig. 7(d), The contribution of phonon drag to the Seebeck coefficient is comparable to that of the diffusive component at a doping concentration of $10^{19}$ cm$^{-3}$. In additional, they proposed an ideal phonon filter to enhance the figure of merit (*ZT*) of Si at room temperature by a factor of 20 to ~0.25. The cubic BAs is also a material where drag thermopower at high hole densities of $10^{21}$ cm$^{-3}$ significantly exceeds diffusive contributions, driven by the weak phonon-phonon and phonon-carrier scattering.[154] Protik et al.[155] also demonstrated that electron-phonon drag significantly enhances thermoelectric properties, such as thermopower, electron mobility, and thermal conductivity, across a wide doping range, in *n*-doped 3C-SiC at room temperature. These findings challenge the traditional view that increased carrier concentration significantly weakens the phonon drag, a phenomenon known as the saturation effect,[148] leading to a new paradigm in enhancing thermopowers through electron-phonon drag.

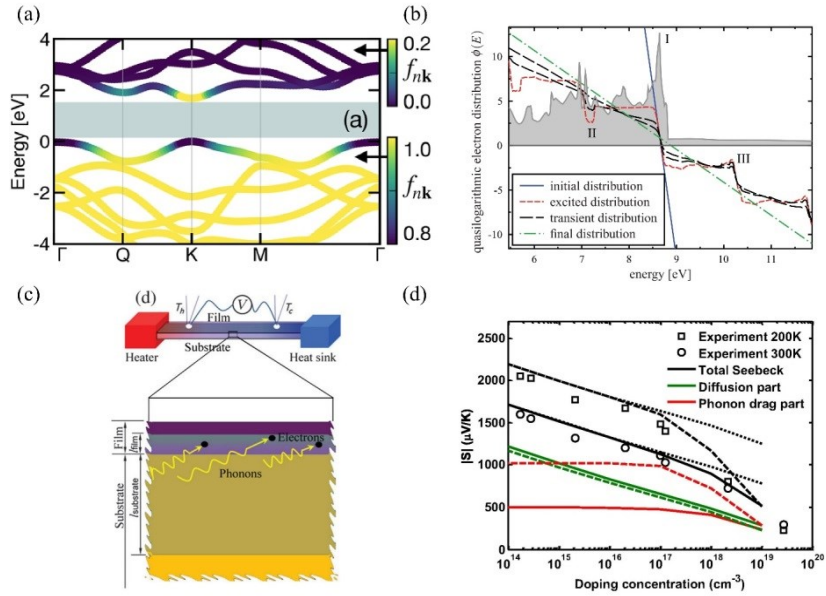

FIG. 7. (a) The electron distribution function overlaid on the band structure of monolayer MoS$_2$. The shaded area represents the band gap, and distinct color scales are applied for the conduction and valence states. Bright regions in the conduction band signify the initial population of excited electrons, whereas dark regions in the valence band denote the hole population,[143] Copyright 2021, with permission from ACS Publications. (b) Quasilogarithmic representation of the transient distribution function $\phi(E)$ for nickel after exposure to a laser pulse with a fluence of 0.12 mJ/cm$^2$. The lines represent the initial distribution, the excited distribution, the transient distributions at two different times (80 fs and 160 fs), and the final distribution, respectively.[144] Copyright 2013, with permission from APS. (c) Diagram illustrating the phonon-drag mechanism in a thin film,[152] Copyright 2013, with permission from APS. (d) Calculated Seebeck coefficient of *n*-type silicon at temperatures of 300 K (Solid lines) and 200 K (dashed lines) with varying doping concentrations compared to experimental results.[153]

The TBTE method can describe with great accuracy the thermalization of electrons and phonons in nonequilibrium conditions; however, the computations required for the electron and phonon distribution functions for each state are highly complex. To simplify the TBTE, the following assumptions are made: (1) the electron and phonon systems are assumed to have independent temperatures, $T_e$ and $T_p$, respectively. At each time step throughout the dynamics, electrons are assumed to populate the electronic bands according to the Fermi-Dirac distribution function at an effective temperature $T_e$. The lattice is assumed to remain in thermal equilibrium, with all boson occupations described by the Bose-Einstein distribution at an effective temperature $T_p$. Through a Taylor expansion, the distribution function is expanded to first order in both phonon modes and electronic degrees of freedom. (2) The electron-phonon energy coupling constant, $G_{ep,i}$, is introduced to represent the rate of energy exchange between electrons and phonons. This constant can be interpreted as the rate of cooling or heating of electrons due to their interaction with phonons. (3) With the local equibrium temperatures of electrons and each phonon branch defined, a macroscopic description is used to consider heat transport in terms of the temperature gradients of the electrons and phonons. Specifically, the energy transport is now described by the terms $\nabla(k_e \nabla T_e)$ and $\nabla(k_{p,i} \nabla T_{p,i})$, which follow the Fourier's law of heat conduction. By incorporating these simplifications, the thermalization kinetics of electrons and phonons can be reformulated into the simpler multiple temperature model (MTM). The MTM model simplifies the complex energy transfer process, allowing for more convenient and faster calculations. The governing equations can be expressed as:

$$C_e \frac{\partial T_e}{\partial t} = \nabla(k_e \nabla T_e) - \sum G_{ep,i}(T_e - T_{p,i}), \tag{2.8}$$

$$C_{p,i} \frac{\partial T_{p,i}}{\partial t} = \nabla(k_{p,i} \nabla T_{p,i}) + G_{ep,i}(T_e - T_{p,i}) + G_{pp,i}(T_{lat} - T_{p,i}), \tag{2.9}$$

where $G_{ep,i}$ denotes the coupling factor between electrons and phonon branch $i$, and $G_{pp,i}$ is the coupling factor between different phonon branches. The MTM model conceptualizes the lattice as a collection of N distinct thermal reservoirs, each coupled to the electron system, allowing for a granular representation of the thermal dynamics by treating each phonon branch as an independent thermal entity, as shown in Fig. 8(a). Therefore, the MTM can capture the distinct behaviors of different phonon branches, providing a comprehensive framework that reflects the complexity of real material systems under laser or electronic excitation conditions.[156,157] For instance, in laser-irradiated graphene, the MTM effectively captures the significant nonequilibrium between electrons, optical phonons, and flexural acoustic phonons, accurately predicting the nonequilibrium thermal conductivity in graphene.[158] This underscores the necessity of considering nonequilibrium phonon states in experimental assessments to improve the accuracy of thermal conductivity measurements in 2D materials. The research on steady-state and transient thermal transport processes under laser irradiation for single-layer graphene have confirmed the significant nonequilibrium among different phonon branches, suggesting weak coupling between the flexural modes and hot electrons or other phonon modes.[159,160] This phenomenon has some implications for the thermal management in graphene-based devices, as the high intrinsic thermal conductivity of graphene is primarily attributed to contributions from these flexural phonons.

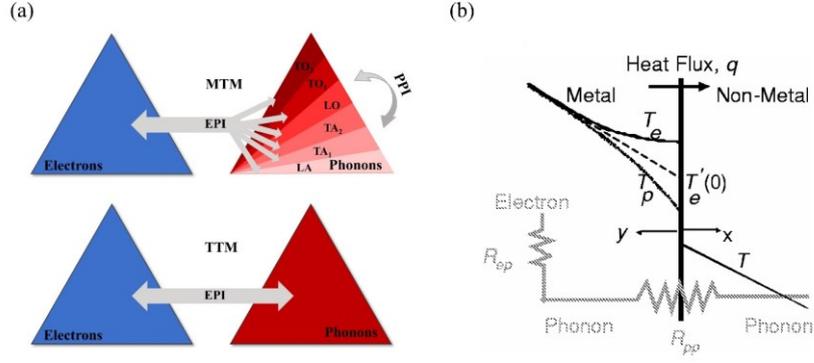

FIG. 8. (a) Schematic representation of the TTM and MTM models. The MTM diagram differentiates itself by segmenting the phonon system into distinct branches, where each interaction is distinctly evaluated. Notably, the diagram exemplifies this approach using six phonon modes: $TO_2$, $TO_1$, LO, $TA_2$, $TA_1$, and LA. Schematic representation of the heat transfer pathways from a metal to a nonmetal across an interface. (b) In TTM model, energy is transferred from electrons to phonons within the metal, encountering a resistance denoted as $R_{ep}$, followed by the phonons transferring energy across the interface, facing a resistance labeled $R_{pp}$,[161] Copyright 2004, with permission from AIP Publication.

The TTM model can be further simplified by assuming that energy transport between individual phonon branches is negligible as shown in Fig. 8(a). This assumption leads to the simplification of the phonon temperature field by reducing the temperatures of individual phonon branches to a single overall effective temperature $T_p$. Consequently, the EPI is simplified to a single coupling term between the electron temperature $T_e$ and the phonon temperature $T_p$. The dynamics of electron and phonon thermalization can be reduced to its most classical form, the two-temperature model (TTM). The governing equations can be expressed as follows:

$$C_e \frac{\partial T_e}{\partial t} = \nabla(k_e \nabla T_e) - G_{ep}(T_e - T_p), \quad (2.10)$$

$$C_p \frac{\partial T_p}{\partial t} = \nabla(k_p \nabla T_p) + G_{ep}(T_e - T_p). \quad (2.11)$$

The TTM offers a straightforward approach to model the thermalization process in nonequilibrium electron and phonon subsystems, and it has been extensively applied to elucidate ultrafast phenomena in solid materials.[162-164] Majumdar and Reddy[161] analyzed a nonequilibrium situation at metal-nonmetal interfaces, focusing on EPI-driven exchanges between the electrons and phonons in the metal, followed by coupling between the phonons in the metal and in the nonmetal, as illustrated in Fig. 8(b). Additionally, the study finds that the total thermal conductance, accounting for both electron-phonon and phonon-phonon contributions, agrees well with experimental data for TiN/MgO interfaces. The TTM has undergone numerous improvements and has found a wide range of applications. For instance, Chen et al.[165] proposed a semiclassical TTM to investigate thermal transport in metals subjected to ultrashort laser pulse heating. Their semiclassical TTM accounts for the effects of electron drift velocity, which significantly influences the electron and lattice temperature responses under high-intensity, short-pulse laser heating. Numerical simulations conducted on gold films showed that when compared with experimental data and the traditional TTM model the semiclassical model provides more accurate predictions of the thermal response, particularly under 600 fs and 800 ps pulse conditions.

To characterize changes in lattice structure, such as defect dynamics and other structural alterations, a two-temperature molecular dynamics (TTM-MD) method was developed. By combining molecular dynamics with the TTM, it becomes possible to account for the effects of electrons while retaining the advantages of MD in modeling lattice change and temperature. The fundamental concept involves modeling the EPI as an additional damping force acting on the atoms within the simulation domain. This model incorporates electronic stopping and electron-ion interactions into MD simulations, thereby enhancing the accuracy of radiation damage simulations.[166] Validated through simulations of 10 keV cascades in iron, the model effectively captures the dynamics of electronic and atomic systems, demonstrating the necessity of considering the cooling effect of electrons in materials with strong EPI. TTM-MD simulations exhibit distinct advantages for studying the mechanisms of defect formation and atomic evolution induced by short-pulse irradiation. Lin et al.[167] employed a combined model in which the MD method replaces the TTM equation to determine the lattice temperature in the Cr target surface, demonstrateing that femtosecond laser pulses induce rapid temperature changes and significant thermoelastic stresses in Cr targets. The transient surface melting, epitaxial resolidification, and the formation of high-density stacking faults and point defects are captured by this simulation. Utilizing nonequilibrium MD simulations combined with TTM, Koči et al.[168] examine how dynamic melting occurs at high pressures and temperatures. The findings indicate that shock-induced melting in nickel occurs at approximately 6400 K and 275 GPa, with electronic heat conduction and EPI reducing the temperature behind the shock front, thereby increasing the melting pressure. Phillips et al.[169] introduced a coarse-grained model that extends the TTM-MD framework's applicability from gapless metallic crystals to radiation events in the insulator α-quartz ($SiO_2$). By comparing various electronic subsystem configurations, the study highlights the significant impact of electron-ion coupling strength on defect annealing and permanent damage reduction.

## E. Remaining challenges and outlooks

### 1. Coupled electron-phonon transport across interfaces

At material interfaces, the mismatched phonon spectra and discontinuities in electronic states complicate energy transfer scenarios. Interfacial transport presents a non-equilibrium, dynamically nonlinear problem, making it difficult to resolve accurately. Modeling of coupled electron-phonon transport across interfaces requires a comprehensive understanding of the interfacial properties and the interactions between phonons and electrons from different materials. Developing advanced models that can accurately predict EPI behavior at interfaces remains an essential task and is vital for designing materials with tailored thermal properties, particularly in the context of heterostructures and nanocomposites.

### 2. Topological band structure effects

Topological materials, characterized by the Berry curvatures and chiral charges in the band structures, present another complexity in studying EPIs. For example, charge carriers have an anomalous velocity in addition to the group velocity, since the Berry curvatures act like an effective magnetic field in the **k**-space. Furthermore, strong EPIs can induce lattice instability at certain high-symmetry points, resulting in phenomena that cannot be treated by the perturbation theory, such as Kohn anomaly and charge-density-wave transitions, just to name a few. The presence of protected edge states, high mobility in semimetals, and the potential for high Seebeck coefficients and low thermal conductivity make topological materials promising for applications in electronics and thermoelectrics.

In addition to charge carriers, leveraging the topological effects of phonons can also open possibilities for manipulating heat generation and transport. For example, topologically protected phonon states can provide robust transport channels that are immune to scattering by defects and impurities, leading to highly efficient thermal conduction in specific directions. Different from excitations of electrons and holes within the Fermi window, the phonon excitation is broadband, such that topological phonons are often buried by the trivial ones. Selectively exciting, detecting, and manipulating topological phonons requires developing both advanced computational and experimental tools.

**3. Long-range electron-phonon interaction in low-dimensional materials**

In low-dimensional materials, long-range EPI plays a critical role in accurately describing material behavior, as the electron-phonon matrix elements are largely dependent on electrostatic interactions that decay more slowly than in bulk materials. Unlike in 3D systems, where Coulomb interactions typically exhibit a more uniform decay, in low-dimensional systems such as monolayer 2D materials or 1D chains, the interactions span a larger spatial range and decay non-uniformly. This results in sharp, spatially varying long-range contributions to the EPI, thereby complicating the modeling of these interactions. In low-dimensional materials, DFT and DFPT methods encounter challenges due to the inherent anisotropy in the electronic and vibrational properties. For example, in 2D materials, the interaction between in-plane and out-of-plane directions is markedly different, with in-plane interactions often exhibiting stronger coupling due to the 2D confinement of both phonons and electrons. This anisotropy complicates the treatment of EPIs, as conventional models relying on isotropic dielectric responses fail to capture the effects of spatial confinement adequately.

Moreover, in low-dimensional systems, direct DFPT calculations are computationally demanding, necessitating the use of alternative strategies to manage the associated computational costs. Interpolation techniques, such as linear interpolation or Wannier interpolation, are commonly employed to perform calculations on dense **q**-grids, which are required to capture the long-range nature of EPI. However, the complexity of Coulomb interactions in low-dimensional materials exacerbates this challenge, as these interactions exhibit sharp kinks (*e.g.*, Fig. 5(a)) at low phonon momentum values, in contrast to the smoother decay observed in 3D systems. This behavior arises from the unique electronic structure of low-dimensional materials, where the Fröhlich interactions from long-wavelength phonons converges to a finite value, further complicating the application of interpolation methods in EPI calculations.

Accurately modeling screening effects in low-dimensional systems is also essential. In 3D materials, screening effects are relatively straightforward; however, in low-dimensional systems, the dielectric properties governing screening are strongly influenced by the material's geometry and dimensionality. The lack of a simple, universally applicable screening model capable of capturing the full spectrum of dielectric responses in low-dimensional systems remains a major obstacle to effective EPI modeling. As a result, new theoretical approaches are essential to address these complex screening effects, often requiring the development of specialized models tailored to the unique electronic structure of each material.

### III. COUPLED THERMAL AND SPIN TRANSPORT

In addition to charges, angular momentum is another degree of freedom in multi-carrier thermal transport. According to quantum mechanics, the total angular momentum (**J**) can include spin angular momentum (**S**) and orbital angular momentum (**L**). Spins are intrinsic features of quasiparticles originated from rotational symmetry, while the orbital angular momentum is generated by the electronic orbitals around the nucleus. The spin and orbital degrees of freedom provide an effective way to store information and energy. Coupled thermal and spin transport is a fundamental problem in the relaxation process in magnets,[170,171] ultrafast demagnetization,[172] thermal management,[173,174] spintronics,[175,176] spin caloritronics,[29,177] and thermoelectric effect in strongly correlated electron systems.[178] The interplay of the spin, charge, and phonon in materials leads to a variety of intriguing physical phenomena. For example, the spin entropy and magnetic field-dependent thermopower have intrigued the research to improve the thermoelectric performance in quantum materials such as Curie-Weiss metal Na$_x$CoO$_2$;[179-181] a colossal anisotropic magnetoresistance (AMR) effect was discovered in the antiferromagnetic material EuMnSb$_2$;[182-184] a large magneto-thermopower (MTEP) in EuMnSb$_2$ was observed;[185] and a photo-Nernst effect in magnetic Weyl semimetal Co$_3$Sn$_2$S$_2$. It should be mentioned that the coupled thermal and orbital transport is not discussed in this review because the recently developed field of orbitronics is still in its infancy.

In this section, we review the recent progress of spin transport driven by temperature gradient and accompanied by thermal transport carried by spins. An introduction of coupled thermal and spin transport is given in Section III.A. Section III.B reviews the progress of the spin Seebeck and spin Nernst effects due to conduction electrons. Section III.C reviews the progress of the spin Seebeck effect due to magnons. Section III.D discusses the magnonic thermal conductivity and magnon effect on phonon thermal conductivity. Section III.E presents the remaining challenges.

## A. Spin thermoelectric effect due to conduction electrons

Nonequilibrium spin transport driven by external electric field, spin accumulation, and temperature gradient leads to abundant physical phenomena. In analogous to the conventional charge Seebeck effect and charge Nernst effect, spin current or spin accumulation generated by temperature gradient is called spin Seebeck effect and spin Nernst effect. Such spin thermoelectric effect has different physical origins in metals and in insulators because of different spin current carriers. In metals and semiconductors, the conduction electrons and holes carry charge and spin simultaneously. Spin current is usually accompanied by a charge current. It is obvious that the spin current is be strongly affected by phonons via EPIs. The spin Seebeck effect, the spin Peltier effect, and the spin Nernst effect have been experimentally observed in various magnetic materials.[25,186-190] There have been several excellent review papers on this topic.[29,177,191,192]

### 1. Spin Seebeck effect

In metals and semiconductors, spin currents carried by conduction electrons and holes can be described by the two-current model. For the electron case, considering two currents with opposite z-components of spin (noted as ↑ for $m_s = 1/2$ and ↓ for $m_s = -1/2$), the charge current is written as $\mathbf{J}_c = j_\uparrow + j_\downarrow$ and the spin current is $\mathbf{J}_{s,z} = j_\uparrow - j_\downarrow$. $j_\uparrow$ and $j_\downarrow$ are particle currents carrying the spin upward and downward, respectively. A charge current can be driven by an external electric field, which is described by electrical conductivity, and by temperature gradient, which is related to the Seebeck coefficient. Similarly, spin current and heat current can also be driven by various driving forces. In linear response and Sommerfeld approximation,[29] the response tensor is written in the Onsager's reciprocal relation as

$$\begin{pmatrix} \mathbf{J}_c \\ \mathbf{J}_{s,z} \\ \mathbf{Q} \end{pmatrix} = \sigma \begin{pmatrix} 1 & P & ST \\ P & 1 & P'ST \\ ST & P'ST & \frac{\kappa_e T}{\sigma} \end{pmatrix} \begin{pmatrix} \frac{\nabla \mu_c}{e} \\ \frac{\nabla \mu_{s,z}}{2e} \\ -\frac{\nabla T}{T} \end{pmatrix}. \quad (3.1)$$

The total electrical conductivity is $\sigma = \sigma_\uparrow + \sigma_\downarrow$ and the charge Seebeck coefficient is $S = (S_\uparrow \sigma_\uparrow + S_\downarrow \sigma_\downarrow)/\sigma$. $\sigma_{\uparrow,\downarrow}$ is the spin-dependent electrical conductivity. $S_{\uparrow(\downarrow)}$ is the spin-dependent Seebeck coefficient. $P = (\sigma_\uparrow - \sigma_\downarrow)/\sigma$ represents the spin polarization of conductivity and $P'$ is its energy derivative. $\mu_c = (\mu_\uparrow + \mu_\downarrow)/2$ is the electrochemical potential and $\mu_{s,z} = \mu_\uparrow - \mu_\downarrow$ denotes the spin accumulation. Eq. (3.1) shows that spin current can not only be driven by spin accumulation, $\nabla \mu_{s,z}$, but also by applied temperature gradient, $\nabla T$. The former phenomenon is spin diffusion and the latter phenomenon is called spin Seebeck effect as shown in Fig. 9. Moreover, heat current $\mathbf{Q}$ carried by electrons is shown in Eq. (3.1) where $\kappa_e$ is the electronic thermal conductivity. The lattice thermal conductivity carried by phonon is not considered because phonons do not carry spin. Besides the external electric field and temperature gradient, heat current can also be driven by spin accumulation. Spin current driven by temperature gradient and heat current driven by spin accumulation satisfy reciprocal relation.

The spin Seebeck coefficient is defined as $S_{SSE} = S_\uparrow - S_\downarrow$ [193] which can be calculated by using the Boltzmann transport equation in a similar way to the calculation of the conventional change Seebeck coefficient by taking the spin index into account: [194]

$$S_{\uparrow(\downarrow)} = \frac{1}{eT}\left[\frac{L_{\uparrow(\downarrow)}^2}{L_{\uparrow(\downarrow)}^1} - \mu_{\uparrow(\downarrow)}\right]. \tag{3.2}$$

The integrals in Eq. (3.2) are

$$L_{\uparrow(\downarrow)}^n = \int \frac{d^3k}{(2\pi)^3} v_{\uparrow(\downarrow),x}^2(\mathbf{k}) \tau_{\uparrow(\downarrow)}(\mathbf{k}) \varepsilon_{\uparrow(\downarrow)}^{(n-1)}(\mathbf{k}) \left[-\frac{\partial f_{\uparrow(\downarrow)}^0(\mathbf{k})}{\partial \varepsilon_{\uparrow(\downarrow)}(\mathbf{k})}\right], \tag{3.3}$$

with integer $n=1, 2$. Here $v_{\uparrow(\downarrow),x}(\mathbf{k})$ is the $x$-component of the velocity of the electron with wave vector $\mathbf{k}$, $v_{\uparrow(\downarrow),x}(\mathbf{k})$ is the energy of electron, and $\tau_{\uparrow(\downarrow)}(\mathbf{k})$ is the relaxation time determined by various scatterings such as electron-impurity scattering, electron-phonon scattering, *etc.* $f_{\uparrow(\downarrow)}^0(\mathbf{k})$ is the Fermi-Dirac distribution function. Once the band structure and parameters of scatterings are given, Eq. (3.3) can be numerically calculated.

There is a simple way to estimate the spin Seebeck coefficient under the Sommerfeld approximation[193]

$$S_{\uparrow(\downarrow)} \approx -\frac{\pi^2 k_B^2}{3e} T \left.\frac{\partial \ln\sigma_{\uparrow(\downarrow)}}{\partial \varepsilon_{\uparrow(\downarrow)}}\right|_{\varepsilon_F}, \tag{3.4}$$

where $\varepsilon_F$ is the Fermi energy. Then we have

$$S_{SSE} = -\frac{\pi^2 k_B^2}{3e} T \left(\left.\frac{\partial \ln\sigma_\uparrow}{\partial \varepsilon_\uparrow}\right|_{\varepsilon_F} - \left.\frac{\partial \ln\sigma_\downarrow}{\partial \varepsilon_\downarrow}\right|_{\varepsilon_F}\right). \tag{3.5}$$

It is straightforward that $S_e = 0$ when two spin states are degenerate. When the signs of $S_\uparrow$ and $S_\downarrow$ are the same, *i.e.*, $S_\uparrow \cdot S_\downarrow > 0$, the contributions from $S_\uparrow$ and from $S_\downarrow$ cancel with each other which leads to a reduction of $S_e$. In contrast, when the signs of $S_\uparrow$ and $S_\downarrow$ are opposite, *i. e.* $S_\uparrow \cdot S_\downarrow < 0$, the value of $S_{SSE}$ could be large and even be larger than the charge Seebeck coefficient, $|S_{SSE}| > |S|$. Equation (3.5) further shows that the energy dependence of $\sigma_{\uparrow(\downarrow)}$ around Fermi energy plays an important role in enhancing spin Seebeck coefficient. Therefore, strong energy-dependent DOS $N_{\uparrow(\downarrow)}$ and relaxation time $\tau_{\uparrow(\downarrow)}$ are preferred. These energy dependences must be different for two spins. The aforementioned conditions are not satisfied in nonmagnetic metals. Pure magnetic metals like nickel and iron partially satisfy these conditions. Magnetic alloys such as CuNi alloy[195] and permalloy $Ni_{80}Fe_{20}$[193] are good choices for large spin Seebeck effect due to their remarkable spin-splitting band structure and energy-dependent scattering mechanisms.

Besides bulk materials, mesoscopic systems or magnetic molecule junctions provides another opportunity to achieve a large spin Seebeck coefficient. In mesoscopic systems, $\sigma_\uparrow$ and $\sigma_\downarrow$ are proportional to the spin-dependent electron transmission probabilities $\mathcal{T}_\uparrow$ and $\mathcal{T}_\downarrow$, respectively. Then Eq. (3.5) can be rewritten as

$$S_{SSE} \sim -\frac{\pi^2 k_B^2}{3e} T \left(\left.\frac{\partial \ln\mathcal{T}_\uparrow}{\partial \varepsilon}\right|_{\varepsilon_F} - \left.\frac{\partial \ln\mathcal{T}_\downarrow}{\partial \varepsilon}\right|_{\varepsilon_F}\right). \tag{3.6}$$

It is clear that strong energy dependence of electron transmission and significant difference between the two spin transmission channels are preferred. For example, Dubi and Di Ventra[196] studied the spin Seebeck effect in a system composed of a quantum dot in contact with ferromagnetic leads. Rejec *et al.*[197] found that the spin Seebeck coefficient in interacting quantum dots is as large as $k_B/e$. Wang *et al.*[198] pointed out that the spin Seebeck coefficient across single molecule magnetic junctions could be larger than the charge Seebeck coefficient.

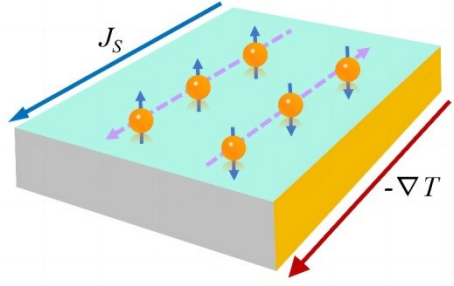

FIG. 9. Schematic illustration of the spin Seebeck effect of conduction electrons. Spin current is parallel to the direction of temperature gradient.

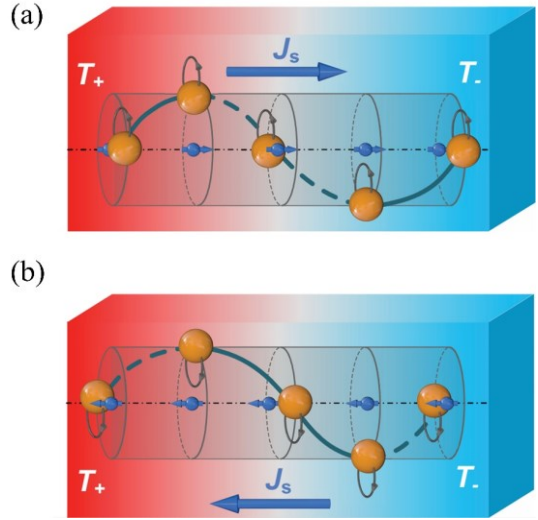

FIG. 10. Schematic diagram of chiral-phonon activated Seebeck effect with two opposite chiralities. The axis of rotation and the direction of spin current is (a) parallel and (b) anti-parallel to the direction of temperature drop.

  The aforementioned spin Seebeck effect requires spin splitting in magnetic materials. Recently, a new type of spin Seebeck effect in nonmagnetic chiral materials has been reported.[30,194] When a temperature gradient is applied on a chiral material, nonzero phonon angular momentum and nonzero phonon magnetic moment are generated due to the local rotation of ions around their equilibrium positions as shown in Fig. 10.[199,200] As a result, the spin degeneration of conduction electrons could be lifted by this temperature gradient generated phonon magnetic moment. Kim *et al.*[30] observed the spin current generated in nonmagnetic 2D chiral hybrid organic-inorganic perovskite after ultrafast laser heating by using the time-resolved magneto-optical Kerr effect measurements and using the spin transfer torque. It is interesting that the spin current generated by this chiral-phonon-activated spin Seebeck effect is proportional to $(\nabla T)^2$ rather than to $\nabla T$ as shown in Eq. (3.1) because the spin splitting induced by phonon magnetic moment is also proportional to the temperature gradient. In other words, it is possible to achieve large spin Seebeck coefficient by increasing temperature gradient. This feature is very different from previous spin Seebeck effect with $\nabla T$-independent spin Seebeck coefficient in Eq. (3.5).

## 2. Spin Nernst effect

In spin Seebeck effect, the direction of spin current is along the direction of the temperature gradient ($\mathbf{J}_{s,z} \parallel \nabla T$). As shown in Fig. 11, when an external magnetic field (**H**) is applied perpendicular to the direction of the temperature gradient, moving electrons with different spins feel Lorentz force perpendicular to their direction of velocity and deflect to opposite sides. In this case, a transverse spin current perpendicular to both the temperature gradient and the magnetic field can be obtained. This phenomenon is called spin Nernst effect which is analogous to the conventional charge Nernst effect.[186] In linear response and Sommefeld approximation, the response tensor of spin Nernst effect is different from the response tensor of spin Seebeck effect in Eq. (3.1):

$$\begin{pmatrix} \mathbf{J}_c \\ \mathbf{J}_{s,\beta} \\ \mathbf{Q} \end{pmatrix} = \sigma \begin{pmatrix} 1 & \theta_{SH}\mathbf{e}_\beta \times & ST \\ \theta_{SH}\mathbf{e}_\beta \times & 1 & ST\theta_{SN}\mathbf{e}_\beta \times \\ ST & ST\theta_{SN}\mathbf{e}_\beta \times & \frac{\kappa_e T}{\sigma} \end{pmatrix} \begin{pmatrix} \frac{\nabla \mu_c}{e} \\ \frac{\nabla \mu_{s,\beta}}{2e} \\ -\frac{\nabla T}{T} \end{pmatrix}, \qquad (3.7)$$

where $\theta_{SH}$ is the spin Hall angle and $\theta_{SN}$ is the spin Nerst angle which characterizes the charge-to-spin conversion efficiency. The cross-product symbols in Eq. (3.7) describe the spin-dependent deflection motion of electrons. The spin polarization is along $\beta$-direction ($\beta = x, y, z$) and $\mathbf{e}_\beta$ is a unit vector along $\beta$-direction. Eq. (3.7) shows that a transverse spin current perpendicular to the heat current can be generated by a longitudinal temperature gradient. Moreover, a transverse heat current can also be driven by the longitudinal spin accumulations. It should be noted that both Eq. (3.1) and Eq. (3.7) are applicable for single particle picture which breaks down in strongly correlated materials because the contribution from quasi-particle is not adequate to explain the Nernst signals in high-$T_c$ superconductors[201] and other quantum materials such as 2D 2M-WS$_2$.[202] Meyer *et al.*[186] observed the spin Nernst angle in Eq. (3.7) by measuring the change of conventional Seebeck coefficient with magnetic field, *i. e.* magneto-thermopower. They found that the value of $\theta_{SN}$ in Pt on YIG is close to the value of $\theta_{SH}$ and their signs are opposite. Sheng *et al.*[203] have reported similar findings in W in W/CoFeB/MgO heterostructures.

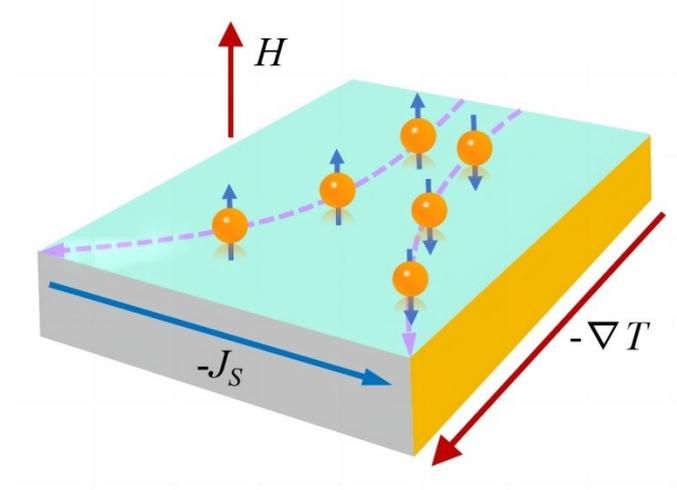

FIG. 11. Schematic illustration of the spin Nernst effect when a magnetic field (*H*) is applied. A spin current perpendicular to both temperature gradient and the magnetic field is generated.[186]

**B. Spin thermoelectric effect due to magnons**

In magnetic insulators, there are neither conduction electrons nor conduction holes. In magnetically ordered insulators without conduction electrons, spin currents are carried by magnons which are low-lying excited states in magnetically ordered solids such as ferromagnet, antiferromagnet, and ferrimagnet.[204,205] Typically, magnons are described by the Heisenberg Hamiltonian $H_H = -\sum_{i,j} J_{ij} \mathbf{S}_i \cdot \mathbf{S}_j$, where $\mathbf{S}_i$ and $\mathbf{S}_j$ are spins on lattice sites $\mathbf{R}_i$ and $\mathbf{R}_j$, respectively, and $J_{ij}$ are the exchange parameters which are functions of distance between two lattice sites, $|\mathbf{R}_i - \mathbf{R}_j|$. Figure 12(a) shows the ground state of a simple ferromagnet in which each $S_i^z$ attains its maximum value of $S$ when $J_{ij} = J > 0$, in other words, all spin directions are parallel. At low temperatures, if $S_j^z$ of arbitrary site $\mathbf{R}_j$ becomes $S\text{-}1$, the total energy of the system will be enhanced. The corresponding elementary excitations, *i.e.* magnons, have wavelike form and describe the oscillation in the relative orientation of spins as shown in Fig. 12(b).[204,205]

When a temperature gradient is applied, the population of magnons at the hot end is larger than the population of magnons at the cold end which means the polar angle of spins at the hot end is larger than the polar angle at the cold end as shown in Fig. 12(c). Consequently, a spin current $\mathbf{J}_m$ and a heat current $\mathbf{Q}_m$ are carried by magnons where the former phenomenon is the spin Seebeck effect and the latter phenomenon is magnonic thermal conduction. The spin Seebeck effect was also found near the ferromagnetic insulator (FI)/nonmagnetic metal (NM) interface as shown in Fig. 13.

The discovery of the spin Seebeck effect in a large number of magnetic insulators provides a powerful tool to probe the magnon transport, the magnetic order and domains, spin correlations, and magnon polarizations.

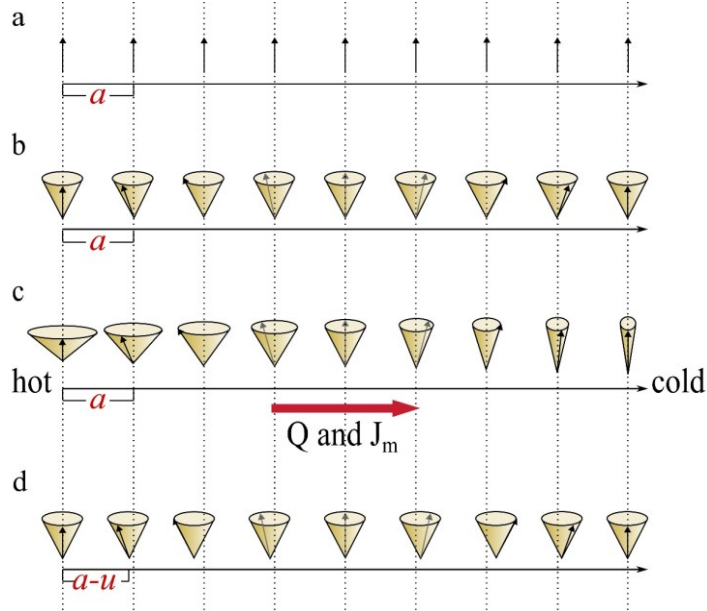

FIG. 12. (a) The ground state of a simple ferromagnet with parallel spin alignment. The exchange parameters $J_{ij}(|\mathbf{R}_i - \mathbf{R}_j|) = J(a) = J$ are constants where $a$ is the lattice constant. (b) Orientations of spins with wavelike azimuthal angles in a spin wave state. (c) Heat current and spin current carried by magnons are generated by temperature gradient where the polar angles of spins at the hot end are different from that at the cold end. (d) Illustration of magnon-phonon interaction when ions are not in their equilibrium positions. The exchange parameters become $J_{ij}(|\mathbf{R}_i - \mathbf{R}_j + \mathbf{u}_i - \mathbf{u}_j|) = J + \delta J(\mathbf{u}_i - \mathbf{u}_j)$ where $\mathbf{u}_i$ and $\mathbf{u}_j$ are the displacements of ions on sites $i$ and $j$, respectively.

## 1. Spin Seebeck effect in ferromagnetic insulators

In the diffusive transport regime, coupled transport of spin and heat by magnons can be described linearly by the Onsager relation:[206]

$$\begin{pmatrix} \frac{2e}{\hbar}\mathbf{J}_m \\ \mathbf{Q}_m \end{pmatrix} = -\begin{pmatrix} \sigma_m & \frac{S_m}{T} \\ \frac{\hbar S_m}{2e} & \kappa_m \end{pmatrix} \begin{pmatrix} \nabla \mu_m \\ \nabla T_m \end{pmatrix}. \tag{3.8}$$

Here $T_m$ is the effective temperature of magnons, $T$ is the environmental temperature, and $\mu_m$ is nonequilibrium magnon spin accumulation. $\mathbf{J}_m$ and $\mathbf{Q}_m$ are spin current and heat current carried by magnons, respectively. The expectation value of magnon spin current is

$$\mathbf{J}_m = -\frac{1}{2}\sum_k \mathbf{v}_k \left( \langle b_k^\dagger b_k \rangle - \langle b_{-k}^\dagger b_{-k} \rangle \right), \tag{3.9}$$

where $v_k$ is the magnon velocity, $b_k^\dagger$ is the magnon creation operator with wave vector k. $\sigma_m \propto e^2 J\tau/(\hbar^2 \Lambda^3)$ is the magnon spin conductivity which is proportional to the magnon spin diffusivity due to the Einstein relation,[206] $S_m \propto eJk_B T\tau/(\hbar^2 \Lambda^3)$ is the bulk spin Seebeck coefficient, $\kappa_m \propto Jk_B^2 T\tau/(\hbar^2 \Lambda^3)$ is the magnonic thermal conductivity, where $\Lambda = \sqrt{4\pi J/(k_B T)}$ is the magnon de Broglie wavelength and $\tau$ is the magnon relaxation time due to various scattering including magnon-magnon scattering, magnon-phonon scattering, and magnon-impurity scattering, *etc*. It is obvious that all three transport coefficients in Eq. (3.8) are related with each other and the key physical parameter is the magnon relaxation time. We will discuss the role of magnon relaxation time in Sec. III C.3.

From an experimental point of view, the spin current $\mathbf{J}_m$ and magnon temperature gradient $\nabla T_m$ are not easy to measure. One must inject a spin current into a nonmagnetic metal and detect the electrical signals induced by the spin current. Instead, the magnonic thermal conductivity, $\kappa_m$, in Eq. (3.8) can be calculated in the framework of the kinetic theory, similar to the lattice thermal conductivity:[207]

$$\kappa_m = \frac{1}{3}\sum_\mathbf{k} C_m(\mathbf{k}) v_m(\mathbf{k}) l_m(\mathbf{k}), \tag{3.10}$$

where $C_m(\mathbf{k})$ is the model contribution of magnons with wave vector $\mathbf{k}$ to the specific heat, $v_m(\mathbf{k})$ is the group velocity of magnons, and $l_m(\mathbf{k})$ is the magnon mean free path.

## 2. Spin Seebeck effect at ferromagnetic insulator/nonmagnetic metal interface

Figure 13 shows the schematic diagram of the spin Seebeck effect near the ferromagnetic insulator (FI)/nonmegnatic metal (NM) interface detected by the inverse spin Hall effect (ISHE).[208] In the presence of temperature gradient along the *x*-direction of a ferromagnetic insulator, magnons are driven to be out-of-equilibrium, which results in a magnon spin current and a magnonic heat current as shown in Eq. (3.8). The exchange leads to a coupling between the electrons in NM and magnons in FI near interface a spin current injected from FI into NM:[209]

$$J_s^{int} = -\frac{\hbar g^{\uparrow\downarrow}}{2e^2\pi s}\int d\epsilon D(\epsilon)(\epsilon - e\mu_z)\left[n_B\left(\frac{\epsilon - e\mu_m}{k_B T_m}\right) - n_B\left(\frac{\epsilon - e\mu_s}{k_B T_e}\right)\right]. \tag{3.11}$$

where $g^{\uparrow\downarrow}$ is the real part of interfacial spin-mixing conductance, $s$ is the equilibrium spin density, $D(\epsilon)$ is the DOS of magnons, $\mu_s$ is the electron spin accumulation, and $T_e$ is the electron temperature. $n_B\left(\frac{\epsilon - e\mu_m}{k_B T_m}\right)$ is the Bose-Einstein distribution function of magnons in FI and $n_B\left(\frac{\epsilon - e\mu_s}{k_B T_e}\right)$ is the Bose-Einstein distribution function describing electron spin accumulation in NM. Using proper approximations, Eq. (3.11) becomes

$$J_s^{int} = -\frac{3\hbar g^{\uparrow\downarrow}}{4e^2\pi s\Lambda^3}\left[e\zeta\left(\frac{3}{2}\right)(\mu_s - \mu_m) + \frac{5}{2}k_B\zeta\left(\frac{5}{2}\right)(T_e - T_m)\right]$$

$$= \frac{\sigma_s^{int}}{\hbar\Lambda}(\mu_s - \mu_m) + \frac{S_{SSE}^{int}}{\Lambda}(T_e - T_m), \tag{3.12}$$

where $\zeta(3/2)$ and $\zeta(5/2)$ are zeta functions. $\sigma_s^{int}$ is the interfacial spin conductivity and $S_{SSE}^{int}$ is the interfacial spin Seebeck coefficient

$$S_{SSE}^{int} = \frac{15g^{\uparrow\downarrow}\zeta\left(\frac{5}{2}\right)}{4e^2\pi s\Lambda^2}. \tag{3.13}$$

Due to the inverse spin Hall effect in NM like Pt, an electric field along $y$-direction ($E_{ISHE}$) is generated by $J_s^{int}$, i. e. $E_{ISHE} \propto \theta_{SH} J_s^{int}$ where $\theta_{SH}$ is spin Hall angle defined in Eq. (3.7). Therefore, the spin Seebeck effect near FI/NM interface can be measured by the inverse spin Hall effect as shown in Fig. 13(a). It is convenient to define the ratio between the electric field and temperature gradient as spin Seebeck coefficient which characterizes the strength of spin Seebeck effect[210]

$$S_{xy} = \frac{E_{ISHE}}{\nabla_x T} \propto S_{SSE}^{int}. \tag{3.14}$$

This spin Seebeck coefficient is different from the bulk spin Seebeck coefficient, $S_m$, defined in Eq. (3.8) because the information of interface and inverse spin Hall effect in Pt, such as spin Hall angle, are included in $S_{xy}$. Experimentalists prefer using $S_{xy}$ since it can be directly measured by a simple experimental setup like Fig. 13(b). The unit of $S_{xy}$ is μV/K which is the same as the conventional charge Seebeck effect although the absolute magnitude of $S_{xy}$ is much smaller.

It should be emphasized that the above discussions are applicable to FI/NM structure. In contrast, the case of ferromagnetic metal/NM is more complicate because the spin current carried by magnons is accompanied by the spin current carried by conduction electrons in ferromagnetic metals. Then one should consider the coupled magnon-electron-phonon transport.[211,212] Therefore, in order to exclude the influence of conduction electrons, insulators like yttrium iron garnet (YIG) are commonly used to study the spin Seebeck effect. In addition, the spin Seebeck effect in many other magnetic materials such as antiferromagnet α-$Fe_2O_3$[213] and van der Waals 2D magnetic materials $Cr_2Si_2Te_6$ and $Cr_2Ge_2Te_6$ [214] were also reported.

Besides measuring $E_y$ and $S_{xy}$ in Eq. (3.14), other methods such as inelastic neutron scattering[215] have been used to investigate the spin Seebeck effect. Theoretically, besides the Boltzmann transport theory,[206,216] other theories and numerical methods such as the Green's function method,[216] the atomistic spin dynamics[217] have been used to calculate the spin Seebeck effect. Readers are referred to Ref. 216.[216]

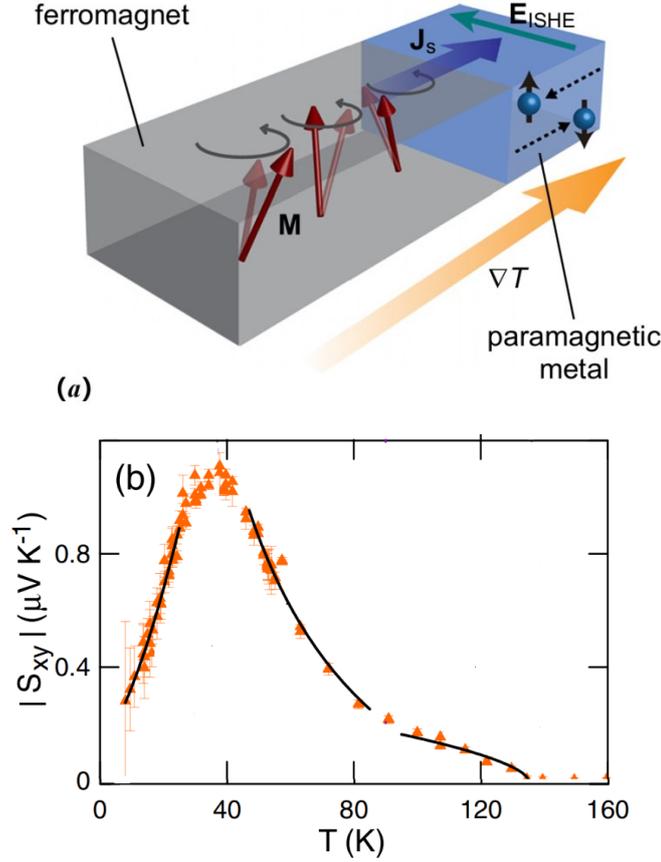

FIG. 13. (a) Schematic illustration of the spin Seebeck effect of FI/NM interface structure. A spin voltage is generated via magnetization (M) dynamics in FI when a temperature gradient ($\nabla T$) is applied. A spin current $J_s$ is injected into the attached metal which is converted into an electric field $E_{ISHE}$ due to the inverse spin Hall effect.[208] (b) Measured temperature dependence of $S_{xy}$ in GaMnAs/Pt structure.[206]

### 3. Magnon-phonon interaction and its effect on spin Seebeck effect

From the expression of the spin Seebeck coefficient in Eq. (3.8), one can see that the relaxation time of magnons ($\tau$) is the key adjustable parameter to manipulate the spin Seebeck effect. In this section, we review the theoretical and experimental studies of magnon-phonon interaction and its effect on spin Seebeck effect.

In the diffusive magnon transport theory, there are four types of scattering processes for magnons: (i) elastic magnon scattering by imperfections such as impurities and boundaries as shown in Fig. 14(a); (ii) magnon-phonon interaction which emits or absorbs one magnon as shown in Fig. 14(b) and Magnon-phonon interaction in which one phonon generates two magnons or two magnons merge into one phonon as shown in Fig. 14(d); (iii) magnon-phonon interaction where the number of magnons does not change as shown in Fig. 14(c); (iv) the momentum conserved magnon-magnon scattering. According to the Matthiessen's law, the total relaxation time of magnon satisfies

$$\frac{1}{\tau} = \frac{1}{\tau_{el}} + \frac{1}{\tau_{mp}} = \frac{1}{\tau_{el}} + \frac{1}{\tau_{mp}^{(2)}} + \frac{1}{\tau_{mp}^{(3)}} + \frac{1}{\tau_{mp}^{(4)}}. \quad (3.15)$$

Here $\tau_{el}$ is the relaxation time of process (i); $\tau_{mp}^{(2)}$ is the relaxation time of magnon-phonon scattering in the process (ii) as shown in Fig. 14(b) and (d); $\tau_{mp}^{(3)}$ and $\tau_{mp}^{(4)}$ are the relaxation times of magnon-phonon scattering due to one phonon process (Fig. 14(c)) and two-phonon processes, respectively, which conserves the number of magnons in the process (iii). It should be emphasized that the momentum conserved magnon-magnon scattering, *i.e.* process (iv), is neglected in this approach because it does not affect magnon transport directly.

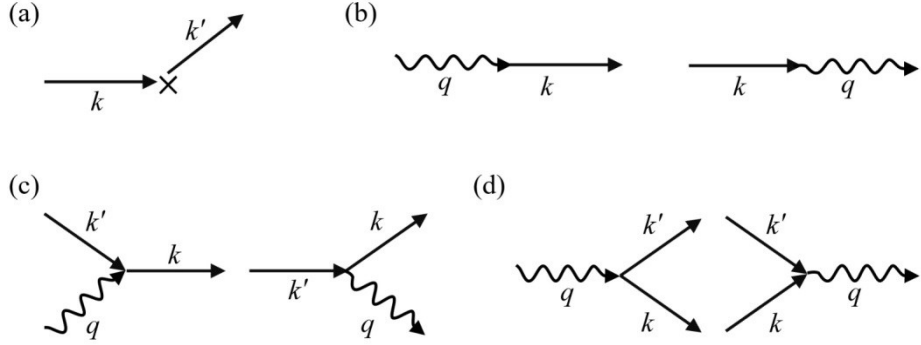

FIG. 14. Typical magnon scattering processes. (a) Elastic magnon scattering by impurities. (b) Directly conversion between magnon and phonon. (c) Magnon-phonon interaction in which one phonon is absorbed or emitted by magnons. (d) Magnon-phonon interaction in which one phonon generates two magnons or two magnons merge into one phonon.

The calculation of $\tau_{el}$ is relatively easier compared to the calculation of $\tau_{mp}^{(2)}$, $\tau_{mp}^{(3)}$ and $\tau_{mp}^{(4)}$. Therefore, we only present here the calculation of magnon-phonon relaxation time $\tau_{mp}$. Qualitatively, a rough estimation of magnon-phonon relaxation time can be written as

$$\tau_{mp} \sim \hbar/\alpha_G k_B T, \quad (3.16)$$

where $\alpha_G$ is the Gilbert damping parameter[212] which is nearly temperature-independent. The value of $\alpha_G$ should be obtained from experiment such as ferromagnetic resonance measurement, for example, $\alpha_G \sim 10^{-4}$ in YIG.[218] In order to calculate $\tau_{mp}$ comprehensively, we should start from the Hamiltonian of magnon-phonon interaction:[171,219]

$$H_{mp} = \int \left[ \gamma^{ij}(\mathbf{M}) + \gamma^{klijmn}(\mathbf{M}) \frac{\partial M^m}{\partial r^k} \frac{\partial M^n}{\partial r^l} \right] u^{ij} d\mathbf{r}, \quad (3.17)$$

where $u^{ij} = \frac{1}{2}\left(\frac{\partial u^i}{\partial r^j} + \frac{\partial u^j}{\partial r^i}\right)$ is the strain tensor, $\mathbf{u}$ is the atomic displacement, $\gamma^{ij}$ and $\gamma^{klijmn}$ are the tensors of the magnetic elastic coupling, and $\mathbf{M}$ is the magnetization of material. Notations, *j, k, l, m, n*=*x, y, z*. In ferromagnetic materials, Eq. (3.17) can be written into several parts where the major ones are

$$H_{mp} \approx H_{mp}^{(2)} + H_{mp}^{(3)} + H_{mp}^{(4)}. \quad (3.18)$$

The first term on the right side of Eq. (3.18) describes the direct conversion between magnon and phonon, which is also called the magnetoelastic coupling. It includes two processes: $a_{\mathbf{q}\lambda}^\dagger b_{\mathbf{k}}$ and $a_{\mathbf{q}\lambda} b_{\mathbf{k}}^\dagger$, which are shown in Fig. 14(b). Here $a_{\mathbf{q}\lambda}(a_{-\mathbf{q}\lambda}^\dagger)$ is the phonon annihilation (creation) operator with phonon wave vector $\mathbf{q}$ and phonon branch index $\lambda$, $b_{\mathbf{k}}(b_{\mathbf{k}}^\dagger)$ is the magnon annihilation (creation) operator with magnon wave vector $\mathbf{k}$. $H_{m-p}^{(2)}$ describes the direct conversion between one magnon and one phonon with the same wave vectors and energy. As a result, $H_{mp}^{(2)}$ is important only in a very small region where the phonon dispersion curve and magnon dispersion curve cross.[171,220-223] Therefore, there is no considerable direct contribution from $H_{mp}^{(2)}$ to magnon-phonon relaxation time $\tau_{mp}^{(2)}$.

The second term on the right side of Eq. (3.18) includes four types of one-phonon processes: $a_{\mathbf{q}\lambda} b_{\mathbf{k}}^\dagger b_{\mathbf{k}'}$, $a_{\mathbf{q}\lambda}^\dagger b_{\mathbf{k}}^\dagger b_{\mathbf{k}'}$, $a_{\mathbf{q}\lambda} b_{\mathbf{k}}^\dagger b_{\mathbf{k}'}^\dagger$, and $a_{\mathbf{q}\lambda}^\dagger b_{\mathbf{k}} b_{\mathbf{k}'}$. The first two processes describe the phonon absorption and phonon emission by a magnon where the magnon number is conserved as shown in Fig. 14 (c). The last two processes describe the transformation between one phonon and two magnons where the magnon number is not conserved. Usually, $\tau_{mp}^{(3)}$ is mainly determined by the first two processes.

The third term on the right side of Eq. (3.18) includes two-phonon Raman process $a_{\mathbf{q}\lambda}^\dagger a_{\mathbf{q}'\lambda'} b_{\mathbf{k}}^\dagger b_{\mathbf{k}'}$ that determines $\tau_{mp}^{(4)}$. Other terms like $a_{\mathbf{q}\lambda}^\dagger a_{\mathbf{q}'\lambda'}^\dagger b_{\mathbf{k}}^\dagger b_{\mathbf{k}'}$ and $a_{\mathbf{q}\lambda} a_{\mathbf{q}'\lambda'} b_{\mathbf{k}}^\dagger b_{\mathbf{k}'}$ are relatively unimportant in magnon-phonon relaxation.[224]

The Hamiltonian of magnon-phonon interaction in antiferromagnetic materials is similar to that in ferromagnetic materials in Eq. (3.18). In a two-sublattice antiferromagnet, there are two kinds of creation and annihilation operators: $\alpha_{\mathbf{k}}$, $\beta_{\mathbf{k}}$, $\alpha_{\mathbf{k}}^\dagger$, and $\beta_{\mathbf{k}}^\dagger$,[224,225] correspond to two types of magnons, respectively. Therefore, aforementioned magnon scattering must be changed through $b_{\mathbf{k}} \to \alpha_{\mathbf{k}}, \beta_{\mathbf{k}}$ and $b_{\mathbf{k}}^\dagger \to \alpha_{\mathbf{k}}^\dagger, \beta_{\mathbf{k}}^\dagger$. The detailed Hamiltonian and mathematical expressions of relaxation times for both ferromagnets and antiferromagnets can be found in Refs. [171, 224, 226, 227].[171,224,226,227]

The magnon-phonon relaxation time has been experimentally measured by several methods. The first method is fitting the magnon-phonon TTM[174] where the effective temperature of magnons ($T_m$) and effective temperature of phonons ($T_p$) are assigned[197,203] and $g_{mp} = \frac{C_m C_p}{C_T} \frac{1}{\tau_{mp}}$ is defined as the magnon-phonon coupling constant in analogous to the electron-phonon coupling constant in electron-phonon TTM. $C_p$ and $C_m$ are heat capacities of phonons and magnons, respectively, and $C_T = C_p + C_m$. $\kappa_m$ and $\kappa_p$ are the magnonic thermal conductivity and lattice thermal conductivity, respectively. The characteristic length of magnon-phonon interaction is noted as $\lambda_{mp} = \sqrt{\kappa_m/g_{mp}}$. Once $g_{mp}$ and $\lambda_{mp}$ are measured, one can easily obtain $\tau_{mp}$. Agrawal et al.[228] measured the spatial dependent magnon temperature in YIG on Gadolinium Gallium garnet (GGG) substrate by using the spin Seebeck effect. They fitted the value of $\lambda_{mp}$ by calculating the magnon temperature when the phonon temperature is fitted with a Boltzmann sigmoid function. The fitting results are shown in Fig.15(a). They found that the fitted value of $\lambda_{mp}$ is smaller than the calculated value by Xiao et al.[229] Hohensee et al.[230] used the TTM to analyze the time-domain thermoreflectance (TDTR) data of $Ca_9La_5Cu_{24}O_{41}$ as shown in Fig. 15(b). They found that $g_{mp}$ increases by two orders of magnitudes with increasing temperature.

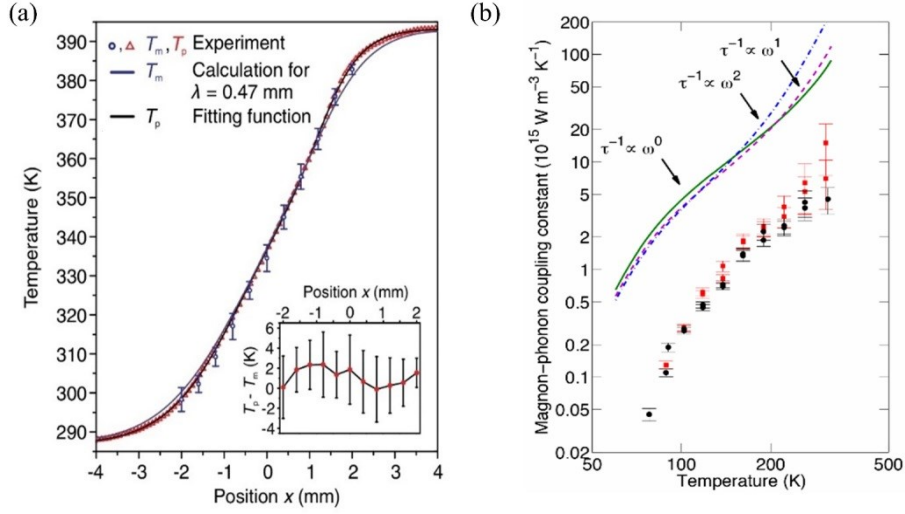

FIG. 15. (a) The measured phonon temperature $T_p$ (open triangles) and magnon temperature $T_m$ (open circles) temperatures. The inset shows the difference between $T_p$ and $T_m$.[228] (b) The measured the magnon-phonon coupling parameters $g_{mp}$ of $Ca_9La_5Cu_{24}O_{41}$ as a function of temperature for two different samples (red squares, black circles). Lines are maximum $g_{mp}$ assuming all scattering events are thermalizing with different scattering rates.[230]

Another method was proposed by Chen et al.[231] through analyzing the thermal conductivities near the ferromagnetic/antiferromagnetic transition temperatures. They used the thermal bridge method to measure the temperature dependence of thermal conductivity of antiferromagnet $MnPSe_3$ nanoribbons whose critical temperature, or Néel temperature, is around 68 K. Clear kinks were found around $T_N$ due to the magnon-phonon interaction. When the temperature is far above $T_N$, for example above 180 K, the measured thermal conductivity is completely determined by phonons. The authors used $(a+bT)^{-1}$ to fit the lattice thermal conductivity and extrapolated the fitting line to low temperatures. Then the difference between the inverse of measured thermal conductivity ($\kappa_{exp}^{-1}$) and the inverse of fitted curve ($\kappa_{fit}^{-1}$) reveals the strength of magnon-phonon interaction. It is reasonable to assume that the ratio of scattering ratio satisfies $\frac{\tau_{mp}^{-1}}{\tau_{p-p}^{-1}+\tau_{p-b\&i}^{-1}} = \frac{\kappa_{exp}^{-1}-\kappa_{fit}^{-1}}{\kappa_{exp}^{-1}}$ where $\tau_{p-p}^{-1}$ is the scattering rate of phonon-phonon scattering and $\tau_{p-b\&i}^{-1}$ is the scattering rate of phonon-boundary and phonon impurity scattering. Figure 16 shows that the ratio of the scattering rate of magnon-phonon coupling is one order of magnitude smaller than the scattering rate of the total phonon scattering rate. A nonmonotonic sample thickness dependence of magnon-phonon relaxation time was also observed.

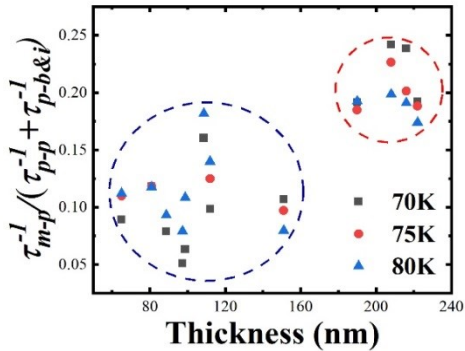

FIG. 16. Ratio of scattering rates versus thickness at 70 K, 75K, and 80K.

The Brillouin light-scattering technique has also been used for the measurement.[232-234] Bozhko et al.[235] used the wave-vector resolved Brillouin light-scattering to measure the magnon-phonon spectra of YIG. The Brillouin light-scattering intensity maps are shown in Fig. 15 with different magnetic fields and pumping conditions. The population near the intersection of magnon and phonon dispersion curves is found to be large due to the strong magnon-phonon coupling and hybridization. Comparing Fig. 17(a) and 17(b), one can find that the magnon dispersion curve and the intersection region move to higher energies when the magnetic field is enhanced.

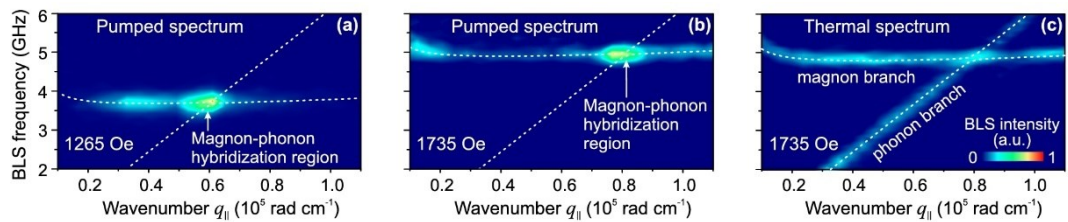

FIG. 17. Magnon-phonon spectra measured by the wave-vector resolved Brillouin light-scattering and their population under different pumping conditions. White dashed lines represent the calculated dispersion relation for the lowest magnon branch, hybridized with a transversal acoustic mode. The bias magnetic fields are (a) 1265 Oe and (b) 1735 Oe, respectively, when the pump power is 2.6 W. Brillouin light-scattering intensity map without pump power is also shown in (c).[235]

In addition, other methods such as the inelastic neutron scattering,[236] the Raman scattering,[237] and the dynamical transport experiments[238] were also proposed to measure the magnon-phonon relaxation time.

The spin Seebeck effect depends strongly on the magnon relaxation time because $S_m \sim \tau$. Therefore, the quality of fabricated magnetic material is important since the relaxation time of magnon-impurity scattering, $\tau_{el}$, in Eq. (3.15) is longer in clean samples. Another way to enhance the spin Seebeck coefficient is by enhancing the relaxation time of magnon-phonon interaction, $\tau_{mp}$, which depends on both magnon dispersion and phonon dispersion. Thanks to the magnetic field dependence of magnon dispersion in ferromagnetic insulators and antiferromagnetic insulators, $\tau_{mp}$ and spin Seebeck coefficient can be changed by applying magnetic field[208] as shown in Fig. 18. The sign of spin Seebeck coefficient reverses when the direction of the magnetic field is reversed.

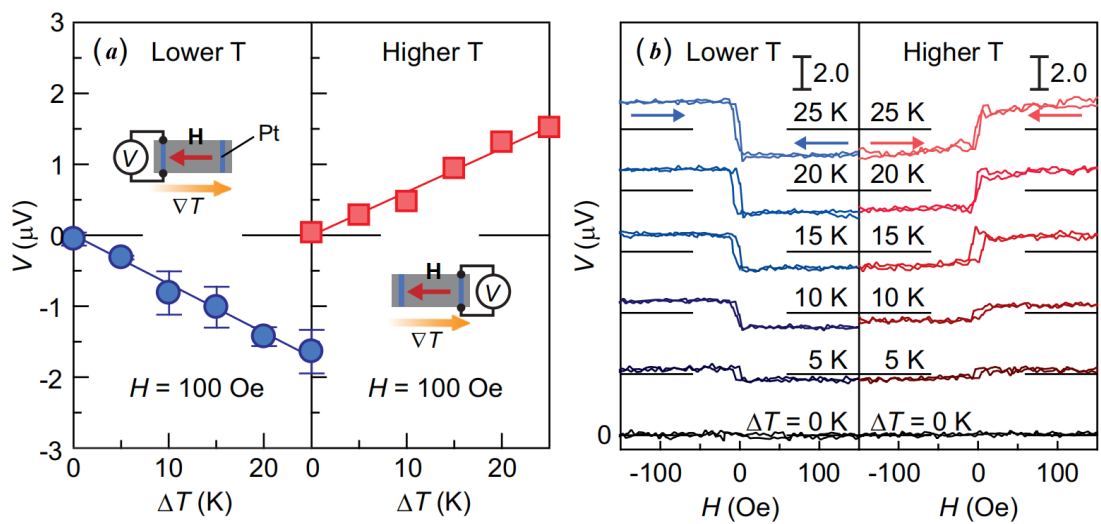

FIG. 18 (a) Voltages generated in Pt wires attached to the low-temperature end (300 K) and to the hot end (300 K+$\Delta T$) with different $\Delta T$ in La:YIG/Pt structure. An external magnetic field $H$ is applied along the direction of the temperature gradient. (b) Voltages versus magnetic field strength with different $\Delta T$.[208]

Recently, a spin Seebeck effect anomaly was discovered due to the formation of magnon polaron which is a magnon-phonon hybrid state. Magnon polaron origins from $H_{mp}^{(2)}$ in Eq. (3.18) which is usually negligible compared to the contributions from $H_{mp}^{(3)}$ and from $H_{mp}^{(4)}$. Magnons and phonons are quantized spin waves and mechanical waves, respectively. If their energies and wavevectors are the same, these two waves would be coherent. However, the wavevector dependence of dispersion curves is different for magnons and phonons where it is linear for phonons and quadratic for magnons as shown in Fig. 19 (a). The coherence effect is important when there are intersections between their dispersion curves. When the magnon dispersion curves are shifted upwards to be tangent with the phonon dispersion in the presence of a magnetic field as shown in Figs. 19(a) and (e), a magnon-phonon hybridization near the intersection region leads to the formation of a new state, called magnon polaron.[239,240] Kikkawa et al.[240] found an anomaly in the spin Seebeck effect signal when magnon polaron is formed in YIG as shown in Fig. 19 (b), (d), (f), (g). Further studies showed that the shape of anomalies is temperature dependent and reveals the underlying physics of magnon scattering strength.[241-243] Moreover, Xi et al.[244] found a significant enhancement of magnon-phonon scattering rate ($\tau_{mp}^{-1}$) when the magnon polaron is formed. These findings provide an effective way to manipulate the magnon relaxation time and to enhance the spin Seebeck coefficient.

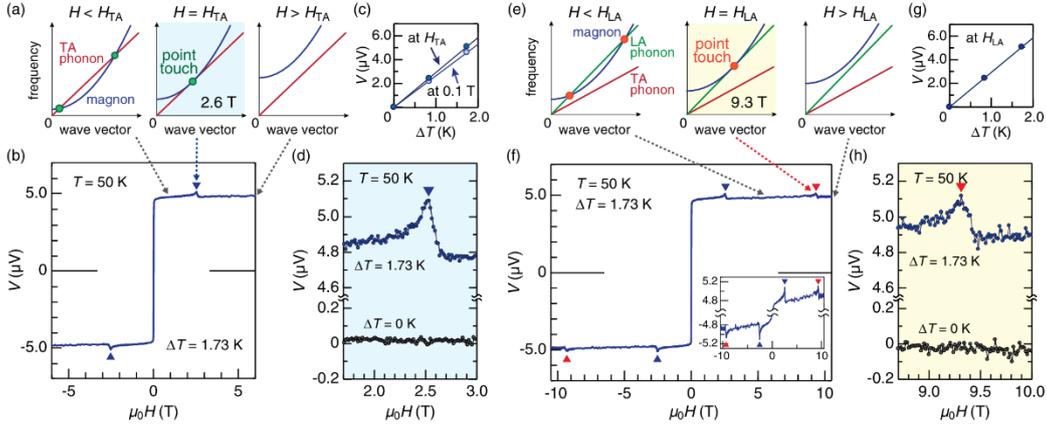

FIG. 19. (a,e) Magnon, TA-phonon, and LA-phonon dispersion relations for YIG in the presence of different magnetic fields. Magnon and TA-phonon (LA-phonon) dispersions touch when $H_{TA}$=2.6 T ($H_{LA}$=9.3 T). (b,f) Magnetic field dependence of voltage due to inverse spin Hall effect (V). (c,g) Voltage versus temperature difference. (d) Magnified view of V around $H_{TA}$. (h) Magnified view of V around $H_{LA}$.

## C. Effect of magnons on thermal conductivity

The progress on the thermal conductivity of magnetic materials is reviewed in this section. The case is simple in magnetic metals where the heat current is mainly carried by electrons and the contributions from magnons and phonons are negligible. The electronic thermal conductivity is approximately $\kappa_e \approx \kappa_{e,\uparrow} + \kappa_{e,\downarrow} = L_0(\sigma_\uparrow + \sigma_\downarrow)T$ where $L_0$ is the Lorenz number and the Wiedemann-Franz law is still satisfied. Therefore, the electronic thermal conductivity in magnetic metal weakly depends on the spin degree of freedom.

In magnetically ordered insulators, the overall thermal conductivity is not simply the summation of magnonic thermal conductivity, $\kappa_m$, in Eqs. (3.8) and (3.10) and lattice thermal conductivity $\kappa_p$. The magnon-phonon coupled transport can be understood by solving the two-temperature model. Considering a temperature gradient is applied on a given sample with length $L$, an effective thermal conductivity ($\kappa_{\text{eff}}$) can be obtained with adiabatic boundary conditions $\frac{dT_m}{dx}\big|_{x=-L/2} = \frac{dT_m}{dx}\big|_{x=L/2}=0$:[173]

$$\kappa_{\text{eff}} = (\kappa_m + \kappa_p)\left[1 + \frac{\kappa_m}{\kappa_p}\frac{\tanh\left(\frac{L}{2l_{mp}}\right)}{\frac{L}{2l_{mp}}}\right]^{-1}, \quad (3.19)$$

where $l_{mp} = \left(g_{mp}\frac{\kappa_m+\kappa_p}{\kappa_m\kappa_p}\right)^{-1/2}$ is the characteristic length of the magnon-phonon interaction. If the interaction is strong and the length of the sample is long, one can easily find that $\lim_{L \gg l_{mp}} \kappa_{\text{eff}} = \kappa_m + \kappa_p$. In contrast, if the magnon-phonon interaction is weak and the length of the sample is short, we have $\lim_{L \ll l_{mp}} \kappa_{\text{eff}} = \kappa_p$. Therefore, the effective thermal conductivity depends on the ratio between $L$ and characteristic length, $l_{mp}$.

Eq. (3.10) shows that $\kappa_m$ depends on magnonic heat capacity, magnon group velocity, and magnon mean free path. In the absence of an external magnetic field and at low temperatures, the magnonic heat capacity of the ferromagnet is proportional to $T^{3/2}$ and that of antiferromagnet is proportional to $T^3$.[171] The magnon group velocity can be obtained from its dispersion relations. The magnon mean free path is determined by various scatterings[245] such as magnon-magnon scattering, magnon-phonon scattering, magnon-impurity scattering, magnon-boundary scattering, and magnon-electron scattering, *etc*. Typical values of the magnon mean free path measured in YIG[246] strongly depend on temperature as shown in Fig. 20. It is clear that the magnon mean free path decreases rapidly from millimeters to micrometers with increasing temperature due to stronger scattering strength.

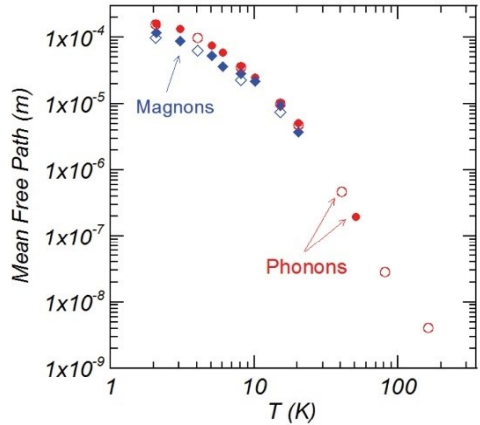

FIG. 20. Measured magnon mean free paths at different temperatures in YIG. The phonon mean free paths are also plotted for comparison. Reprinted with permission from APS.[246]

Theoretically, Chernyshev and Brenig[247] have employed the Boltzmann transport equation to calculate the magnonic thermal conductivity in a 2D antiferromagnet. All possible scatterings have been considered in their calculations. Rozhkov and Chernyshev[248] calculated the magnonic thermal conductivity in $Sr_2CuO_3$.

They find that $\kappa_m \propto T^2$ at low temperatures, $\kappa_m \propto T$ at intermediate temperatures, and $\kappa_m$ saturates to a constant value at high temperatures.

A direct measurement of magnonic thermal conductivity is challenging because of its coupling with phonons. One possible way is to measure the magnetic field dependence of total thermal conductivity.[249,250] In contrast to phonons, magnons are sensitive to the applied magnetic field ($H$). Consequently, both heat capacity and thermal conductivity depend on the external magnetic field. This phenomenon provides an effective way to distinguish magnonic thermal conductivity from phononic thermal conductivity.[249] When $H$ is large enough, the magnonic thermal conductivity would be reduced. The measured saturated value of thermal conductivity at a high field is believed to be the lattice thermal conductivity. Then the difference between the zero-field thermal conductivity and the high-field thermal conductivity can be used to estimate the magnonic thermal conductivity when the effect of magnons on the lattice thermal conductivity is ignored as an approximation. Figure 21 shows the measured field-dependent thermal conductivity in YIG at two different temperatures.[246] The measured thermal conductivity decreases with increasing magnetic field and eventually saturates at a high field. This proves the suppression of magnonic thermal conductivity. Furthermore, the magnonic thermal transport has also been measured in low-dimensional magnetic materials such as spin -1/2 ladder system $Sr_{14-x}Ca_xCu_{24}O_{41}$,[251] 2D quantum magnet $La_2CuO_4$[252] and $Nd_2CuO_4$[253,254], *etc*. It should be emphasized that this method is not applicable for antiferromagnets,[255] because of the nonmonotonic field dependence of magnon dispersion of antiferromagnets.

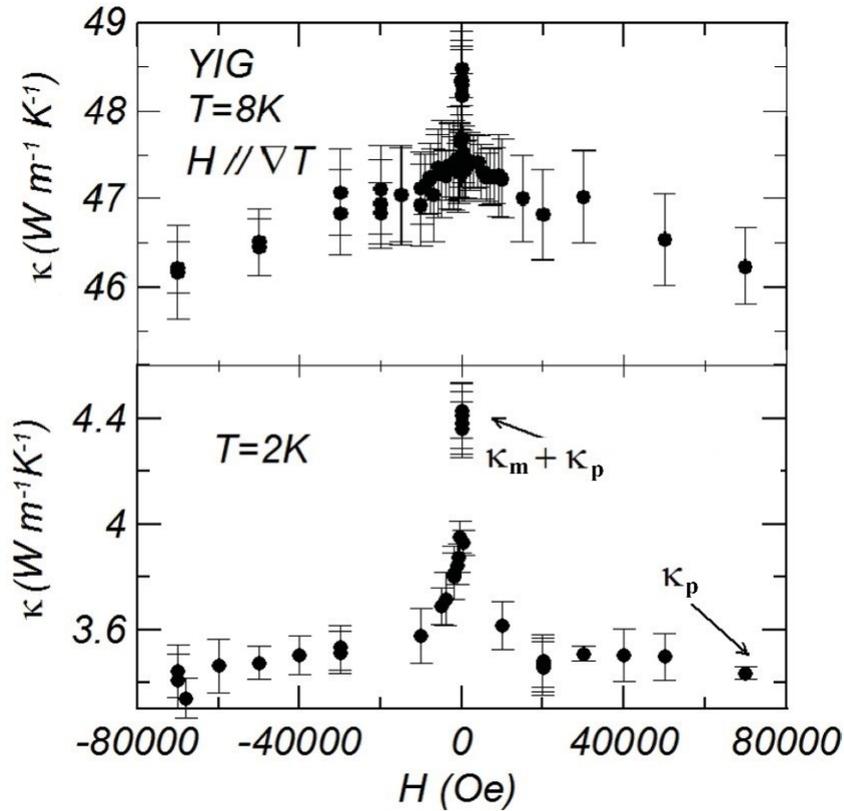

FIG. 21. Magnetic field dependence of thermal conductivity in YIG when $T$=2 K and 8 K. The magnonic thermal conductivity and lattice thermal conductivity are distinguished. Reprinted with permission from APS.[246]

The lattice thermal conductivity has been found to be strongly affected by the presence of magnon in many kinds of magnets.[256-258] Theoretically, the lattice thermal conductivity in a magnet cannot be easily calculated because phonons properties can be strongly affected by magnons. For example, the phonon dispersion could be modified which changes the phonon heat capacity ($C_p$) and phonon group velocity ($v_p$). It is a practical approximation by ignoring the variations of $C_p$ and $v_p$ and merely considering the change of magnon-phonon in phonon mean free path ($l_p$). In the presence of magnon-phonon interaction, the phonon mean free path would be shortened because $l_p^{-1} = l_{pp}^{-1} + l_{pt}^{-1} + l_{pm}^{-1}$ where $l_{pp}$ is the mean free path of phonon-phonon scattering, $l_{pt}$ is the mean free path of phonon-impurity, phonon-boundary, and phonon-electron scatterings, $l_{pm}$ is the mean free path of magnon-phonon scattering. As a result, the existence of magnon-phonon interaction suppresses the contribution of phonons, whose energy and wave vector are close to the intersection of magnon and phonon dispersion curves, to lattice thermal conductivity. Dixon[259] theoretically calculated the relaxation times of phonons due to the two-magnon one-phonon scattering process (see $H_{mp}^{(3)}$ in Eq. (3.18)) in the presence of an external magnetic field. The magnon dispersion relation can be shifted to higher energies by the magnetic field. As a result, the strength of magnon-phonon interaction as well as the lattice thermal conductivity can be changed by magnetic field.[249,260] Rives et al.[249] found that the temperature-dependence of thermal conductivity ($\kappa_H$) of GdCl$_3$ at different temperatures as shown in Fig. 22. When $T$=0.46 K and 0.77 K, $\kappa_H$ monotonously increases with increasing magnetic field. When $T$=1.3 K, $\kappa_H$ first decreases and then increases with increasing magnetic field. The reason was attributed to the comparison between the phonon energy at the dispersion intersection ($\hbar\omega_{int}$) and $4k_BT$ around which the phonon contribution is peaked. Moreover, the saturated value of thermal conductivity at high field is found to be larger than the zero-field value, i.e. $\lim_{H\to\infty} \Delta\kappa > 0$. This implies that the method to determine magnonic thermal conductivity by using a magnetic field is not applicable to GdCl$_3$.

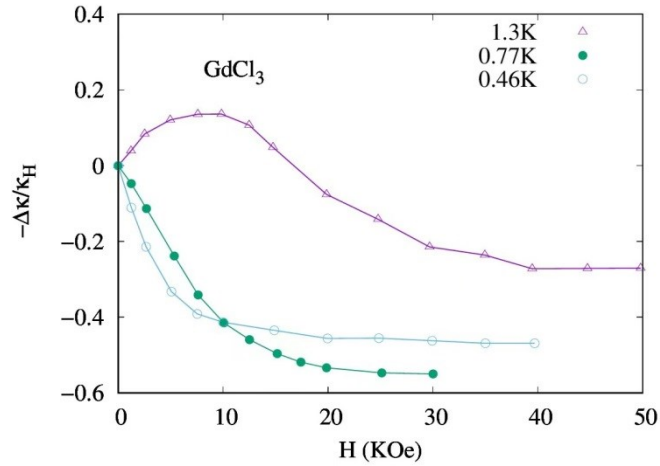

FIG. 22. The change of magnetic field-dependent thermal conductivity in GdCl$_3$ for three different temperatures. $\kappa_H$ is the measured thermal conductivity and $\Delta\kappa=\kappa_H-\kappa_{H=0}$. Reprinted with permission from AIP.[249]

Another effect of the magnon-phonon interaction on thermal conductivity has been observed near the Curie temperature ($T_c$) in ferromagnet and near the Néel temperature ($T_N$) in antiferromagnet. Yang et al. [261] found there exists a minimum of thermal conductivity near $T_c$ in van de Waals layered ferromagnet Cr$_2$Si$_2$Te$_6$ as shown in Fig. 23(a). A kink of thermal conductivity near $T_N$ in antiferromagnet has been widely found in various materials such as MnBi$_2$Te$_4$,[262] MnPSe$_3$ nanoribbons,[260] Ba$_3$CoSb$_2$O$_9$,[263] and FeCl$_2$.[264] Figure 23(b) shows the kink of lattice thermal conductivity in MnPSe$_3$ nanoribbons with different film thickness around 68 K. These phenomena originate from the sudden appearance of magnon modes when the temperature is below the transition temperatures.

In addition, there are many exotic phenomena caused by magnon-phonon interaction. For example, a magnetic field-dependent thermal conductivity dip near the Morin temperature in antiferromagnet α-$Fe_2O_3$ was found by Xi et al.[265] We would like not to discuss these works in detail.

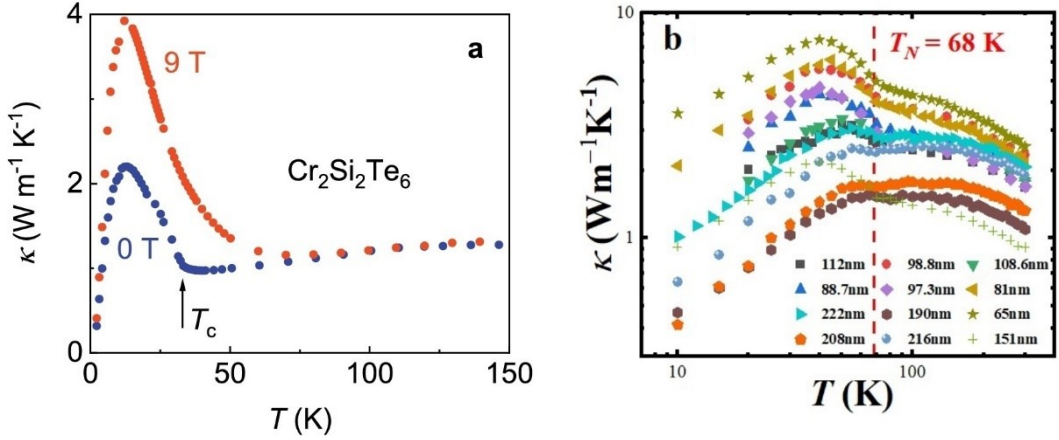

FIG. 23. (a) Thermal conductivity of $Cr_2Si_2Te_6$ with and without external magnetic field. (b) Thermal conductivity of $MnPSe_3$ nanoribbon with different thicknesses.[200,260]

## D. Remaining challenges and outlooks

### 1. Complexity of coupled spin-heat transport in magnetic metals and semiconductors

While in magnetic insulators, the angular momentum is dominantly transported by spin waves (magnons), there are different channels for angular momentum transport in magnetic metals and semiconductors. In magnetic metals, both conduction electrons and magnons are spin current carriers and heat carriers. Then one needs to consider coupled electron- spin-magnon-phonon. It is challenging to study the interplay among the electron-phonon coupling, the magnon-phonon coupling, and the electron-magnon coupling. It is a significant challenge to distinguish the contributions of each carrier to thermal and spin transport. In magnetic semiconductors, both electrons and holes are charge current carriers and spin current carriers. Then the two-current model must be extended to the six-current model because holes possess a spin of one-half and a spin of three-halves. The situation is more complicated when these six channels are not parallel due to the spin-flipping scatterings and the spin-orbital coupling.

### 2. Theoretical limits in spin Seebeck effects

Currently, the measured values of the spin Seebeck coefficient are always smaller than the value of the charge Seebeck coefficient. Finding a s spin Seebeck coefficient larger than the charge Seebeck coefficient could be an interesting direction, which would significantly boost the performance of spin thermoelectric devices. Another interesting limit case would be zero charge Seebeck coefficient but finite spin Seebeck coefficient, which will enable the experimental observation of a pure spin Seebeck effect.

### 3. Thermally driven transport of orbital angular momentum

The orbital angular momentum driven be temperature gradient could be a new research direction in the future. In experimental observation of the spin Seebeck effect (in Fig. 13), strong spin-orbit coupling in metals with heavy atomic mass such as Pt and W are required while other light metals like Cu and Al are useless. Therefore, the orbital angular momentum must play an important role in thermal and spin-coupled transport. One may ask: is it possible to generate and utilize the orbital current in a similar way to the spin Seebeck effect? Coupled thermal and orbital angular momentum transport could be an interesting topic beyond thermal and spin coupled transport.

## IV. COUPLED TRANSPORT OF HEAT AND IONS/MOLECULES

The transport of chemical species in a nonisothermal system is a fundamental manifestation of nonequilibrium thermodynamics. Such coupled transport of heat and ions/molecules is of great importance in energy conversion and storage,[10,266] water production and desalination,[267,268] manipulation of colloidal particles,[269-271] isotope enrichment,[272] radioactive waste disposal,[273] and so on. In particular, this section focuses on two closely related phenomena of temperature gradient-driven ion/molecule transport, ionic thermoelectric effects, and thermo-osmosis, due to their promising applications in low-grade heat energy harvesting. We first discuss the nonequilibrium thermodynamic principles of coupled thermal and ion/molecule transport in Section IV.A, including both the phenomenological and microscopic description of ionic thermoelectric and thermo-osmotic effects. We then review the strategy for tuning ionic Seebeck coefficients in Section IV.B through solvent structural effects, unipolar thermodiffusion, nanoconfinement effects, and the synergistic contribution to thermal voltage through thermodiffusion and redox reactions. In Section IV.C, methods for tailoring thermo-osmotic effects are discussed, including surface hydrophilicity effects, size effects of nanochannels, and charge regulation effects. Section IV.D concludes this section with remaining challenges and outlooks.

### A. Nonequilibrium thermodynamic principles

The diffusion of chemical species is commonly driven by the chemical potential due to the nonuniform concentration profile, which is well-described by Fick's law. The chemical flux can also be driven by the temperature gradients. A notable example is the Ludwig-Soret effect (also known as thermodiffusion) where a concentration gradient is established by the temperature gradient, first discovered in the late 19th century.[274] When the Soret effect takes place in electrolytes, a measurable thermal voltage can be generated by the Soret separation of cations and anions, due to the breakdown of local charge neutrality. This effect is also referred to as the ionic Seebeck effect, resembling the thermoelectricity of electrons observed in metals and semiconductors. While the thermodiffusion effect can occur in bulk mixtures, the thermo-osmosis effect, *i.e.* the creep fluid flow induced by the temperature gradient, can only be observed near a confining surface parallel to the temperature gradient, which was first discovered by Lippman in early 20th century.[275] Both thermal-induced transport phenomena can be described by the phenomenological Onsager transport theory:

$$\begin{bmatrix} J_i \\ \mathbf{Q} \end{bmatrix} = \begin{bmatrix} L_{ii} & L_{iq} \\ L_{qi} & L_{qq} \end{bmatrix} \begin{bmatrix} -\nabla_T \bar{\mu}_i / T \\ \nabla(T^{-1}) \end{bmatrix}, \qquad (4.1)$$

where $J_i$ and $\mathbf{Q}$ are the chemical flux of species $i$ and the heat flux respectively, $L_{ii}$, $L_{iq}$, $L_{qi}$, and $L_{qq}$ are linear transport coefficients, and $\nabla_T \bar{\mu}_i$ denotes the gradient of electrochemical potential at constant temperature.[276] According to Onsager reciprocity, the cross-transport coefficients $L_{qi} = L_{iq}$. From the thermodynamic relations: $dG = -SdT + Vdp + \sum_i \bar{\mu}_i dn_i$, the isothermal gradient of electrochemical potential can be written as:

$$\nabla_T \bar{\mu}_i = \nabla_T \mu_i + z_i e \nabla \phi + v_i \nabla p, \qquad (4.2)$$

where $v_i$ is the volume occupied by chemical species $i$, $\phi$ is the electrostatic potential, $\mu_i$ is the chemical potential, respectively. In a bulk system with no pressure gradient, by substituting Eq. (4.2) and $\nabla_T \mu_i = k_B T \nabla \ln n_i$ into Eq. (4.1), the Nernst-Plank equation of the chemical flux can be derived:

$$J_i = -D_i \left( \nabla n_i + \frac{z_i e n_i}{k_B T} \nabla \phi + n_i \frac{Q_i^*}{k_B T} \nabla T \right), \tag{4.3}$$

where $D_i = L_{ii} k_B / n_i$ is the diffusivity, $Q_i^* = L_{iq}/L_{ii}$ is termed as the heat of transport. The heat flux can be also simplified as:

$$\mathbf{Q} = \sum_i Q_i^* \mathbf{J}_i - \kappa \nabla T, \tag{4.4}$$

where $\kappa = \left(L_{qq} - \sum_i L_{iq} L_{qi}/L_{ii}\right)/T^2$ is the thermal conductivity of the mixture. The heat of transport $Q_i^*$ can be understood by setting the system to isothermal conduction $\nabla T = \mathbf{0}$, such that the total heat flux is written as $\mathbf{Q} = \sum_i Q_i^* \mathbf{J}_i$. $Q_i^*$ is therefore similar to the Peltier heat carried along with the ionic flux $\mathbf{J}_i$, which determines the magnitude of Soret separation:

$$\nabla \ln n_i = \frac{Q_i^*}{k_B T} \nabla \ln T, \tag{4.5}$$

The concept of "heat of transport" was proposed as early as 1926 by Eastman.[277] This quantity can be understood microscopically as shown in Fig. 24(a). Due to the temperature gradient, the local free energy density profile becomes non-uniform, and when an ion migrates from the region at a temperature $T_{x+dx}$ to the region at temperature $T$, the associated free energy change is expressed as $[g_i(T_x) - g_i(T_{x+dx})]v_i$, where $g_i$ is the local free energy density and $v_i$ can be viewed as the volume occupied by ion $i$. The derivative of local $g_i$ at nonequilibrium with respect to temperature can be viewed as the "entropy" carried with the ion transport, known as Eastman entropy of transfer:

$$S_i^* = \frac{[g_i(T_x) - g_i(T_{x+dx})]}{dT} v_i = -\frac{\partial g_i}{\partial T} v_i. \tag{4.6}$$

with $-\partial g_i / \partial T$ the entropy density around the ion $i$. The entropy flux can be expressed as $\mathbf{J}_S = \mathbf{J}_Q/T$, thus Eastman entropy of transfer can be expressed as:

$$S_i^* = \frac{Q_i^*}{T}. \tag{4.7}$$

In the simplest case, all thermodiffusive ions are not redox-reactive, and the thermal voltage generated by the temperature gradient is non-Faradaic. In this case, an internal electric field $\mathcal{E}$ is established when the cations and anions are mismatched Soret concentration profiles resulting in the breaking of local charge neutrality. The Seebeck coefficient, $S_{td}$, is defined as the proportionality between the electric field and the temperature gradient: $\mathcal{E} = S_{td} \nabla T$. By linear-order perturbation to the Poisson-Nernst-Planck equations, the ionic Seebeck coefficient is expressed as:

$$S_{td} = \frac{k_B}{e}(w_+ S_+^* + w_- S_-^*), \tag{4.8}$$

where the weighting coefficients $w_\pm$ are determined by the boundary conditions, and the subscript $td$ indicates the ionic Seebeck coefficient is due to ionic thermodiffusion. For a closed system with an ion-blocking boundary, the weighting coefficients $w_\pm = \pm/(z_+ - z_-)$ are only determined by the valance charges $z_\pm$ of cations and anions. On the other hand, when the boundaries allow for ionic exchange with the thermal and chemical reservoirs, then weighting coefficients are $w_\pm = \pm \mu_\pm / (z_+ \mu_+ - z_- \mu_-)$, which are simultaneously affected by the valence charges and the ionic mobilities. Eqs. (4.7-4.8) point out that a large Seebeck coefficient is related to the weighted mismatch in Eastman entropy of transfer. In fact, giant Seebeck coefficients are often observed in systems with extremely asymmetrical ions, such as the polyelectrolyte in which the poly-ions are also most immobile while the thermodiffusion of the mobile counterion carriers is responsible for the induced thermopower.

From the microscopic point of view, ion migration along the temperature gradient would result in the exchange of solvent molecules with the surroundings. Agar adopted this kind of picture and derived an expression Eastman entropy of transfer using a hydrodynamic method, by integrating the entropy density upon the flow field surrounding the ion due to ionic thermodiffusion.[276] When the ion is sufficiently small such that the perturbative flow field can be neglected, the Eastman entropy of transfer at the dilute limit is reduced to the solvation entropy,[278] which can be qualitatively described by the Born solvation model:

$$S_i^* = -\frac{(z_i e)^2}{8\pi\epsilon R_i}\frac{d(\ln \epsilon)}{dT}, \quad (4.9)$$

where $z_i$ is the ionic valence, $\epsilon$ is the dielectric constant and $R_i$ is the effective radius of the ion. A more comprehensive theory of $Q_i^*$ and $S_i^*$ derived by Helfand and Kirkwood by analyzing the manybody correlation functions based on the Bogolyubov–Born–Green–Kirkwood–Yvon (BBGKY) hierarchy theory,[279] which highlighted the importance of ion-solvent interaction and the microscopic arrangement of molecules around the thermodiffusive ion.

When the system contains redox pairs $O + ne \leftrightarrows R$, the electrodes can exchange electrons with the electrolyte. In this case, the measured thermal voltage is simultaneously contributed by the thermodiffusion of the redox ions and the redox entropy change,[280] known as the thermogalvanic effect (Fig. 24(b)). The thermopower $S_{tg}$ of a thermogalvanic cell operating under a temperature gradient is written as:

$$S_{tg} = \frac{(S_O + S_O^*) - (S_R + S_R^*)}{nF} + S_{td}, \quad (4.10)$$

where $S_i$ is the partial molar entropy of the species $i(= O, R)$, $n$ is the electron transfer number and $F$ is the Faraday constant, and $S_{td}$ is the ionic Seebeck coefficient due to the thermodiffusion of all ions. Such heat-to-electric conversion device containing redox pairs is known as the thermogalvanic cell or thermo-electrochemical cell. In most liquid redox electrolytes, the thermodiffusive contribution to the measured $S_{tg}$ is negligible, and the thermopower is reduced to the reaction entropy:

$$S_{tg} \approx -\frac{\Delta S_{rxn}}{nF} = -\frac{S_R - S_O}{nF}. \quad (4.11)$$

However, in systems with strong ion-ion or ion-solvent coupling, the thermodiffusion effects ($S_{td}$) can also make a significant contribution to the thermopower, such as in gel electrolytes and ionic liquids.[281-283]

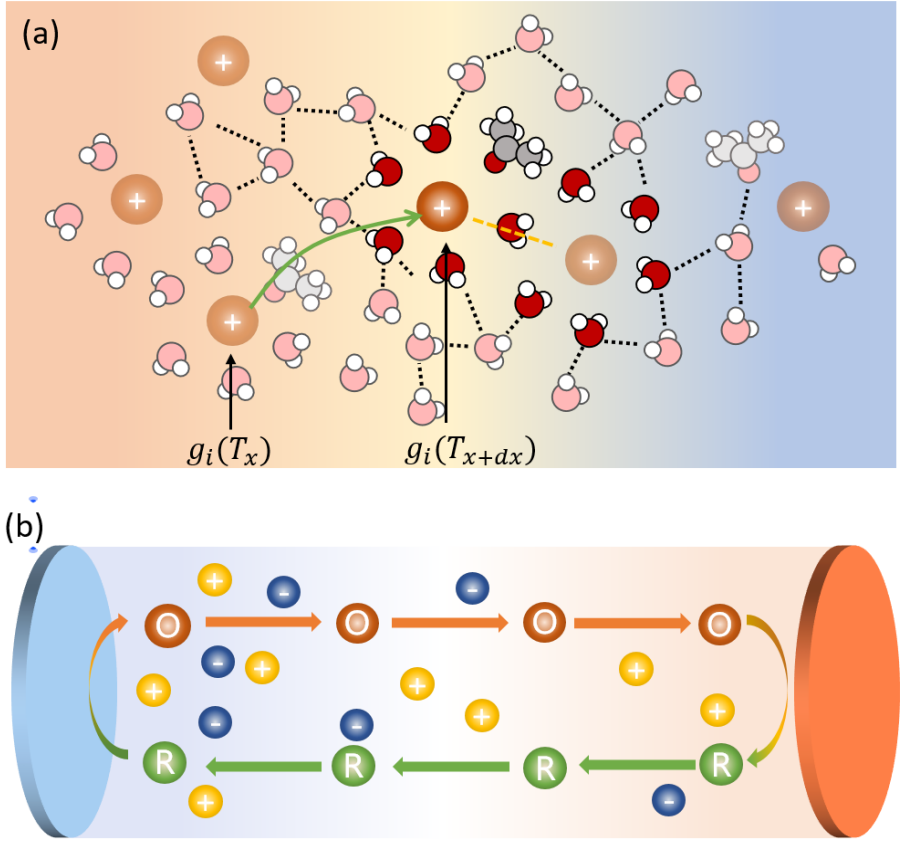

FIG. 24. (a) Schematic of ionic thermodiffusion under a temperature gradient. (b) A non-isothermal thermogalvanic cell containing both thermodiffusive supporting electrolytes and redox pairs.

Now we turn to the discussion on the thermo-osmosis effect in confined liquids. Consider a liquid near a confining boundary under a temperature gradient, as shown in Fig. 25(a). In addition to the thermal force $\nabla(T^{-1})$, the electrochemical potential gradient can be written as $\nabla_T \bar{\mu} = v \nabla p$. The Onsager relation for the volumetric flow rate and heat flux is written as:

$$\begin{bmatrix} J_v \\ \mathbf{Q} \end{bmatrix} = \begin{bmatrix} M_{vv} & M_{vq} \\ M_{qv} & M_{qq} \end{bmatrix} \begin{bmatrix} -\nabla p \\ -\dfrac{\nabla T}{T} \end{bmatrix}, \tag{4.12}$$

where $J_v$ is the volumetric flow rate $J_v = v J_n$ with $J_n$ the chemical flux and the new Onsager transport coefficients are $M_{vv} = v L_{nn}/T$, $M_{qq} = L_{qq}/T$, $M_{vq} = L_{nq} v/T$, $M_{qv} = L_{qn} v/T$ with the Onsager coefficients $L$ relating the flux with thermodynamic forces $-\nabla_T \bar{\mu}/T$ and $\nabla(T^{-1})$. Clearly, the Onsager reciprocity still holds with $M_{qv} = M_{vq}$. The creep flow rate can also be expressed as $J_v = n v \mathbf{u}_s$, with $\mathbf{u}_s$ the hydrodynamic velocity of thermo-osmotic flow, resulting in the interfacial velocity field as shown in Fig. 25.

To determine the thermo-osmosis coefficient $M_{vq}$, one should consider the hydrodynamics of the boundary layer near the confining surface. According to the Gibbs-Duhem relation $n d\mu = -S dT + V dp$, the pressure gradient is related to the temperature gradient through:

$$\frac{\partial p}{\partial x} = \rho(s - s_b)\frac{\partial T}{\partial x} = \rho \frac{h - h_b}{T} \frac{\partial T}{\partial x}, \tag{4.13}$$

where $s$ is the specific entropy of the fluid in the boundary layer, and $s_b = -\partial \mu / \partial T$ is the specific entropy in the bulk fluid; $h$ and $h_b$ are correspondingly the specific enthalpy in the boundary layer and the bulk fluid, and $\rho$ is the density. Therefore, a pressure gradient is induced once there is a temperature gradient $\partial T/\partial x$, , due to the excess enthalpy density in the boundary layer $\delta h(z) = \rho(z)(h(z) - h_b)$:

$$\frac{\partial p}{\partial x} = \frac{\delta h(z)}{T}\frac{\partial T}{\partial x}. \tag{4.14}$$

The thermo-osmotic flux can then be calculated as:

$$J_v = \frac{1}{\mu}\int_0^\infty (z - z_s + b)\frac{\delta h(z)}{T}\left(-\frac{\partial T}{\partial x}\right)dz, \tag{4.15}$$

where $z_s$ is the position of the shearing plane and $b$ is the slip length defined as $u(z=0) = b\left(\frac{\partial u(z)}{\partial z}\right)_{z=0}$, which is also illustrated in Fig. 25(a). The thermo-osmotic coefficient can therefore be determined by integrating the excess enthalpy across the boundary layer:

$$M_{vq} = M_{qv} = \frac{1}{\mu}\int_0^\infty (z - z_s + b)\frac{\delta h(z)}{T}dz. \tag{4.16}$$

This interfacial expression for the thermo-osmosis coefficient was first derived by Derjaguin and Sidorenkov[284] using the nonslip boundary condition in which $b = 0$. The first microscopic observation of the microscopic thermo-osmotic flow field was made by Bregulla *et al.*[285] in 2016 by tracking the migration of tracer nanoparticles. Fu *et al.* and Chen *et al.* further introduced the slippery length $b$ into the thermo-osmosis theory and investigated the impact of surface hydrophilicity on the excess interfacial entropy $\delta h(z) = \rho(z)(h(z) - h_b)$ using MD simulations (Fig. 25(b)). Close to the confining surface, the structure and density of the interfacial fluid deviate from the bulk fluid due to the interaction between the molecules with the surface, which is manifested in the enthalpy density distribution $h(z)$ as shown in Fig. 25(c). The enthalpy density can be calculated from atomic energy and virial stress tensors $h = \rho(z)[\langle u(z)\rangle + \langle p(z)\rangle]$ using MD simulations,[286] where $u(z)$ is the averaged atomic energy $\langle u(z)\rangle$ and the pressure can be calculated using $p(z) = \text{tr}[\mathbb{S}]/3$ with $\mathbb{S}$ denoting the atomic virial stress tensor. Eq. (4.15) also manifests the significance of interfaces in thermo-osmosis: the thermo-osmotic flow is absent in bulk fluid since the excess enthalpy is essentially zero. The thermo-osmotic coefficient $M_{vq}$ can be both negative or positive. For positive excess enthalpies near the surface $\delta h(z) > 0$, the fluid flow moves from the hot side to the cold side, while for negative excess enthalpy the fluid flow is driven toward the hot region.

After discussing the fundamental nonequilibrium thermodynamic principles of thermodiffusion and thermo-osmosis, we summarize recent advances in manipulating the coupled transport of heat and ions/molecules in the following sections.

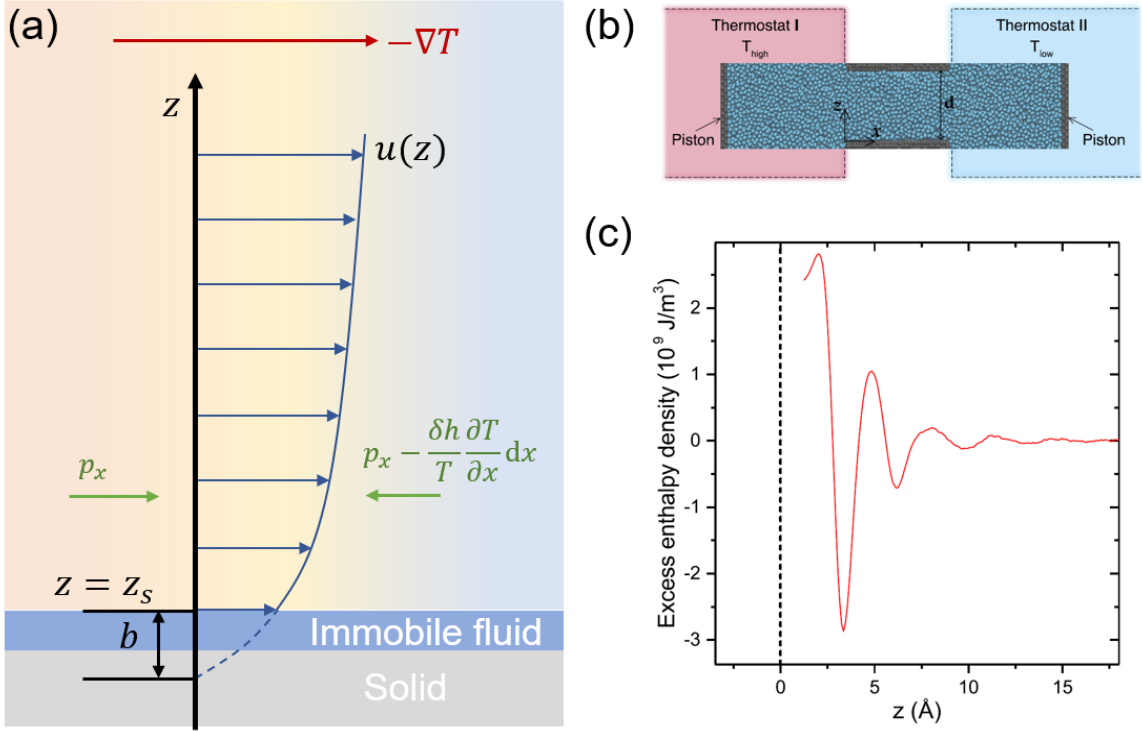

FIG. 25. Thermo-osmosis phenomenon. (a) Schematic of thermo-osmotic flow near a confining surface, where a pressure gradient is generated in response to the temperature gradient due to the excess enthalpy near the surface. The shearing plane position $z_s$ and the slip length $b$ are also indicated. (b) The MD simulation cell for modeling thermo-osmosis across a confining channel. (c) Excess specific enthalpy distribution near the confining graphene surface obtained from MD simulations.

### B. Strategies for enhancing ionic thermopower

Significant progress has been made in the past decade in developing ionic thermoelectric materials with high thermopowers for low-grade heat harvesting or temperature sensing, such as nonaqueous solutions,[287] nanoconfined electrolytes,[288-290] polyelectrolytes containing immobile poly-ions and mobile counterions,[282,291,292] salts or ionic liquids in gel solvents,[281,282,293-297] electrolytes containing complex ions,[298] *etc*. This section summarizes major means for obtaining high thermopower in electrolytes, including (1) solvent structural effects, (2) polyelectrolytes with unipolar thermodiffusion, (3) nanoconfinement effects, and (4) synergistic thermodiffusion-thermogalvanic effects.

#### 1. Solvent structural effects

The Soret thermopower is closely related to the ionic solvation entropy, the solvent structure can therefore significantly affect $S_{td}$. A notable example is the discovery of the huge Soret thermopower as large as 7 mV/K in nonaqueous tetrabutylammoniumnitrate (TBAN), tetradodecylammonium nitrate (TDAN) and tetraoctylphosphonium bromide (TOPB) electrolytes.[287] In electrolytes, the ion field would disturb the intermolecular hydrogen bond network formed among solvent molecules, either facilitating the interconnection of hydrogen bonds (structure-making effect) or breaking apart the network into fragmented clusters with fewer molecules (structure-breaking effect), as illustrated in Fig. 26(a). Such disturbance in the local chemical environment is not captured in the Born solvation model in which the electrolyte is treated as an effective media with dielectric screening effects. As a result, the Born model severely underestimated the dependence of Soret thermopower on the dielectric constant of solvents, compared with experimental measurements of TBAN and TDAN in different solvents (Fig. 26(b)). Such large organic ions can facilitate the formation of a hydrogen bonding network, resulting in a larger heat flux and thereby a higher thermopower.

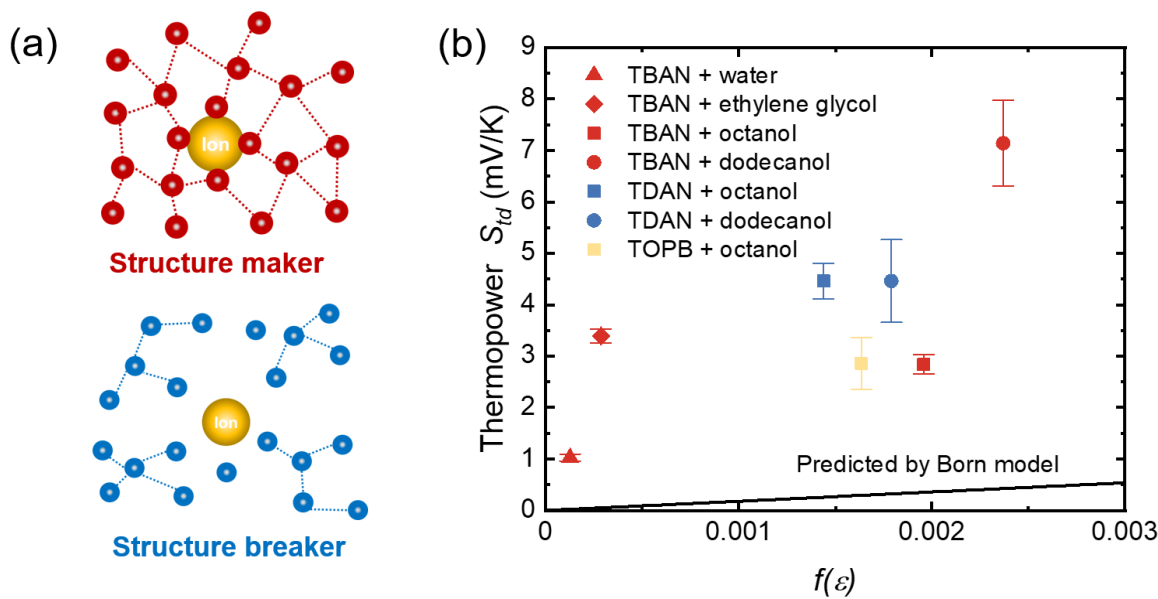

FIG. 26. Structure making/breaking effects for Soret thermopower $S_{td}$ improvement. (a) Soret thermopower $S_{td}$ as a function of the square root of ion concentrations $c$ (mol/L) of TBAN and tetradodecylammonium nitrate (TDAN) in dodecanol. (b) Soret thermopower $S_{td}$ of TBAN salts in various solvents, as a function of dielectric constant $f(\epsilon) = -(r_i\epsilon^2)^{-1}d\epsilon/dT$. Readapted from Ref. 287[287], Copyright 2011 with the permission of AIP Publishing.

For liquid redox electrolytes such as $Fe^{3+}/Fe^{2+}$ and $Fe(CN)_6^{3-}/Fe(CN)_6^{4-}$, the measured thermopower is dominantly the solvation entropy difference between the oxidized state and reduced state. Recent experiments showed that the solvation entropy $\Delta S_{solv}$ can be enhanced by introducing nonaqueous solvents. For instance, Kim et al.[299] and Liu et al.[300] demonstrated that the addition of organic solvents to aqueous $Fe(CN)_6^{3-}/Fe(CN)_6^{4-}$ and $Fe^{3+}/Fe^{2+}$ electrolytes can lead to high $S_{tg}$ values of 2.9 mV/K and -2.5 mV/K, respectively. Inoue et al.[301] systematically studied the solvent effect on the thermopower of $FeCl_3/FeCl_2$ and found that $Fe^{3+}/Fe^{2+}$ in acetone solvents can reach 3.6 mV/K, nearly three times the thermopower of the pristine aqueous solution. This dependence of thermopower of $Fe^{3+}/Fe^{2+}$ on solvents is related to the $D_{4h}$-type deformation of the $FeL_6$ octahedra (L denotes ligands). Chen et al.[302] recently provided a molecular-level insight into the solvent effect on $S_{tg}$. Using the free energy perturbation method and solvation structure analysis, the authors found that the larger difference in probability distribution of solvent dipole orientations would result in a larger $S_{tg}$. Huang et al.[303] found that the redox thermopower can also be affected by the structure-making and structure-breaking effects of supporting ions (Fig. 27(a)). A structure-making ion can enhance the proton network and attract water molecules to form a large solvation shell,[304] where water molecules are more structured than those in pure bulk water. Oppositely, a structure-breaking ion interacts weakly with water molecules, making solvent molecules in the solvation shell of redox ions more disordered. Fig. 27(b) shows the linear correlation between the change in reaction entropy (proportional to the redox thermopower) and the structural entropy of solvents after adding supporting ions.

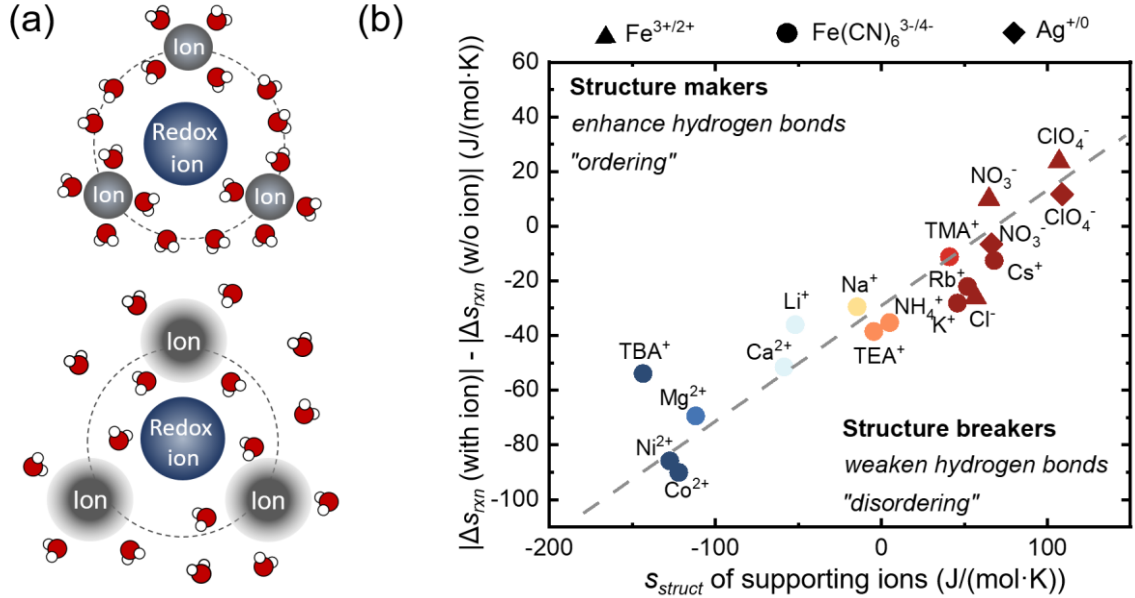

FIG. 27. Effect of supporting electrolytes on the $S_{tg}$. (a) Solvation shell structure changes by adding structural making (upper) and breaking ions (lower). (b) Change of redox reaction entropy of hydrated $Fe^{3+}/Fe^{2+}$ after adding supporting ions. Readapted from Ref. 303 [303] under the Creative Commons Attribution 3.0 Unported Licence, Copyright 2018 by Royal Society of Chemistry.

### 2. Polyelectrolytes with unipolar thermodiffusion

According to Eq. (4.8), asymmetry in $S^*$ of cations and anions is pivotal for enhancing thermopower $S_{td}$. Polyelectrolyte is an extreme case of cation-anion asymmetry, with the poly-ion almost immobile to external driving forces and the mobile counterions participating in unipolar thermodiffusion. For example, a large thermopower of 11 mV/K was reported by Zhao et al.[291] in polyethylene oxide (PEO)-NaOH polyelectrolyte in which the ionic charge transport is dominantly by $Na^+$. However, simply neglecting the thermodiffusion of poly-ions still cannot explain the large thermopower observed experimentally. For example, the Eastman entropy of transfer for protons in water is 44.3 J/(mol·K)[276] only corresponds to a maximum $p$-type $S_{td}$ of 0.23 mV/K at the unipolar limit, which is still an order of magnitude lower than the measured ~ 5.5 mV/K in proton conductive membranes.[292] In polyelectrolytes and gel solutions, the polymer chains contain a large amount of charged or polar functional groups[293] whose interaction with the mobile ions contributes significantly to $S_i^*$. Interestingly, the counterion-polyion interaction can be even leveraged to tune $S_{td}$ in a bidirectional manner. Recent work by Huang's group discovered that ion coordination between $Li^+$ and $BF_4$/TFSI anions in gel electrolytes can be used to tune the ionic thermopower from the $n$-type -15 mV/K to the $p$-type 17 mV/K.[305]

Nanoconfinement could be another effective method to induce asymmetric thermodiffusion between cations and anions. When the distance between confining boundaries approaches Debye length, the electric double layers (EDLs) become overlapped, resulting in the breakdown of local charge neutrality and selective transport of cations or anions. Dietzel and Hardt [288] theoretically proposed to improve $S_{td}$ by confining electrolytes in a slit open channel. Experimentally, Li et al. observed a large thermopower of 24 mV/K of sodium polyethylene oxide (Na-PEO) electrolytes confined in cellulosic membranes.[289] Note that the experimental measurements for confined electrolytes are performed in a closed system with ion-insulating electrodes, whereas the theory by Dietzel and Hardt neglected the effects of ion-insulating boundaries.[288] Qian et al.[290] derived an analytical model to describe the nanoconfinement effect on $S_{td}$ in both open and closed systems (Figs. 28(a)-28(b)). For symmetrical electrolytes with equal valences $z = z_+ = z_-$, the thermopower of an open system $\tilde{S}_{td}$ and a closed system $S_{td}$ can be written as:

$$\tilde{S}_{td} = \frac{1}{e} \frac{S_+^* \mu_+ - S_-^* \mu_- - f_\psi(S_+^* \mu_+ + S_-^* \mu_-)}{\mu_+ + \mu_- - f_\psi(\mu_+ - \mu_-)}, \quad (4.17)$$

$$S_{td} = \frac{1}{2e}\left[(S_+^* - S_-^*) - f_\psi(S_+^* + S_-^*)\right], \quad (4.18)$$

where $H$ is the gap of the nanochannel, the tilde symbol $\tilde{S}_{td}$ denotes the thermopower in an open system. The coefficient $f_\psi = \int_{-H/2}^{H/2} \sinh(z\Psi)\,dy / \int_{-H/2}^{H/2} \cosh(z\Psi)\,dy$ is determined by the dimensionless EDL potential $\Psi = e\psi/k_B T$, where $\psi$ denotes the electric potential in EDL. The sign of $f_\psi$ is the same as the sign of surface charges. As shown in Figs. 28(b)-28(c), a more negative surface potential $\psi_0$ and decreasing channel widths $H$ result in a more unipolar ion transport and an increased p-type thermopower. Interestingly, whether the thermopower of polyelectrolytes can be enhanced depends on whether the system is open or closed. For the Na-PEO polyelectrolytes, the mobility of the cations is much larger than the anions $\mu_+ \gg \mu_-$. In an open channel allowing ion-exchange with the reservoirs, the contribution to thermopower from the poly-anions can be negligible because $\mu_- \approx 0$, and Eq. (4.17) is reduced to $\tilde{S}_{td} \approx S_+^*/e$, independent of the EDL factor $f_\psi$. This means that nanoconfinement cannot improve the thermopower of polyelectrolytes in an open channel. For closed systems, however, the overlapping effect of EDLs can effectively enhance thermopower (Figs. 28(e)-28(f)), which is consistent with observations of large thermal voltage measured from the nanocellulose capacitors.[289] While the above modeling treats ions as point charges, recent work by Zhang et al.[306] showed that ion steric effects due to the finite size of ions can couple with the nanoconfinement effect and induce a large enhancement in $S_{td}$ by nearly one order of magnitude compared to the case without ion steric effects, pointing new directions for achieving giant thermopower in ionic liquids or concentrated electrolytes.

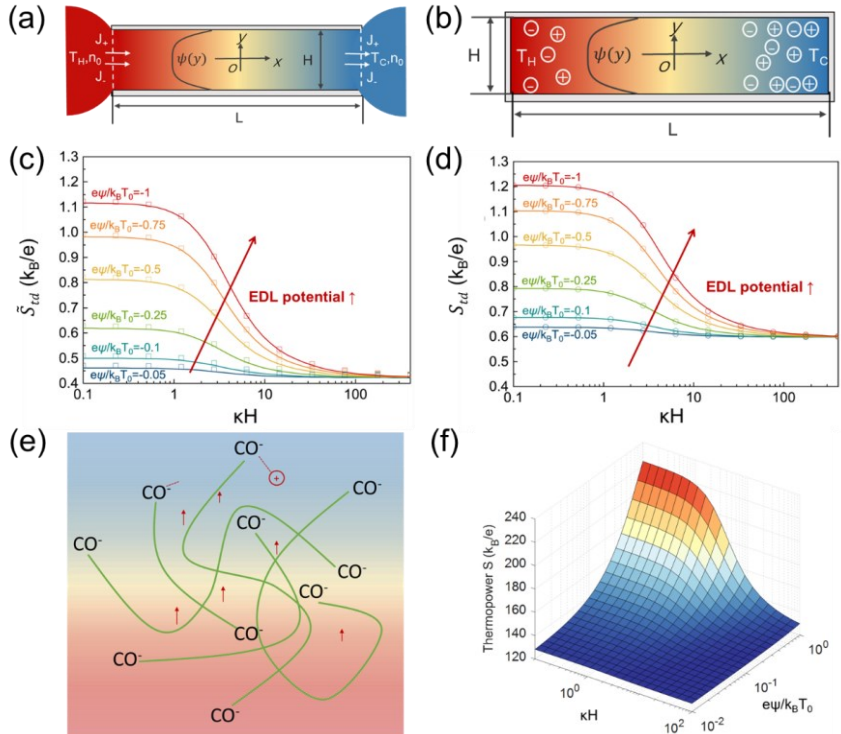

FIG. 28. Nanoconfinement effect on thermodiffusion. Schematic of nanoconfinement in (a) open channel and (b) closed capacitor systems. (c-d) Dependence of thermopower on the potential of electric double layer (EDL) and channel widths of (c) closed capacitor and (d) open channel systems. (e) Schematic of Na-PEO polyelectrolyte with mobile $Na^+$ and immobile $PEO^-$ poly-ions, and (f) Thermopower in a closed system with various EDL potentials and channel widths. Reprinted from Ref. 290[290] Copyright 2022, with permission from Elsevier.

**3. Synergistic thermodiffusion-thermogalvanic effects**

In any non-isothermal electrochemical cell, the redox ions can migrate, driven by the temperature gradient and induce an extra thermal voltage in addition to the contribution by the redox entropy, $\Delta S_{rxn} = S_R - S_O$, as illustrated in Eq. (4.10). Such synergistic thermodiffusion-thermogalvanic effect recently gained intensive interests,[293,307] which can be dated back to theoretical work Agar as early as 1957.[308] It has been well-received that one must correct measured thermal voltages by subtracting the contribution from the temperature dependence of the electrodes[270], namely the thermogalvanic effect. However, this synergistic effect has been largely neglected by researchers in ionic thermoelectrics, because the thermodiffusion contribution to the effective thermopower in liquid thermogalvanic cells is usually negligible compared with that from redox entropies. Until recently, Han and Qian *et al.*[9] discovered that both the thermodiffusion from redox-inert ions and the thermogalvanic effect of redox ions contribute to the total thermopower in ionic gelatin. The synergistic thermodiffusion-thermogalvanic effect is illustrated in Fig. 29(a). The thermodiffusion of redox-inert ions results in an accumulation of net charges at the hot and cold electrode surfaces, and an internal electric field is induced. For a *p*-type electrolyte, the electric field points from the cold side to the hot side. On the other hand, the thermogalvanic effect generates a voltage due to the shift of electrochemical potential in response to the temperature. Fig. 29(b) shows the thermogalvanic effect of the *p*-type $Fe(CN)_6^{3-}/Fe(CN)_6^{4-}$ pair with negative redox entropy change $\Delta S_{rxn} < 0$. Such electrochemical potential difference is in the same direction as the thermodiffusion effect of *p*-type electrolytes. Therefore, if the electrolyte system contains both the redox pair and the thermodiffusive ions with the same direction of thermal voltage, then they will synergistically contribute to the total thermopower and generate a large thermal voltage. Han and Qian *et al.*[281] realized such a synergistic effect and observed a giant thermopower of 17 mV/K in ionic gelatin containing both redox active $Fe(CN)_6^{4-}/Fe(CN)_6^{3-}$ and thermodiffusive KCl. The giant thermopower enabled the generation of a voltage as high as 2 V from body heat using only 25 i-TE unipolar legs serially connected in an S-shape. Later, Li *et al.*[307] proposed an aqueous electrolyte-based asymmetrical TIC-TGC device using $Zn(CF_3SO_3)_2$ electrolyte, with Zn as the anode but $VO_2$ as the cathode (Fig. 29(b)). The thermopower of the asymmetric hybrid TIC-TGC device achieved a high thermopower of 12.5 mV/K and an efficiency of 7.25% relative to the Carnot limit.

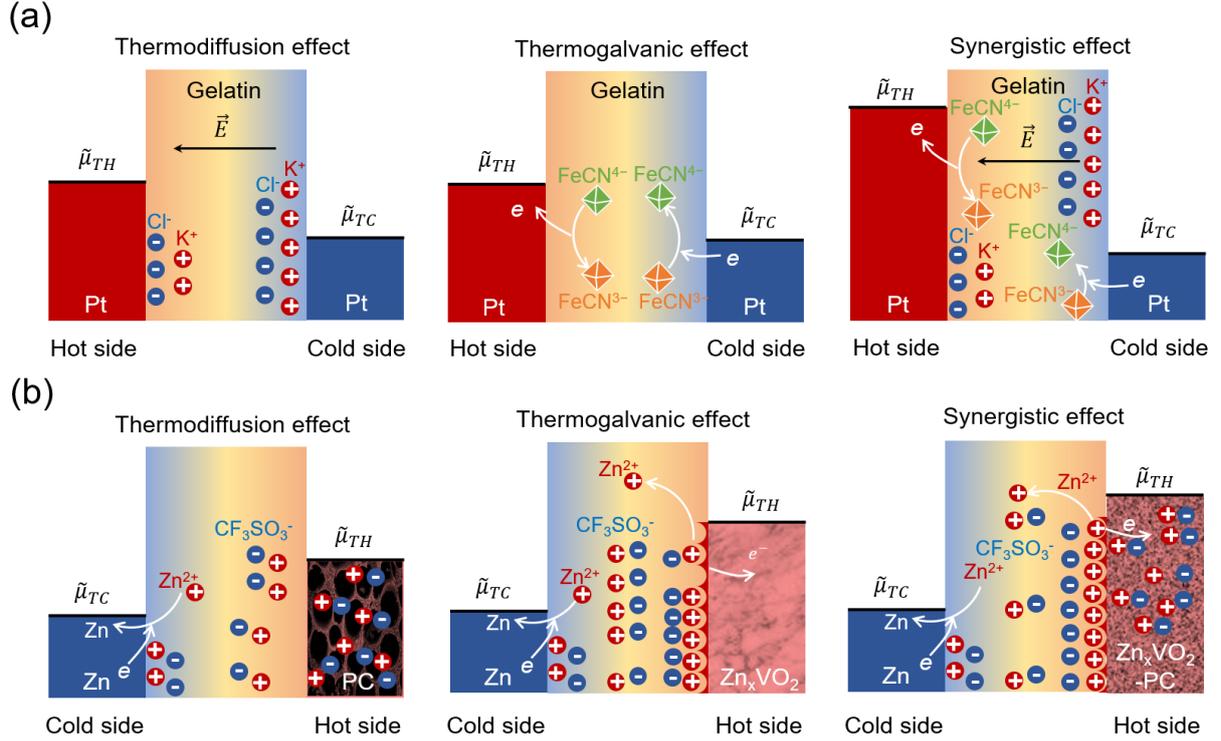

FIG. 29. Hybrid TIC/TGC devices. (a) Synergistic contributions to the thermopower due to the thermodiffusion of $K^+$ and $Cl^-$ and the thermogalvanic effect of the redox pair $Fe(CN)_6^{3-}/Fe(CN)_6^{4-}$ (denoted as $FeCN^{3-/4-}$).[293] (b) Asymmetric TIC/TGC device, with $Zn^{2+}/Zn$ as the anode, $Zn^{2+}/Zn_xVO_2$ as the cathode, and $Zn(CF_3SO_3)$ as the ion provider for thermodiffusion effect,[307] with PC denoting Porous Carbon. Reapated from Ref. 10,[10] Copyright 2024 by American Chemical Society.

## C. Manipulating thermo-osmotic flow

Despite the thermo-osmosis was discovered long ago, it was only until the recent decade that research interests in manipulating thermo-osmotic flow reemerged due to its important role in a wide range of engineering processes and technologies, such as manipulation of nano-object[309] and nanofluids,[310] the thermo-osmotic energy conversion heat recover,[311,312] membrane desalination,[268] fuel cell operation,[313] just to name a few. This section summarizes recent advances in understanding and controlling thermo-osmotic flows through tuning surface hydrophilicity, leveraging the size effects of nanochannels, and the charge regulation effects.

### 1. Effects of surface hydrophilicity

As an interfacial cross-transport phenomenon[284], thermo-osmosis is sensitive to the surface chemistry of the confining surface. For a non-wetting surface with low liquid-solid interaction energy, the excess enthalpy is found to be positive everywhere across the boundary layer (Fig. 30(a)-30(b)). As a result, the obtained thermo-osmotic coefficient obtained by integrating excess enthalpy across the boundary layer is positive, indicating the flow moves towards the cold side (thermophobic), consistent with experiments in hydrophobic membranes from different vendors such as Fluoropore, Hipore, and Yumicron.[314] The large thermo-osmotic coefficient is also found to be largely affected by the dimensionless slip length $b/L$, with $b$ the slip layer thickness and $L$ is characteristic length of thermo-osmotic flow field, defined as $L = \int_0^\infty (z - z_s)\delta h(z)dz / \int_0^\infty \delta h(z)dz$. On hydrophilic surfaces with large $\varepsilon_{ls}$, the excess enthalpy shows a complex oscillatory behavior away from the surface, resulting in a small integral $\int z\delta h(z)dz$ which is proportional to $M_{vq}$. With strong liquid-solid interaction energy, the thermo-osmotic coefficient $M_{vq}$ becomes negative, indicating a thermophilic behavior. The large amplification of thermo-osmosis on hydrophobic surfaces is later confirmed by tracking the drift velocity of fluorescent polystyrene nanoparticles in a microchannel coated with hydrophobic polymers[315] (Fig. 30 (c)-30 (e)). The nanoparticles in the non-isothermal microchannel are simultaneously driven by the thermophoretic velocity, $u_T$, and the thermo-osmotic creeping velocity $v(z)$. In comparison to the thermo-osmotic drift velocity above a hydrophilic surface, the hydrophobic surface has an extra contribution from the intrinsic slip flow $v_s^*$ caused by the hydrophobic polymers. Such effect is manifested in the large $v(z)$ with small distance $z$ from the surface with hydrophobic coatings.

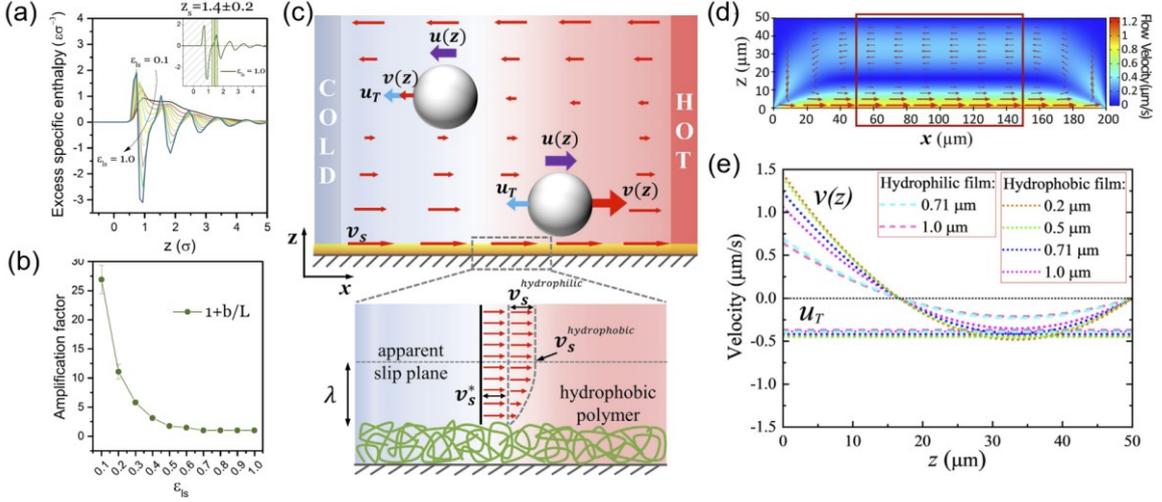

FIG. 30. (a) Excess enthalpy distribution $\delta h(z)$ affected by the liquid-solid interaction energy $\varepsilon_{ls}$. (b) The amplification factor of the thermo-osmotic coefficient compared with the no-slip limit $M_{vq}/M_{vq,b=0}$. (a-b) is repeated from Ref. 19[19] Copyright 2017 by APS. (c) Thermo-osmotic flow and the motion of nanoparticles in the non-isothermal microchannel. The bottom panel shows the velocity profile due to the thermo-osmosis above a hydrophobic surface. (d) The overall creep flow field in the microchannel. (e) Thermophoretic velocity and the thermo-osmotic flow velocity profile. (c-e) repeated from Ref. 315 [315], Copyright 2023 by Elsevier.

### 2. Size effects of nanoconfinement

The channel widths are found to have a complicated influence on the thermo-osmotic slip velocity. As shown in Fig. 31(a)-31(b), Fu *et al*. investigated the size effect of thermo-osmotic flow velocity across a CNT using MD simulations.[316] The thermo-osmotic flow velocity oscillates and can even change signs with the CNT radius below ~ 6 Å. In a confined channel, the thermo-osmotic coefficient is expressed as an integral of excess enthalpy over the cross-section:

$$J_v = \frac{1}{\mu}\int_0^R 2\pi r \frac{\delta h(r)}{\bar{T}}\left(-\frac{\partial T}{\partial x}\right)dr, \quad (4.19)$$

where $R$ is the CNT radius. In the limit of long slipping length, the thermo-osmotic flow velocity is determined by the averaged excess enthalpy $\overline{\delta h}$ in the channel:

$$u_s = \frac{R\overline{\delta h}}{2\lambda_f L + \pi\mu C}\frac{\Delta T}{\bar{T}}\left(1 - \frac{\bar{T}\Delta\Pi}{\Delta T\overline{\delta h}}\right), \quad (4.20)$$

where $\lambda_f$ is the friction coefficient of the liquid-solid interface, $C = \Delta p R/(\pi \mu u_s)$ is a constant determined by hydrodynamic boundary conditions. $\Delta \Pi$ is the osmotic pressure difference between the two reservoirs. The averaged excess enthalpy is calculated as an integration over the cross-section of the thermo-osmotic channel, corrected by the slipping length $b$:

$$\overline{\delta h} = \frac{1}{\pi R^2} \int_0^R \delta h(r) \cdot \left(1 - \frac{r^2}{R^2 + 2bR}\right) 2\pi r dr. \quad (4.21)$$

As shown in Fig. 31(c), the boundary layers would overlap and interfere when the CNT radius is smaller than 6 Å, resulting in large negative dips in $\delta h(r)$ and thereby a negative flow velocity (thermophilic). For large CNT radius, such overlapping effect of excess enthalpy becomes negligible, and the flow velocity is reduced to the case of flow above a single confining surface. According to a more detailed analysis of water structure by Chen et. al.,[317] the spatial probabilistic distribution of water molecules in the second solvation shell is different from bulk water under nanoconfinement when channel widths lower than 1.5 nm. Such distinct water structure is responsible for the excess enthalpy oscillation near the surfaces.

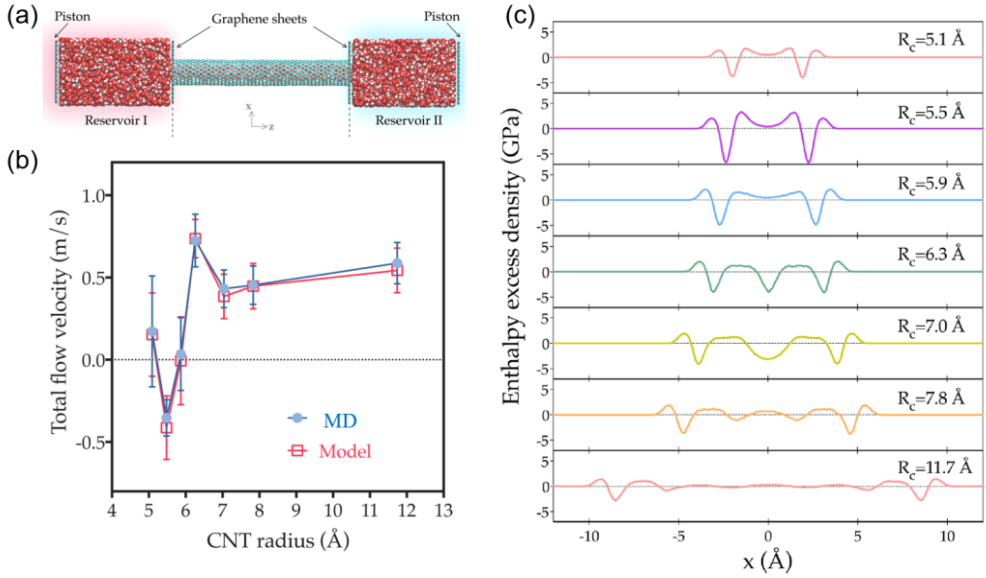

FIG. 31. Size effect of thermo-osmotic flow velocity. (a) Thermo-osmotic flow across a CNT. (b) dependence of flow velocities on the radius of CNTs. (c) Excess enthalpy density across the CNT. Readapted from Ref. 316[316], Copyright 2018 by American Chemical Society (ACS).

### 3. Charge regulation of thermo-osmosis

In practical applications such as desalination and thermo-osmotic energy conversions, one has to consider the effects of ionic charges and the electric double layer (EDL) at the liquid-solid interfaces, as shown in Fig. 32(a). Near the charged interface, the electrostatic field contributes to extra excess enthalpy: $\delta h_{el}(z) = \rho_e(z)\phi(z) + \frac{\epsilon}{2}|\nabla\phi|^2$, where $\rho_e$ is the local charge density, $\phi$ is the electrostatic potential in the EDL, and $\epsilon$ is the dielectric constant of the solution.[318] Note that such excess enthalpy is a combined result of dielectric saturation, interfacial water structuring, and excess stresses due to electrostatic interaction.[319] As a result, the thermo-osmotic coefficient $M_{to}$ is separated into water contribution $M_{to}^{wat}$ and the electrostatic contribution $M_{to}^{el*}$. The sign of $M_{to}^{wat}$ is strongly determined by surface wetting (Fig. 32(b)), while the electrostatic term $M_{to}^{el*}$ is affected by both the surface charge density $\Sigma$ and the Debye length. In hydrophobic cases with contact angle $\theta > 110°$ and a large effective slipping length $b_{eff} = b - z_s$, the thermo-osmosis is dominated by $M_{to}^{wat}$, while for the less hydrophobic case with $\theta \sim 110°$, electrostatic contribution $M_{to}^{el*}$ can dominate at large Debye lengths $\lambda_D$, which is understandable because a large $\lambda_D$ means a slow decay in the electrostatic field close to the boundaries. For the intermediate wetting case, there is a sign change from a thermophobic behavior ($M_{to} > 0$) to a thermophilic behavior ($M_{to} < 0$), corresponding to the dips Fig. 32(d).

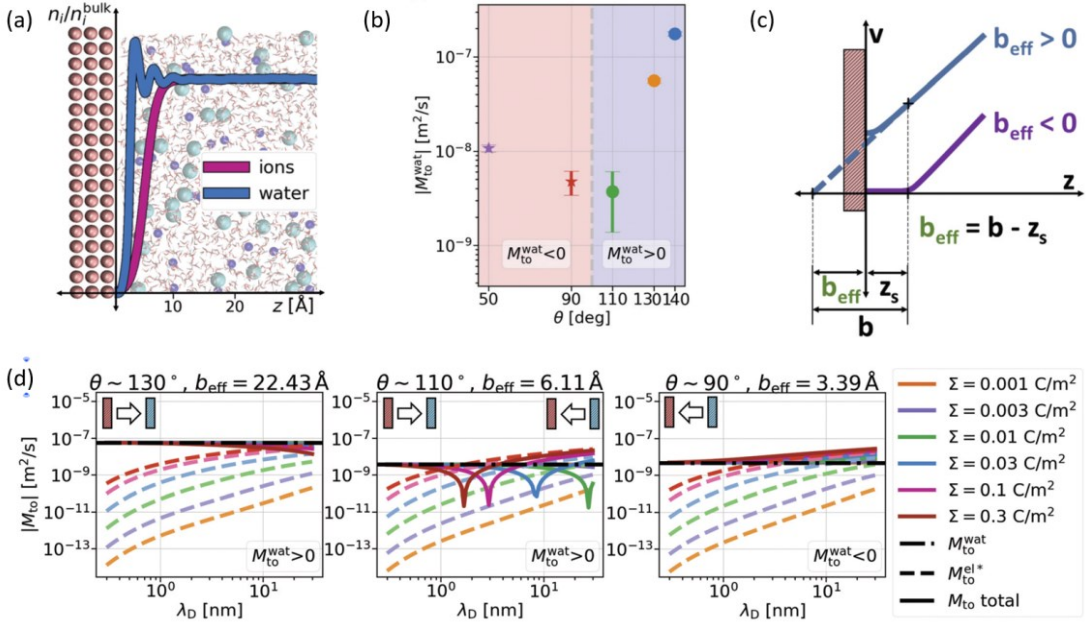

FIG. 32. (a) Schematic of ions and water density at the liquid-solid interface. (b) Water contribution to the osmotic coefficient $M_{to}^{wat}$ as a function of contact angle. (c) Illustration of the effective slip length $b_{eff}$. (d) Thermo-osmotic coefficient $M_{to}$ (solid lines) as a function of the Debye length $\lambda_D$ for different surface wettability (surface angle $\theta$ and surface charges $\Sigma$). The dips in solid lines indicate the change in the thermo-osmotic flow direction. Replot from Ref. 318 [318], with permission from Royal Society of Chemistry.

In addition to using electrolytes to regulate the ion transport in thermos-osmosis, light can also actively trigger and manipulate thermo-osmotic flows. An amplified thermo-osmotic flow is experimentally observed by Yeom et al.[320] in nanosheets with a photothermal responsive ionic channel composed of plasmonic gold nanostars, cellulose nanofibers, and MXene flakes. As illustrated in FIG. 33, a nonuniform temperature profile can be effectively activated under near-infrared illumination, due to the plasmonic resonance effect of gold nanostars and MXene. Such local temperature rise induces a thermo-osmotic flow, carrying the mobile ionic charges across the channel. Such ionic current is amplified by the selective cation transport in the channel, due to the negative charges of the surface functional groups on MXene and cellulose nanofibers. Such demonstration can have promising applications in nanofluidic circuits as photo-ionic switches.

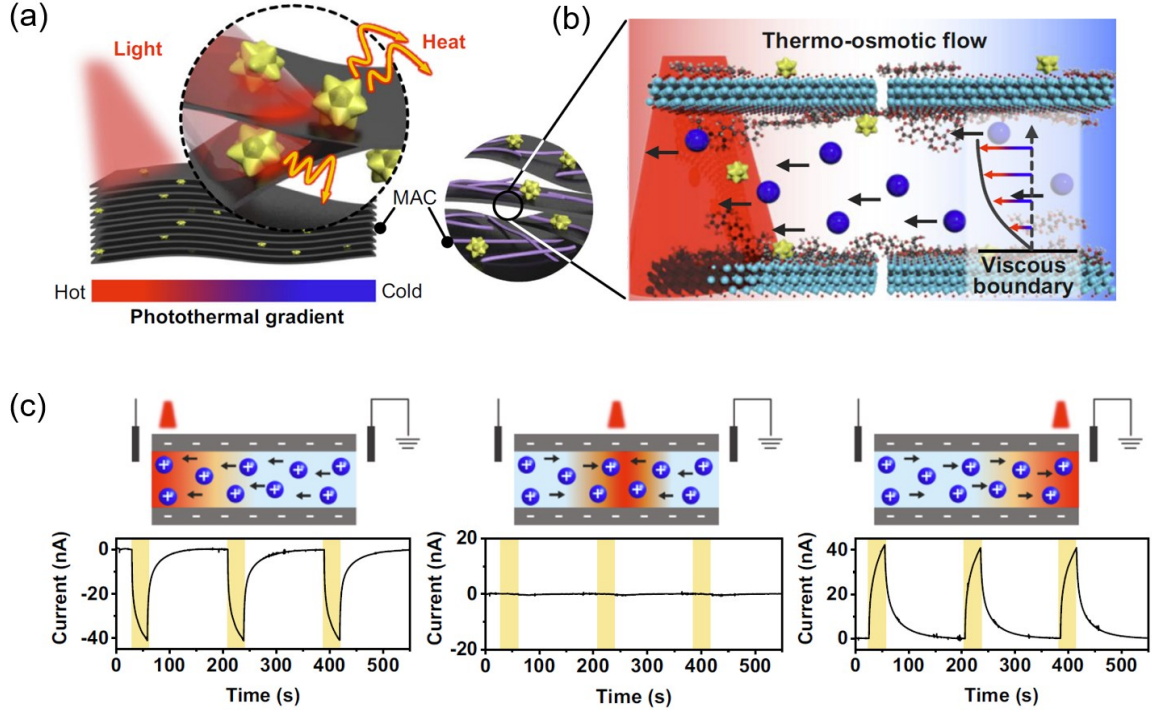

FIG. 33. (a) Photo-induced temperature gradient and (b) thermo-osmotic flow and ion transport. (c) Ionic current measured with optical-heating on the left, center, and right of the ionic channel. Replot from Ref. 320[320] with permission from Springer Nature.

### D. Remaining challenges and outlooks

#### 1. Theoretical prediction of ionic thermopower

Although great progress has been made in developing ionic thermoelectric materials and devices, there remain several fundamental questions to be resolved. First, methods of predicting the ionic Seebeck coefficients are in urgent need for the development of high-thermopower i-TE materials, but remain highly challenging due to the complicated and long-range feature of ion-ion and ion-solvent interactions. Although MD simulations have been recently applied to predict $S_{tg}$ of $Fe^{3+}/Fe^{2+}$ and $Fe(CN)_6^{3-}/Fe(CN)_6^{4-}$ redox pairs, it is only applicable to a dilute solution of simple redox pairs with negligible Eastman entropy of transfers, and cannot predict thermopower of intercalation or plating electrodes. In complex electrolytes such as polyelectrolytes, ionic liquids, and gels, clarifying the relation between ion-matrix interaction and $S_{td}$ remains an open question. In addition to the thermopower, the coupling between ionic thermopower $S_{tg}$ or $S_{td}$ and ionic conductivity/diffusivity needs further study. For example, polyelectrolytes or ionogels show giant $S_{td}$ typically at the sacrifice of ion transport properties due to the increased hopping barrier. For TGCs, the $S_{tg}$ can also couple with the specific conductivities of different ion species, because regulating the solvation structure of redox species might also affect the effective ionic radii or introduce ionic couplings, which could affect the ion transport.

#### 2. Rational design of i-TE devices

Current research mostly focuses on improving the thermopower of either *n*-type or *p*-type materials. However, the overall performance of an i-TE pair depends on the thermopower matching between *n*-type and *p*-type materials. For example, in contrast to the significant advances in *p*-type TGCs, such as the $Fe(CN)_6^{4-}/Fe(CN)_6^{3-}$ system, their *n*-type counterparts with higher effective thermopower needs to be developed. On the device side, there are massive opportunities in making hybrid i-TE devices with multi-functionality, such as hybrid i-TE and solid-state TE devices to simultaneously achieve high voltage and continuous power output, [321] flow TGC devices for simultaneous heat harvesting and cooling purposes,[322-324] integrated TGC with reverse electrodialysis devices for co-generation of fresh water and electric power,[267] and TGC embedded with photocatalysts for hydrogen-electricity cogeneration.[325] Finally, heat transfer between the i-TE device and the heat reservoirs plays a pivotal role in improving the overall energy conversion efficiency. Improving the coupling between the i-TE device with the heat reservoirs is crucial.

## 3. Characterization of microscopic thermo-osmotic flow

For the transport of solvent molecules through thermo-osmosis, many recent advances in manipulating thermo-osmotic flow rely on theoretical modeling and molecular dynamics simulations of simple electrolytes across nanochannels. In contrast, the majority of experiments of thermo-osmosis were performed across a porous membrane by measuring the thermo-osmotic pressure $\Delta p$ at a constant temperature difference,[326-329] which cannot provide insights into the microscopic flow fields. It was not until the recent decade that the thermo-osmotic flow was probed at the microscale by tracking the thermophoresis of plasmonic nanoparticles under optothermal heating.[330] However, characterization of thermo-osmotic flow confined in nano-slits is still absent, despite it being the most studied configuration in theoretical simulations. As a result, it remains challenging to directly compare the modeling results with the experimental results. For example, MD simulations predicted that the higher electrolyte concentration can generally enhance the thermo-osmotic flux, but collodion membranes show no concentration dependence on KCl below 0.1 mol/kg.[331-333] It is also hard to explain the reasons why the direction of thermo-osmosis across cation exchange membranes depends on the cation types. For example, Suzuki et al.[334,335] observed that the thermo-osmotic flux of $H^+$ and $Na^+$ electrolytes in hydrocarbon-sulfonic acid-type cation-exchange membranes was from the cold to the hot side, whereas the direction was from the hot side to the cold side with the $Li^+$ and $K^+$. To resolve these questions, on one hand, modeling efforts towards more realistic configurations and different types of thermo-osmotic electrolytes are necessary; quantitatively controlling the pore sizes, and surface polarizations of nanochannels or membranes on the experimental side is also important. The recent development of covalent-organic-frameworks (COFs) with controllable charge density, pore sizes, and ionic selectivities[336] can offer a playground for testifying theoretical and computational studies.

## 4. Optimization of thermo-osmotic energy conversion devices

Besides the fundamental questions, developing advanced membranes for thermos-osmotic energy conversion (TOEC) also faces challenges in balancing the trade-offs between high thermo-osmotic flux and other requirements. For example, TOEC can be performed by applying a hydraulic pressure on the cold electrolyte side, such that the thermo-osmotic vapor flow driven by temperature difference across the membrane becomes pressurized and can output extra power through turbines. This method is referred to as pressure retarded membrane distillation (PRMD). PRMD membranes face challenges in simultaneously achieving high vapor transport flux, a large liquid entry pressure to prevent membrane wetting, and a low thermal conductivity to maintain the temperature difference.[337] A smart solution to mitigate these trade-offs is to fabricate membranes with asymmetrical pore structure,[338] in which a thin active layer with small pore sizes can maintain the non-wetting state and the thermo-osmotic vapor flux, and a thick supporting layer with large pore size for effective mass transfer and thermal insulation. Alternatively, the temperature difference can be directly applied across a reverse electrodialysis (RED) device which harvests the concentration-induced Nernstian potential.[339] The temperature difference further increases the osmotic pressure difference, such that the power output of the non-isothermal RED is enhanced. The synergistic and coupling effects among the thermo-osmotic flow of solvents, thermodiffusion, and electrophoresis of ionic species[340] remains largely unexplored, which is essential for the rational design of the nano-porous structure for simultaneously achieving a high ion selectivity, a large thermo-osmotic flux, and a high power output.

## V. SUMMARY AND OUTLOOK

In summary, recent advances in multi-carrier thermal transport are reviewed in this paper, covering coupled transport of phonons, electrons, spin, ions, and molecules and the multi-carrier interaction mechanisms. EPI is crucial for managing heat generation and dissipation in electronics and the development of high-performance thermoelectric energy conversion devices, which can be tuned by engineering band structures, dopants, polarity, *etc*. EPI causes scattering events of electrons and phonons, which can be tailored for optimizing the thermal, electrical, and thermoelectric transport coefficients. In the strong-coupling regime, however, EPI can no longer be simply studied using perturbation methods, and would result in renormalized band structures of phonons and electrons, and even exotic solid-state phase changes. In addition, most work on EPI is conducted near the thermal equilibrium regime, and the nonequilibrium electron-phonon thermalization remains elusive, despite its significance in heat generation in electronic devices. Spin-phonon interaction (SPI) underpins the developments of spintronics, thermomagnetic energy conversion, and the management of heat generation in memory devices. SPI results in novel transport phenomena such as spin Seebeck and Nernst effects, where temperature gradient can induce nonequilibrium spin currents parallel or perpendicular to the temperature gradients. Since spin can be carried by both electrons and magnons, it remains challenging to distinguish the contributions from different spin carriers in magnetic semiconductors. In addition, the exchange of angular momentum between spin and phonons can be a fruitful future direction, which provides the possibility of manipulating spin using heat. The coupled transport of heat and ions/molecules is of great importance in energy conversion and storage water production and desalination, manipulation of colloidal particles, isotope enrichment, radioactive waste disposal, and so on. Recently, the ionic Seebeck effect, thermogalvnaic effect, and thermo-osmosis effects have gained intensive research interest due to the promising applications in low-grade heat harvesting, motivated by the global climate issue. In the ionic Seebeck effect and thermogalvanic effects, the temperature gradient introduces changes in electric field and electrode potential changes, respectively. While the former can take place in bulk electrolytes, the thermogalvanic effect is only measurable using electrodes. The thermo-osmotic effect, on the other hand, is an interfacial phenomenon where the temperature gradient drives the hydrodynamic flow of liquid near the surface. Engineering the solvent structure, ion-solvent interaction strengths, and nanoconfinements are effective methods for tuning the coupled transport of heat and ion/molecules. Uncovering the intricate multi-species interactions within liquids and at the electrode/electrolyte interfaces remains to be open questions, which requires future developments in both advanced modeling methods and characterization techniques. Finally, implementing multi-carrier modeling at multiscale is a grand challenge but a significant future direction. Most current studies of multi-carrier transport are still focused on unveiling fundamental physics in materials. Integrating the microscopic understanding of multi-carrier transport into the design and optimization of electronic, memory, and energy conversion devices is a fruitful future direction.

**ACKNOWLEDGEMENT**
R.Y. acknowledges the financial support of the National Key Research under Grant No. 2022YFB3803900. T.H.L., J.Z., and X.Q. acknowledge the financial support of the National Natural Science Foundation of China under Grant No. 52076089, 52350088, and 52276065. T.H.L. acknowledges the financial support of

**Nomenclature**

$\hat{a}^{\dagger}_{\mathbf{q}\nu}$ ($\hat{a}_{\mathbf{q}\nu}$) phonon creation (annihilation) operator

$a_{\mathbf{q}\lambda}(a^{\dagger}_{-\mathbf{q}\lambda})$ phonon annihilation (creation) operator

$\alpha_{\mathbf{k}}, \beta_{\mathbf{k}}$ ($\alpha^{\dagger}_{\mathbf{k}}, \beta^{\dagger}_{\mathbf{k}}$) magnon creation (annihilation) operator

$\alpha_G$ Gilbert damping parameter

$b_{\mathbf{k}}(b^{\dagger}_{\mathbf{k}})$ magnon annihilation (creation) operator

$b$ slip length

$b_{eff}$ effective slipping length

$\hat{c}^{\dagger}_{n\mathbf{k}}$ ($\hat{c}_{n\mathbf{k}}$) electron creation (annihilation) operator

$c$ light speed

$C_T$ heat capacity $\quad\quad\quad\quad\quad\quad\quad\quad\quad\quad C_{\mathbf{q}\nu}$ mode-specific heat capacity

$C_p$ ($C_m$) heat capacity of phonons (magnons)

$D_i$ diffusivity

$e_{\kappa\alpha,\nu}(\mathbf{q})$ normal modes of vibration waves

$\mathbf{e}_\beta$ unit vector along $\beta$-direction

$\mathbf{E}$ electric field $\quad\quad\quad\quad\quad\quad\quad\quad\quad\quad E_y$ electric field along $y$-direction

$\mathcal{E}$ internal electric field

$\epsilon$ dielectric constant

$\varepsilon_{n\mathbf{k}}$ electron energy

$\varepsilon_{ls}$ liquid-solid interaction energy

$f_{n\mathbf{k}}$ electron distribution function

$f_{\alpha\mathbf{k}}$ nonequilibrium electron distribution functions

$f^0_{\uparrow(\downarrow)}(\mathbf{k})$ Fermi-Dirac distribution function

$F$ Faraday constant

$f$ generalized force

$f(\epsilon)$ function of dielectric constant

$g_{mn\nu}(\mathbf{k},\mathbf{q})$ EPI matrix element $\quad\quad\quad\quad\quad g_{mp}$ magnon-phonon coupling constant

$g_i$ local free energy density

$G_{ep,i}$ electron-phonon energy coupling constant

$\hbar$ reduced Planck constant $\quad\quad\quad\quad\quad\quad H$ magnetic field

$H$ gap of the nanochannel in Eq. (4.18)

$h, h_b$ specific enthalpy $\quad\quad\quad\quad\quad\quad\quad h(z)$ enthalpy density distribution

$\mathbf{J}_c, J_c$ charge current $\quad\quad\quad \mathbf{J}_s, J_s$ spin current $\quad\quad\quad J_o$ orbital current

$\mathbf{j}_{\uparrow,\downarrow}$ particle current $\quad\quad\quad J_m$ spin current density

$J_L$ angular momentum current

$J_{ij}$ exchange parameter

$\mathbf{J}_i$ chemical flux of species $i$

$J_v$ volumetric flow rate

$\mathbf{J}_n$ chemical flux

$J_{mass}$ mass flow

$\kappa_e$ electronic thermal conductivity $\quad\quad\quad\quad \kappa_p$ lattice thermal conductivity

$\kappa_m$ magnonic thermal conductivity $\quad\quad\quad\quad \kappa_{\text{eff}}$ effective thermal conductivity

$\kappa_H$ temperature-dependence of thermal conductivity

$\kappa$ thermal conductivity

$l_{\mathbf{q}\nu}$ zero-point displacement amplitude

$L_0$ Lorenz number

$l_{mp}, \lambda_{mp}$ characteristic length of magnon-phonon interaction

$l_m$ magnon mean free path

$l_p$ phonon mean free path

$l_{pp}$ mean free path of phonon-phonon scattering

$l_{pt}$ mean free path of phonon-impurity, phonon-boundary and phonon-electron scatterings

$l_{pm}$ mean free path of magnon-phonon scattering

$L_{ii}, L_{iq}, L_{qi}, L_{qq}$ linear transport coefficients

$L$ Onsager coefficient

$M_0, M_\kappa$ nucleus mass

$m_e$ electron mass

$M_{vv}, M_{qq}, M_{vq}, M_{qv}$ Onsager transport coefficients

$M_{to}$ thermo-osmotic coefficient

$M_{to}^{wat}$ water contribution to thermo-osmotic coefficient

$M_{to}^{el*}$ electrostatic contribution to thermo-osmotic coefficient

$N_\mathrm{p}$ number of unit cells

$N_\uparrow(\varepsilon), N_\downarrow(\varepsilon)$ spin-dependent density-of-states

$n$ electron transfer number

$n_c$ nonuniform electron density

$n_s$ spin density

$n_{ion}$ ion density

$\omega_{\mathbf{q}\nu}$ lattice vibration frequency

$\omega_c$ characteristic phonon frequency

$\omega_{int}$ phonon frequency at dispersion intersection

$\Omega_{\mathrm{BZ}}$ volume of the first Brillouin zone

$P$ spin polarization

$p(z)$ pressure

$\mathbf{Q}$ heat current

$\mathbf{Q}_m$ heat current density

$Q_i^*$ heat of transport

$\mathbf{R}_i, \mathbf{R}_j$ lattice site

$R_i$ effective radius of the ion

$R$ CNT radius

$S$ charge Seebeck coefficient

$S_{\uparrow,\downarrow}, S_{xy}, S_e$ spin Seebeck coefficient

$S_m$ bulk spin Seebeck coefficient

$S_{td}$ ionic Seebeck coefficient

$\mathbf{S}_i, \mathbf{S}_j$ spin

$S_i^*$ Eastman entropy of transfer

$S_O, S_R$ partial molar entropy of the species

$s$ specific entropy of the fluid in the boundary layer

$s_b$ specific entropy in the bulk fluid

$\mathbb{S}$ atomic virial stress tensor

$S_{tg}$ thermopower

$\tilde{S}_{td}(S_{td})$ thermopower of an open (closed) system

$\sigma_{\uparrow,\downarrow}$ spin-dependent electrical conductivity

$\sigma_m$ magnon spin conductivity

$T_m$ temperature of magnon

$T_p$ temperature of phonon

$T_c$ Cuire temperature

$T_N$ Néel temperature

$\tau_{\kappa\alpha}$ nuclear displacement

$\tau_{\mathbf{q}\nu}^{ep}$ phonon lifetime

$\tau$ magnon relaxation time

$\tau_\uparrow(\varepsilon), \tau_\downarrow(\varepsilon)$ spin-dependent relaxation time

$\tau_{mp}$ magnon-phonon relaxation time

$\tau_{mp}^{(2)}, \tau_{mp}^{(3)}, \tau_{mp}^{(4)}$ relaxation time of magnon-phonon scattering

$\tau_{el}$ relaxation time of magnon-impurity scattering and magnon-boundary scattering

$\tau_{p-p}^{-1}(\tau_{p-b\&i}^{-1})$ scattering rate of phonon-phonon (phonon-boundary and phonon impurity) scattering

$\theta_{SH}$ spin Hall angle   $\theta_{SN}$ spin Nerst angle   $\theta$ contact angle

**u** atomic displacement   $u_{n\mathbf{k}}$, $u_{m\mathbf{k+q}}$ Bloch-periodic waves

$\boldsymbol{u}_s$ hydrodynamic velocity of thermo-osmotic flow

$\langle u(z) \rangle$ averaged atomic energy

$u_T$ thermophoretic velocity

$V^{\mathrm{KS}}$ KS effective potential   $V^{\mathrm{en}}$ nuclear potential

$V^{xc}$ exchange and correlation potential   $V^{\mathrm{H}}$ Hartree electronic screening

$v_i$ volume occupied by chemical species $i$

$\mu_{ion}$ electrochemical potential   $\mu_i$ chemical potential

$\mu_s$ spin dependent chemical potential

$\mu_m$ nonequilibrium magnon spin accumulation

$v_{\mathbf{q}\nu}$ group velocity

$v_m$ magnon group velocity   $v_p$ phonon group velocity

$v(z)$ thermo-osmotic creeping velocity

$w_\pm$ weighting coefficients

$z_\pm$ valance charges

$z_i$ ionic valence

$z_s$ position of the shearing plane

$\phi$ electrostatic potential

$\Psi$ dimensionless EDL potential

$\psi_0$ surface potential

$\lambda_D$ Debye length

$\lambda_f$ friction coefficient of the liquid-solid interface

$\rho$ density   $\rho_e$ local charge density

$\Delta\Pi$ osmotic pressure difference between the two reservoirs

$\Sigma$ surface charge density

$\Lambda$ magnon de Broglie wavelength

$\Delta_{\mathbf{q}\nu} v^{\mathrm{KS}}$ lattice-periodic function

$\nabla_x T$ temperature gradient along $x$-direction

$\nabla_T \bar{\mu}_i$ gradient of electrochemical potential at constant temperature

$\Xi_{\mathrm{ADP}}$ acoustic deformation potential

$\Pi$ phonon self-energy operator

$\mathcal{T}_\uparrow$, $\mathcal{T}_\downarrow$ spin-dependent electron transmission probability